\documentclass[UTF8,a4paper]{article}
\usepackage{bbold}
\usepackage{booktabs}
\usepackage{amsmath}
\usepackage{ tipa }
\usepackage{graphicx}
\usepackage{subfigure}
\usepackage{epstopdf}
\usepackage{geometry}
\usepackage{stackengine}

\usepackage{ulem}
\usepackage{indentfirst}
\usepackage{amsfonts,amssymb,dsfont}
\usepackage{setspace}
\usepackage[hidelinks]{hyperref}
\usepackage{threeparttable}
\usepackage{amsmath,mathtools,amsthm}
\usepackage{array}
\usepackage{extarrows}
\usepackage{upgreek}

\makeatletter
\renewcommand*\env@matrix[1][\arraystretch]{%
	\edef\arraystretch{#1}%
	\hskip -\arraycolsep
	\let\@ifnextchar\new@ifnextchar
	\array{*\c@MaxMatrixCols c}}
\makeatother
\begin{document}
	
	
	\title{\bf Statistic behaviors of gauge-invariance-dominated 1D chiral current random model}
	\author{Chen-Huan Wu
		\thanks{chenhuanwu1@gmail.com}
	\\College of Physics and Electronic Engineering, Northwest Normal University, Lanzhou 730070, China}

\maketitle
\vspace{-30pt}
\begin{abstract}
	\begin{large} 
		By considering energy flow,
		we construct the one-dimensional (1d) model consisting of the quasiparticles
		caused by asymmetric hopping (in carrier position space) or the complex
		bosonic potential whose varying gradience with a chiral ordering
		plays the role of ingredience of quasiparticles. A bosonic potential can be generated 
		and the chaotic dynamics of chiral excitations after disorder average can be investigated
		in the presence of gauge invariance.
		This feature is also shared by the well-known non-Hermitian systems.
		\\
		\\
		$ Keywords$: Gauge-invariance;
		Gaussian ensemble;
		Wigner-Dyson statistic;
		Many-body localization;
		Chiral system;
		\\
		
	\end{large}

\end{abstract}

\clearpage
\tableofcontents
\clearpage

\begin{large}
	
	\section{Introduction}
	
	The quantum critical behaviors can be found in itinerant electron system due to the quantum fluctuation induced
	quantum phase transition between disordered state and the Fermi liquid state.
	The emergence of Sachdev-Ye-Kitaev (SYK) physics can be observed by the chaotic non-Fermi liquid behaviors
	as well as the conformal symmetry in large-$N$ limit. 
	The SYK model is usually be constructed in zero-dimensional system 
	to removing the momentum-dependence.
	There are also several experimental protocols, for example, 
	using the strong magnetic field\cite{Chen A}
	to obtain the flat Landau levels, by constructing an artificial Kagome-type optical lattice\cite{Wei C},
	using the topological superconducting wire coupled to quantum dot to 
	realizing random Majorana coupling\cite{Chew A},
	and similar implementation has also being discussed in other
	topological flat band systems.
	The generation of maximal thermalization in SYK model
	relys on the randomness of coupling
	where the antisymmetry exchange interactions play the key role in a disordered system.
	Such random couplings has been studied in Thirring model\cite{Berkooz M}, Falicov-Kimball model\cite{Byczuk K},
	random impedence network\cite{Olekhno N A}, and even the Anderson localized integer spin Hall systems\cite{Onoda M}.
	The projection onto the zero-energy states in a flat band is an essential procedure
	during these implementations,
	which connecting the Coulomb phase in real space under perturbations
	to the many-body localization (MBL) protected quantum orders\cite{You Y Z}
	like the spin glass order in flat-band localization system\cite{Chalker J T}.
	
	In conformal perturbation theory,
	the random coupling could be marginally irrelevant in large-$N$ limit
	in a translationally invariant system, e.g., the 1+1 dimensional Thirring model
	in the absence of chirality\cite{Berkooz M}.
	Thus it is important to get further insight into the proble about the random coupling
	regarding their marginal relevance or irrelevance.
	Also, in the presence of replica symmetry
	the dynamics of such random coupling system under a fluctuating U(1)
	gauge field is rarely studied.
	The reason is due to the complexity origin from the gauge fixing restrictions,
	e.g., in Ref.\cite{Patel A A}, the gauge fixing restrictions are being ignored since the
	saddle-point results will not be affected in the large-$N$ limit.
	However, for systems in the absence of replica symmetry or the ones in spin glass order due to the
	strong localization effects,
	the global symmetry will missing and the resulting system can no longer kept invariant.
	For this reason,
	we consider the complex coupling problem in a modified Wishart-SYK
	model which is dominated by a U(1) gauge-field.
	The most outstanding feature is the emergence of three-degrees-of-freedom character.
	We start by considering a conserving current in momentum space
	with a finite energy gradient-induced frustrated hopping,
	which can be treated as an energy current in real space.
	A bosonic potential can be generated 
	and the chaotic dynamics of chiral excitations after disorder average can be invesitigated
	in the presence of gauge invariance.
	
	By considering an energy and particle flow,
	we can constructing the one-dimensional (1d) model consisting of the quasiparticles
	caused by asymmetric hopping (in carrier position space) or the complex
	bosonic potential whose varying gradience with a chiral ordering
	plays the role of of ingredience of quasiparticles.
	This feature is also shared by the well-known non-Hermitian systems\cite{P2}.
	For such a 1d system,
	since all the extended eigenstates will be localized once an infinitesimal disorder 
	is introduced, the statistical behaviors of the system depend more
	on the related (topological) symmetries, instead of the disorder strength.
	As an typical example, the nonlinear response-induced current reads
	\begin{equation} 
		\begin{aligned}
			J_{2\omega}^{z}=\sigma_{z\alpha}(E^{\alpha}_{\omega})^{2},
		\end{aligned}
	\end{equation}
	where $\alpha=(x,y)$ is the planar component of the electric field.
	The effective electric field term can be represented as the gradient of potential in real space, 
	$E(r)=-\nabla \phi(r)$, or in momentum space, $E_{q}=-iq\phi_{q}$,
	where $\phi$ is the scalar potential
	For the latter case, as $q$ is the in-plane component of the electron scattering momentum,
	in long-wavelength limit,
	the summation over $q$ vanishes\cite{Mikhailov S A}.
	Note that for second-order response, the density matrix $\rho$ should be expanded in power of electric field,
	thus we have $\rho\propto (-e\phi(r))^{2}$. This will be important for the following derivations.
	Next we consider the preserved induced current in $z$-direction as a consequence
	of the fermion tunneling in $z$-direction.
	This should be similar to the procedure when considering the screened Coulomb interaction between 
	fermions at different positions,
	where such Coulomb interaction can be of the SYK type by considering that the fermion operators 
	create eigenstates at the corresponding positions
	and then projecting the Coulomb interactions into the highly disordered zero-dimensional system.
	Similarly, we consider the creation fermion operators at positions $z_{1}$ and $z_{2}$ 
	create particles in eigenstates $\psi(z_{1})$ and $\psi(z_{2})$, respectively.
	
	In terms of a 1d chiral quasiparticle system,
	we develop a usefull tool in investigating the many body quantum chaos or ergodicity in large-$N$ limit
	that allows the extension to higher dimensions, and involving more (pseudo) degrees-of-freedom
	that arise after the ensemble average. 
	The numerical evidences for the emergent SYK physics are mainly through the 
	Wigner-Dyson statistic in different ensembles:
	GOE(Gaussian orthogonal ensemble), GUE (Gaussian unitary ensemble) and GSE (Gaussian symplectic ensemble).
	Here we briefly introduce some development on this topic.
	First one is the zero-spacial-dimensional system in non-Fermi liquid picture\cite{Bi Z}
	as well as the Thirring or Gross-Neveu models with non-random interaction term in 1d spacial dimension
	where the left and right moving fermions are exactly diagonal.
	In this case the leading term is quadratic in the four-fermion coupling
	(also the four-point interaction), and the interaction is marginally relevant for one sign of it
	which leads to a transition from chaotic phase to non-chaotic phase.
	A common character for systems of this case is the existence of conserved quantity,
	like the conserved fermion number\cite{Davison}, or other quantities that commutes with the Hamiltonian.
	Note that here time-reversal symmetry (of BDI symmetry class) will be broken after the phase transition,
	while in our article we impose the time-reversal-invariance Majorana chain of BDI symmetry class
	unless specifically mentioned.
	The second one is the nonchiral homogeneous systems where the random couplings are indeed 
	marginally irrelevant in the large $N$ limit\cite{Berkooz M} as the translational invariance 
	in spacial space smoothing out the phase transitions related to the chiral symmetry breaking,
	after the ensenble average is performed, and thus the 
	left and right moving fermions exhibiting scale invariance.
	Different to this case,
	in our model, the chiral symmetry broken is guaranteed (down to the IR) by adding some restriction
	on the statistical aspect which leads to the scale invariance breaking.
	Such restrictions are realized through the gauge field as well as the induced 
	intrinsic fluctuations to the (effective) field operators,
	and the explicit breaking of scale invariance can be seem from the equation of motion of the effective field
	operators, as we illustrate in detail in Secs.2.3-2.7.
	In the calculations therein, we involving the intrinsic property about the gauge invariance
	into the commutation relations between the effective field operator and a functional of it,
	which exhibits great convinience in the statistical-mechanical calculations 
	with the ensemble average.
	The third one is the chiral SYK model in the extented $(1+1)$d\cite{Lian B},
	but different to this case where the conformal symmetry (and Lorentz symmetry) breaking 
	are intrinsic due to the spin structure of chiral Majorana fermions,
	the scaling invariance broken in our model in our model more likes the explicit one,
	which is due to the gauge field after the disorder (ensemble) average,
	instead of the spontaneous or intrinsic one.

	\section{Hopping-induced current model with gauge-invariance}
	
	For this system, we set $z<z'$ in position space,
	where $z=1$ denotes the surface layer while $z=N$ denotes the bottom layer.
	In such a 1D chiral system, it is easy to know that the current carriers (hopping fermions govered by a U(1) gauge field) 
	has momenta $q_{z}>q_{z'}$.
	Instead of considering the decay of the current energy with depth into the calculations,
	we set a conserved current energy, while the energy density is variable.
	Such an energy density $E$ of current could be the effective velocity $v_{z}=E_{q}/q_{z}$ for carriers in momentum space
	or the quasiparticle number operator (will also be called "current density" in this paper)  
	$\rho_{z}=E_{z}/z$ for quasiparticles in position space.
	Throughout this paper, we set a chiral-dependent constant ratio $n_{0}\gtrsim 1$,
	which is the quasiparticle number at initial position,
	while the quasiparticle number operators 
	in any other positions are written as $n_{0}^{z}$.
	Thus the carrier number (which is a quantity in position space) has $n_{z}<n_{z'}$.
	However, since since the quantity which is really one-dimensional is the carrier
	random hopping-induced energy gradient (which is nonzero only in the $z$-direction),
	like a effective electric field originates from the gradient of a boson field.
	Thus in order to investigating the dynamics of 1D randomly interacting model under gauge field,
	we need to using the quasiparticle which can be really one-dimensional insteads of
	the carrier which indeed has a three-dimensional momentum although only its $q_{z}$-component
	takes effect during a current generation.
	We will define the quasiparticle in this paper as the derivation of carrier-number-dependent boson field,
	where the boson field here is defined in spire of the Jordan-Wigner representation.
	The number operator of quasiparticle defined in this way will be inversely proportional to the carrier number (both in position space),
	and it can be regarded as the 1D current density, which follows the chiral structure in the distribution.
	
	\subsection{Current density}
	
	Note that these fermions should indeed have three-dimensional momentum,
	but we just ignore the in-plane components by taking the long-wavelength limit to the effective electric field term,
	while the summation over in-plane components of the momentum are self-cancelled.
	Thus the current can be expressed by the single carrier random hopping term in $z$-direction of momentum space
	which is coupled to a U(1) gauge field 
	\begin{equation} 
		\begin{aligned}
			J=\sum_{i<k}^{N}\sum_{q_{z}>q_{z'}}
			\psi_{i}^{*}(q_{z})c_{i}^{\dag}
			e^{iA_{q_{z}q_{z'}}}
			t_{ik}^{q_{z}q_{z'}}
			c_{k}\psi_{k}(q_{z'}),
		\end{aligned}
	\end{equation}
	where we assume $c_{i}^{\dag}$ creates a fermion in eigenstate $\psi_{i}(q_{z})$ in momentum space.
	The fluctuating U(1) gauge field reads
	$A_{q_{z},q_{z'}}=\frac{1}{2\pi}\int ^{q_{z'}}_{q_{z}}{\bf A}({\bf q})\cdot d{\bf q}$.
	The vector potential ${\bf A}({\bf q})=(0,0,A(q_{z}))$ can be regarded has only the $z$-component due to the in-plane cancellation effects
	as mentioned above,
	the gradient of this vector potential in momentum space is the original of the effective electrical field.
	Due to the chiral structure, there is another gauge constrain on the vector potential
	$\nabla \cdot {\bf A}=\nabla_{xy} A_{q_{xy}}+\nabla_{z}  (A_{q_{z}}+A_{q_{z'}})=0$.
	The phase factor may be changed by the fluctuation of gauge field, but the phase accumulated on a close
	circuit will be gauge invariant, $\sum A_{q_{z},q_{z'}}=2\pi \overline{n}$ where $\overline{n}$ is the number of flux quanta.
	For fermions considered here, the random hopping term $t_{ik}^{q_{z}q_{z'}}$ should be an antisymmetry tensor,
	and produces the eigenvalues when acted by the corresponding eigenstates.
	That means the indices of $t_{ik}^{q_{z}q_{z'}}$ should be summed antisymmetrically due to the intrinsic chiral property
	\begin{equation} 
		\begin{aligned}
			t_{ik}^{q_{z}q_{z'}}=t_{0}(\delta_{\overline{i}q_{z}}\delta_{\overline{k}q_{z'}}-\delta_{\overline{i}q_{z'}}\delta_{\overline{k}q_{z}}),
		\end{aligned}
	\end{equation}
	where we use the notations of fermion indices with an overline to represent
	their $q_{z}$-momentum-related information carried by them.
	$t_{0}$ is a constant in unit of energy.
	
	By decomposing the gauge field to two 
	transformation operators $e^{iA_{q_{z},q_{z'}}}=e^{i\theta_{q_{z}}}e^{-i\theta_{q_{z'}}}$ which acting on the coherent states separately,
	we have
	\begin{equation} 
		\begin{aligned}
			&\langle \psi^{*}_{i}(q_{z})c_{i}^{\dag}|e^{i\frac{1}{2}\theta_{q_{z}}}=\langle c_{i}^{\dag}e^{i\phi(q_{z})}|,\\
			&e^{-i\frac{1}{2}\theta_{q_{z'}}}|\psi_{l}(q_{z'})c_{l}\rangle =|e^{-i\phi(q_{z'})}c_{l}\rangle.
		\end{aligned}
	\end{equation}
	Here $\theta_{q_{z}}=\phi_{i}(q_{z})\sum_{z}q_{z}$
	where $\sum_{z}q_{z}=\sum_{\overline{i}}c^{\dag}_{\overline{i}}c_{\overline{i}}$
	is a conserved quantity for fermions in $q_{z}$ species (although it is no more conserved for the fluctuating gauge field).
	
	Then through the complicated but straightforward process as presented in Appendix.A,
	the above hopping current in terms of 3D carriers can be transformed to
	that in terms of 1D Majorana fermions,
	\begin{equation} 
		\begin{aligned}
			J=\sum_{i<k}^{N}\sum_{\alpha<\beta}^{M}\chi_{i}(\alpha)t_{ik}^{\alpha\beta}\chi_{k}(\beta),
		\end{aligned}
	\end{equation}
	where we use the notations $\alpha$ and $\beta$ to index the chiral current densities (i.e.,
	the quasiparticle number $\rho_{\alpha}\propto n_{z}^{-1}$ which
	is inverse proportional to the carrier number as we shown in Appendix.A).
	
	The chiral current density, which originates from the random hopping in a fermion current,
	is the only original degrees-of-freedom (DOF) in this system,
	and we will using the ordering $\alpha<\beta$ throughout this paper,
	which enforcing in any cases $\{\overline{i}\overline{j}\}<\{\overline{k}\overline{l}\}$.
	Also, in the mean time, 
	the antisymmetry exchanges only exist
	between these two current density sectors,
	as well as the individual Majorana fermions belong to the same sector.
	For example,
	we have $\chi_{i}(\alpha)\chi_{j}(\alpha)=-\chi_{j}(\alpha)\chi_{i}(\alpha)$,
	$\chi_{k}(\beta)\chi_{l}(\beta)=-\chi_{l}(\beta)\chi_{k}(\beta)$,
	$\chi_{i}(\alpha)\chi_{k}(\beta)=\chi_{k}(\beta)\chi_{i}(\alpha)$
	(see Eqs.(\ref{155}-\ref{157}) and Sec.3 for a detailed proof).
		As will be proved below,
		the subgroup $\{\overline{k}\overline{l}\}$ has the same size with that of $\alpha$ or $\beta$,
		which is of order $O(M)$, origin from the infinitely fractionalized current-density, as discussed in next subsection.

		Then we consider an interacting model between two carriers of different species (before and after the random hopping, respectively)
		as described above, under the effect of gauge field.
		\begin{equation} 
			\begin{aligned}
				H
				&=\sum_{q_{z},q_{z'}}\sum_{ijkl}
				c_{i}^{\dag}e^{i\phi_{i}(q_{z})}e^{i\phi_{j}(q_{z})}c_{j}
				T_{ik}^{jl}
				c_{l}^{\dag}e^{i\phi_{l}(q_{z'})}e^{i\phi_{k}(q_{z})}c_{k}\\
				&=-\sum_{i<k;j<l}\sum_{\alpha<\beta}
				\chi_{i}(\alpha)\chi_{j}(\alpha)
				\chi_{k}(\beta)\chi_{l}(\beta),
			\end{aligned}
		\end{equation}
		where gauge fields are decomposed into transformation operators acting on the carrier fermions
		of different species,
		and transform them into complex fermions as we shown in Appendix.A.
		Here we define $\chi_{i}(\alpha)=\varphi_{\overline{i}}(\alpha)\chi_{i}$ as the Majorana fermion operator that can 
		createa a particle in eigenstate $\varphi_{\overline{i}}(\alpha)$.
		Note that here $\overline{i}$ and $\alpha$ are of the mutually independent variable groups.
		The anyons generated here by the boson field could also be a hard core boson or the chiral Majorana fermion
		defined under a normal ordering, or even the ones defined through Dirac-$\gamma$ matrix which satisfy the
		Clifford algebra similar to the Klein factor.
		The antisymmetry tensor $T_{ik}^{jl}$ is important in diagonalizing the current matrix for each species,
		which reads
		\begin{equation} 
			\begin{aligned}
				\label{T4}
				T_{ik}^{jl}=
				(\delta_{\overline{i}\overline{j}}\delta_{\overline{k}\overline{l}}
				-\delta_{\overline{i}\overline{k}}\delta_{\overline{j}\overline{l}}-\delta_{\overline{j}\overline{l}}\delta_{\overline{i}\overline{k}}
				+\delta_{\overline{k}\overline{l}}\delta_{\overline{i}\overline{j}}).
			\end{aligned}
		\end{equation}

		\subsection{Restrictions from gauge invariance}
		
		The invariant gauge field here is induced soly by the chiral structure of the current density .
		(or inverse quasiparticle number) distribution.
		The gauge field $A_{\alpha\beta}$
		can be expressed in terms of the one-dimensional boson fields $\Phi(\alpha)$ defined above antisymmetrically
		\begin{equation} 
			\begin{aligned}
				&A_{\alpha\beta}=\frac{1}{2}(\overline{A}_{\alpha\beta}-\overline{A}_{\beta\alpha})=\theta_{\alpha}-\theta_{\beta},\\
				&\overline{A}_{\alpha\beta}=\theta_{\alpha}-\theta_{\beta},\\
				&\overline{A}_{\beta\alpha}=\theta_{\beta}-\theta_{\alpha}.
			\end{aligned}
		\end{equation}
		Similar to the gauge field coupled to fermion system which satisfies the Coulomb gauge,
		the above gauge field satisfies $\nabla A_{\alpha\beta}=\frac{1}{2}(\overline{A}_{\alpha\beta}+\overline{A}_{\beta\alpha})=0$,
		where $\nabla$ here denotes the gradient with respect to the quasiparticle number $n_{z}(=n_{0}^{z})$
		and it follow the chiral character:
		$\nabla \overline{A}_{\alpha\beta}\approx \overline{A}_{\alpha\beta}$, $\nabla \overline{A}_{\beta\alpha}\approx -\overline{A}_{\beta\alpha}$,
		and this chirality-dependent behavior is similar to the derivative of the $z$-positions in real space,
		$\partial_{z}=-\partial_{z'}$ (see Appendix.A), or the
		derivative of imaginary time, $\partial_{\tau}\psi(\tau)=\psi(\tau')$
		where $\tau'=\tau+i0^{+}$.
		
		Combining the discussions in Sec.2.1 and Appendix.A,
		the phases $\theta$ can be written as
		\begin{equation} 
			\begin{aligned}
				\theta_{\alpha}=\Phi(\alpha)\rho_{\alpha},
			\end{aligned}
		\end{equation}
		where current density (consisted by the quasiparticles in position space) 
		can be expressed in terms of 1D quasiparticle boson field $\Phi(\alpha)$,
		\begin{equation} 
			\begin{aligned}
				\rho_{\alpha}
				=\partial_{\alpha}\frac{\Phi(\alpha)}{2\pi}
				=\partial_{\alpha}\sum_{\gamma}^{\alpha-1}\rho_{\gamma}
				=\partial_{n_{0}^{z}}\frac{-i}{2\pi}{\rm ln}\frac{\phi(z)}{2\pi}
				=\partial_{n_{0}^{z}}\frac{-i}{2\pi}{\rm ln}\frac{n_{z}}{n_{0}-1}
			\end{aligned}
		\end{equation}
		is the always a conserved quantity for a part of supercharge product (with gauge fixing restriction) in a single sector $\alpha$
		(as will be clarified in the next section),
		by matching the notations of anyons (can be the complex fermions or the hard core bosons) with O(M) at level one,
		but it is never a conserved quantity for the fluctuating gauge field $A_{\alpha\beta}$ itself
		as it can changes all the time but keeping the gauge invariance.
		For example,
		if we replace the summing indices of the current density by the Majorana fermon indices,
		the resulting term $\sum_{i<j}\rho_{ij}$ will not be a conserved quantity for each species (or sectors).
		This is due to the gauge fixing requirement which will be clarified below.
		
		Thus the U(1) gauge field can be written in the form which is similar to the
		derivative "kinetic term" of a Lagrangian describing the noninteracting potential
		(but here for the current density, which can be viewed as the quasiparticle charge)
		\begin{equation} 
			\begin{aligned}
				&A_{\alpha\beta}=-i\rho_{\alpha}\sum_{\gamma}^{\alpha-1}\rho_{\alpha}
				-i\rho_{\beta}\sum_{\gamma}^{\beta-1}\rho_{\beta}
				=-i\frac{\Phi(\alpha)}{2\pi}\partial_{\alpha}\frac{\Phi(\alpha)}{2\pi}
				-i\frac{\Phi(\beta)}{2\pi}\partial_{\beta}\frac{\Phi(\beta)}{2\pi}.
			\end{aligned}
		\end{equation}
		
		According to the initial definition of the hopping matrix which induces the current
		by connecting the four fermions of two species, the disorder average process will be gauge invariant under the following transformations\cite{Patel A A}
		\begin{equation} 
			\begin{aligned}
				\label{11191}
				A_{\alpha\beta}\rightarrow A_{\alpha\beta}+\Phi_{\alpha}-\Phi_{\beta},\\
				c^{\dag}_{i}(\alpha)\rightarrow c^{\dag}_{i}(\alpha)e^{i\Phi(\alpha)},\\
				c_{j}(\alpha)\rightarrow c_{j}(\alpha)e^{-i\Phi(\alpha)},\\
				c^{\dag}_{k}(\beta)\rightarrow c^{\dag}_{k}(\beta)e^{i\Phi(\beta)},\\
				c_{l}(\beta)\rightarrow c_{l}(\beta)e^{-i\Phi(\beta)},
			\end{aligned}
		\end{equation}
		where the last four lines are the
		U(1) symmetry acting on fermions through a constraining variable
		and such intrinsic gauge invariance will be incorporated by performing a Jordan-Wigner transformation (Eq.(\ref{JW})).
		Base on the gauge properties,
		we can not simply integrating out all the four fermion indices when we try to integrating out the $O(M^{2})$ gauge field $A_{\alpha\beta}$
		(we consider $M\approx N$ here in large limit),
		since it requires gauge fixing to avoiding overcounting redundant configurations\cite{Patel A A}.
		This is similar to the problem of a intercluster-coupling model where the coupling behaviors are dominated by a fluctuating
		gauge field as reported in Ref.\cite{Patel A A}.
		Specifically in the system described in this paper,
		the above problem can be understood in this way:
		a pair of degrees-of-freedom $(\alpha,\beta)$ with size $O(M^{2})$ is
		initially fully shared by a part of fermions $c^{\dag}_{i}(\alpha)$ and $c_{l}(\beta)$ with degrees-of-freedom $(i,l)$,
		then after the above gauge invariant transformations (we consider no replica symmetry breaking here),
		the $(\alpha,\beta)$ must also be shared by another part of fermions $c^{\dag}_{k}(\beta)$ and $c_{j}(\alpha)$,
		and in this process $A_{\alpha\beta}$ becomes $A_{\alpha\beta}+\Phi_{k}(\beta)-\Phi_{j}(\alpha)$,
		then the $O(M^{2})$ gauge field $A_{\alpha\beta}$ must be constrained by only $O(N)$ constraining fermion degrees-of-freedom
		$\Phi_{k}(\beta)$ or $\Phi_{j}(\alpha)$.
		In other word,
		the space of gauge configurations consist of $\Phi_{k}(\beta)$ and $\Phi_{j}(\alpha)$
		can only be occupied by $O(M)$ gauge configurations of $A_{\alpha\beta}$.
		Thus the unique gauge configurations over the four fermion indices should be of order $O(N^{3})$ (or $O(N^{2}M)$)
		instead of order $O(N^{4})$.
		Note that in the above disorder average process, the Lagrangian multipliers
		in replica diagonal configurations change as
		$G_{\alpha}\rightarrow G_{\alpha}e^{i(\Phi(\alpha)-\Phi(\alpha))},
		\Sigma_{\alpha}\rightarrow \Sigma_{\alpha}e^{i(\Phi(\alpha)-\Phi(\alpha))}$.
		In Ref.\cite{Patel A A},
		this mismathcing of gauge configurations is ignored in large-$N$ limit,
		which will not affect the conformal results,
		and their resulting model is the SYK$_{4}$ one.
		While in this article, we will consider the ensential effects of the invariant gauge field $A_{\alpha\beta}$
		brought initially by the antisymmetry hopping tensor in a 1D chiral current.

		As we mentioned above, the chiral gradient operator acts on the gauge field is similar to the imaginary time derivative 
		of the Majorana fermions,
		thus to representing such a restriction (gauge fixing)-induced suppression to the fermion degrees-of-freedom,
		we can give each Majorana fermion operator an imaginary time component,
		and then defining a bosonic supercharge (bilocal field),
		each one containing both the $\alpha$ and $\beta$ degrees-of-freedoms.
		This additional time-component of each Majorana fermion (which is for more conviniently sign their intrinsic degrees-of-freedom properties)
		should be distinguished from
		the ones relating them and their replicas (and combine them to form the Lagrangian multiplier field).
		The additional time-component thus will not be summed.
		The details of this expression will be clarified in the next section,
		where we show that this antisymmetry algebra guarantees the formation of well-defined weight functions of a single bosonic supercharge. The four Majorana fermions are also classified by four mutually independent degrees-of-freedom $i,j,k,l$,
		which is similar to the normal four-point interacting SYK model.
		However, as the statistical behaviors in our model originate from the current density
		(that being classified into two distinct sectors),
		to considering the intrinsic physical property of the current density in 1D,
		the restriction from the gauge invariance is required,
		which is related to the self-similarity between the two original degrees-of-freedom,
		i.e., the current densities $\alpha$ and $\beta$.
		This can also exhibited by imaginary-time-dependence of the Majorana fermions.
		We will see that, the restriction brought by a finite-size time mapping group
		will reduce the four degrees-of-freedom of the fermion indices $ijkl$ to simply
		three degrees-of-freedom.
		
		\subsection{Gauge-invariance and the effective overlap between fermion bilinear and interacting fermion
			in $\alpha$-sector}
		
		Treating the phase $\Phi(\tau)$ as a continuously differentiable function of the imaginary time,
		we have (in addition to the Eq.(\ref{11191})),
		\begin{equation} 
			\begin{aligned}
				A_{\alpha}(\tau)\rightarrow A_{\alpha}(\tau)-\partial_{\tau}\Phi_{\alpha}(\tau).
			\end{aligned}
		\end{equation}
		From Eq.(\ref{11191}), by introducing another positional coordinate $r$,
		we can represent the fermion field in Heisenberg representation
		as
		$\psi_{s}=e^{r\mathcal{L}}\psi_{s}(r)e^{-r\mathcal{L}}$
		where $s=i,j$ is the gauge-field induced DOF.
		$\mathcal{L}$ is the effective Gor’kov Hamiltonian
		\begin{equation} 
			\begin{aligned}
				\label{L25}
				\mathcal{L}=
				\int d\tau_{1}
				\psi^{\dag}(\tau_{1})\mathcal{L}_{1}\psi(\tau_{1})
				+\frac{1}{2} \int d\tau_{1}\int d\tau_{2}
				V_{|\tau_{1}-\tau_{2}|}
				\psi^{\dag}(\tau_{1})\psi^{\dag}(\tau_{2})\psi(\tau_{2})\psi(\tau_{1}),
			\end{aligned}
		\end{equation}
		where the kinetic term $\mathcal{L}_{1}$ can be described by the 
		kinetic energy operator after minimal substitution 
		$\mathcal{L}_{\tau_{1}}=(i\partial_{\tau_{1}}-A_{\alpha}(\tau_{1}))^{2}$,
		$\mathcal{L}^{*}_{\tau_{1}}=(-i\partial_{\tau_{1}}-A_{\alpha}(\tau_{1}))^{2}$.

		The shiftment of the fermion operators during the gauge transformation can be described by the
		continuously differential result 
		\begin{equation} 
			\begin{aligned}
				\label{201}
				&c^{\dag}_{\alpha}{}^{'}(\tau)=c^{\dag}_{\alpha}(\tau) e^{i\Phi_{\alpha}(\tau)},\\
				&c_{\alpha}'(\tau)=c_{\alpha}(\tau)e^{-i\Phi_{\alpha}(\tau)},\\
			\end{aligned}
		\end{equation}
		where we have the following relation due to the variance of guage phase
		\begin{equation} 
			\begin{aligned}
				&\mathcal{L}_{\tau}[\Phi_{\alpha}]e^{i\Phi_{\alpha}(\tau)}
				=e^{i\Phi_{\alpha}(\tau)}\mathcal{L}'_{\tau}[\Phi_{\alpha}],\\
				&\mathcal{L}_{\tau}^{*}[\Phi_{\alpha}] e^{-i\Phi_{\alpha}(\tau)}
				=e^{-i\Phi_{\alpha}(\tau)}\mathcal{L}^{*'}_{\tau}[\Phi_{\alpha}]
				,
			\end{aligned}
		\end{equation}
		with the functional containing the variation of $(\Phi_{\alpha}'-\Phi_{\alpha})$:
		$\mathcal{L}_{\tau}'[\Phi_{\alpha}']=( -i\partial_{\tau}-(A_{\alpha}-\partial_{\tau}\Phi_{\alpha}(\tau)))^{2}$,
		$\mathcal{L}^{*'}_{\tau}[\Phi_{\alpha}']=( i\partial_{\tau}-(A_{\alpha}-\partial_{\tau}\Phi_{\alpha}(\tau)))^{2}$.

		The most essential relation between the fermion field $\psi(\tau)$
		and the $r$ operator for the following discussion is 
		\begin{equation} 
			\begin{aligned}
				\label{12191}
				[[\psi(\tau),r],r]=0.
			\end{aligned}
		\end{equation}
		Next we start by the commutation relation between the imaginary-time-dependent fermion field $\psi(\tau)$
		with the quantity $r$ in positional space as appears in the traditional functional derivative
		(note that the powers of $r$ could be of arbitrary high-order for a continuously differentiable functional), 
		\begin{equation} 
			\begin{aligned}
				[\psi(\tau),r^{m}]
				&=[\psi(\tau),r]r^{m-1}+r[\psi(\tau),r^{m-1}]\\
				&=[\psi(\tau),r]r^{m-1}+r[\psi(\tau),r]r^{m-2}+r^{2}[\psi(\tau),r^{m-2}]\\
				&\cdots\\
				&=[\psi(\tau),r]r^{m-1}+r[\psi(\tau),r]r^{m-2}+r^{2}[\psi(\tau),r]r^{m-3}
				\cdots+r^{m-1}[\psi(\tau),r]\\
				&=\sum_{\ell=0}^{m-1}r^{\ell}[\psi(\tau),r]r^{m-1-\ell}.
			\end{aligned}
		\end{equation}
		Due to the critical commutation relation mentioned above,
		$[[\psi(\tau),r],r]=0$,
		the above equation becomes
		\begin{equation} 
			\begin{aligned}
				[\psi(\tau),r^{m}]
				&=m[\psi(\tau),r]r^{m-1}
				=m r^{m-1}[\psi(\tau),r],
			\end{aligned}
		\end{equation}
		or equivalently,
		\begin{equation} 
			\begin{aligned}
				[\psi(\tau),r^{n}]
				=m r^{m-1}\psi(\tau)r-nr^{m}\psi(\tau).
				=m\psi(\tau)r^{m}-m r\psi(\tau)r^{m-1}.
			\end{aligned}
		\end{equation}
		Then we have two important relations between $\psi(\tau)r^{m}$ and $r^{m}\psi(\tau)$
		\begin{equation} 
			\begin{aligned}
				&\psi(\tau)r^{m}+r^{m}\psi(\tau)=r^{m-1}\psi(\tau)r+r\psi(\tau)r^{m-1},\\
				&\psi(\tau)r^{m}-r^{m}\psi(\tau)=
				m r^{m-1}\psi(\tau)r-m r^{m}\psi(\tau)=m\psi(\tau)r^{m}-m r\psi(\tau)r^{m-1},
			\end{aligned}
		\end{equation}
		where we can obtain the following expressions
		for the $\psi(\tau)r^{m}$ and $r^{m}\psi(\tau)$
		\begin{equation} 
			\begin{aligned}
				\psi(\tau)r^{m}&
				=m r^{m-1}\psi(\tau)r-(m-1)r^{m}\psi(\tau)
				=\frac{m}{m-1}r\psi(\tau) r^{m-1}-\frac{1}{m-1} r^{m}\psi(\tau))\\
				&=\frac{1}{2-m}r^{m-1}\psi(\tau)r+\frac{1-m}{2-m}r\psi(\tau)r^{m-1},\\
				r^{m}\psi(\tau)&
				=m r\psi(\tau)r^{m-1}-(m-1)\psi(\tau)r^{m}
				=\frac{m}{m-1} r^{m-1}\psi(\tau) r-\frac{1}{m-1}\psi(\tau)r^{m})\\
				&=\frac{1}{2-m}r\psi(\tau)r^{m-1}+\frac{1-m}{2-m}r^{m-1}\psi(\tau)r.
			\end{aligned}
		\end{equation}
		
		There is a useful list,
		\begin{equation} 
			\begin{aligned}
				&\frac{1}{m}[\psi(\tau),r^{m}]\\
				=&[\psi(\tau),r]r^{m-1}=\psi(\tau)r^{m}-r\psi(\tau)r^{m-1}\\
				=&r[\psi(\tau),r]r^{m-2}=r\psi(\tau)r^{m-1}-r^{2}\psi(\tau)r^{m-2}\\
				=&r^{2}[\psi(\tau),r]r^{m-3}=r^{2}\psi(\tau)r^{m-2}-r^{3}\psi(\tau)r^{m-3}\\
				&\cdots\\
				=&r^{m-1}[\psi(\tau),r]=r^{m-1}\psi(\tau)r-r^{m}\psi(\tau),
			\end{aligned}
		\end{equation}
		where we can obtain
		\begin{equation} 
			\begin{aligned}
				r^{m-k}[\psi(\tau),r]r^{k}
				=\frac{k}{m}[\psi(\tau),r^{m}]+r^{m}\psi(\tau).
			\end{aligned}
		\end{equation}
		For example, we would verify by $r^{m-1}[\psi(\tau),r]r$ which satisfies the equality,
		\begin{equation} 
			\begin{aligned}
				r^{m-1}[\psi(\tau),r]r
				=&\frac{1}{m}\psi(\tau)r^{m}-(\frac{1}{m}-1)r^{m}\psi(\tau)\\
				=&\frac{1}{2}r^{m}\psi(\tau)+\frac{1}{4}r^{m-1}\psi(\tau)r+\frac{1}{8}r^{m-3}\psi(\tau)r^{3}+
				\cdots +\frac{1}{2^{m-1}}r^{2}\psi(\tau)r^{m-2}-\frac{1}{2^{m-1}}\psi(\tau)r^{m}\\
				=&\frac{1}{2}r^{m}\psi(\tau)+(\frac{1}{2}-\frac{1}{2^{m-1}})r^{m}\psi(\tau)
				+(1-\frac{m}{2^{m-1}})\frac{1}{m}[\psi(\tau),r^{m}]-\frac{1}{2^{m-1}}\psi(\tau)r^{m}.
			\end{aligned}
		\end{equation}

		Then since $\frac{1}{m}[\psi(\tau),r^{m}]$ can be expressed in terms of the following general form
		\begin{equation} 
			\begin{aligned}
				\label{12111}
				r^{k}[\psi(\tau),r]r^{m-(k+1)}
				=r^{k}\psi(\tau)r^{m-k}-r^{k+1}\psi(\tau)r^{m-(k+1)},
			\end{aligned}
		\end{equation}
		which represent the same quantity for $k=0,\cdots,m-1$,
		we have
		\begin{equation} 
			\begin{aligned}
				2r^{k}\psi(\tau)r^{m-k}
				=r^{k-n}\psi(\tau)r^{m-(k-n)}+r^{k+n}\psi(\tau)r^{m-(k+n)},
			\end{aligned}
		\end{equation}
		which represent the same quantity for integers $k=0,\cdots,m-1$ and $n=0,\cdots, {\rm min}[k,m-k]$,
		and we have 
		\begin{equation} 
			\begin{aligned}
				\label{12112}
				&[[r^{k-n}\psi(\tau)r^{m-(k+n)},r^{n}],r^{n}]=0,\\
				&[[r^{k-n'}\psi(\tau)r^{m-(k+n')},r^{n'-n}],r^{n'+n}]=0,\\
				&[[r^{k-n'}\psi(\tau)r^{m-(k+n')},r^{n'+n}],r^{n'-n}]=0,\\
				&r^{n}\psi(\tau)r^{m-n}-r^{n'}\psi(\tau)r^{m-n'}=[r^{n}\psi(\tau)r^{m-n'},r^{n'-n}]
				=\frac{n'-n}{m}[\psi(\tau),r^{m}]\\
				&=(n'-n)[\psi(\tau),r]r^{m-1}=[r^{n-n'},\psi(\tau)]r^{m-(n-n')},
			\end{aligned}
		\end{equation}
		for $n'> n$.
		
		To related to the topological character as described by Eq.(\ref{1119}),
		we rewrite the commutator $[\psi(\tau),r^{m}]$ in belowing form
		\begin{equation} 
			\begin{aligned}
				&[\psi(\tau),r^{m}]
				=h[\psi(\tau),r]r^{m-1}+r^{h}[\psi(\tau),r^{m-h}]\\
				&=h[\psi(\tau),r]r^{m-1}+\sum_{j=2}^{h} (r^{j}-r^{j-1})[\psi(\tau),r^{m-h}]
				+r[\psi(\tau),r^{m-h}]\\
				&=h[\psi(\tau),r]r^{m-1}+\sum_{j=2}^{h} (r^{j}-r^{j-1})[\psi(\tau),r]
				+r[\psi(\tau),r^{m-h}]+\sum_{j=m-h+1}^{m-1}(r^{j}-r^{j-1})[\psi(\tau),r](m-h-1).
			\end{aligned}
		\end{equation}
		The (correlation) pattern of the topological modes $\chi_{2}\chi_{3}$ can be described by  
		terms $[\psi(\tau),r]r^{m-1}$ while the terms $[\psi(\tau),r]r^{m-1}+(r^{j}-r^{j-1})[\psi(\tau),r]$
		(for $j=m-h+1,\cdots,m-1$) correspond to the mode $\chi_{2}\chi_{3+m-j}$.
		The edge modes $\chi_{1}\chi_{2+h}$ corresponds to the
		term $r[\psi(\tau),r^{m-h}]+\sum_{j=m-h+1}^{m-1}(r^{j}-r^{j-1})[\psi(\tau),r](m-h-1)$.
		Using the relations Eq.(\ref{12111}-\ref{12112}),
		the mode not in the edges can be expressed as
		\begin{equation} 
			\begin{aligned}
				\chi_{2}\chi_{3+m-j}&=[\psi(\tau),r]r^{m-1}+(r^{j}-r^{j-1})[\psi(\tau),r]\\
				&=(r^{j}-r^{j-1}+r^{m-1})\psi(\tau)r -(r^{j+1}-r^{j}+r^{m})\psi(\tau),
			\end{aligned}
		\end{equation}
		and the powers of $r$ act on commutator $[\psi(\tau),r]$ can be expressed in terms of the modes
		(for $j=1,\cdots,m-1$)
		\begin{equation} 
			\begin{aligned}
				\label{13kk}
				r^{m-j}[\psi(\tau),r]
				&=j\chi_{2}\chi_{3}-\chi_{2}\chi_{4}-\chi_{2}\chi_{5}-\cdots -\chi_{2}\chi_{j+2}.
			\end{aligned}
		\end{equation}

		\subsection{exchanging rule between $[\psi(\tau),r]$ and powers of $r$}
		
		Next we discuss the exchanging rule for between the fermion field $\psi(\tau)$ and the powers of $r$ operator,
		which leads to the results of above section.
		Firstly, according to the original relation Eq.(\ref{12191}),
		we have
		\begin{equation} 
			\begin{aligned}
				\label{12192}
				r^{k}(\psi(\tau) r^{k})=(r^{k}\psi(\tau))r^{k},
			\end{aligned}
		\end{equation}
		where $k$ is an arbitrary positive integer.
		According to above subsection, we using the functional $Q$ to replacing the $r$ operator,
		we have 
		\begin{equation} 
			\begin{aligned}
				[[\psi,Q],Q]=0,
			\end{aligned}
		\end{equation}
		by define the $Q$ in the form of supercharge
		\begin{equation} 
			\begin{aligned}
				Q=\frac{1}{2}\int\int d\tau_{1}\tau_{2} \mathcal{C}_{\tau_{1},\tau_{2}}\psi^{\dag}(\tau_{1})\psi^{\dag}(\tau_{2}),
			\end{aligned}
		\end{equation}
		where $\mathcal{C}_{\tau_{1},\tau_{2}}$ is an antisymmetry tensor
		satisfies $\mathcal{C}_{\tau_{1},\tau_{2}}=-\mathcal{C}_{\tau_{2},\tau_{1}}$.
		Then the above relation Eq.(\ref{12192}) reads
		\begin{equation} 
			\begin{aligned}
				Q^{k}(\psi(\tau) Q^{k})=(r^{k}\psi(\tau))Q^{k},
			\end{aligned}
		\end{equation}
		which directly leads to due to the relation Eq.(\ref{12192}),
		\begin{equation} 
			\begin{aligned}
				\label{12192}
				(\psi(\tau)Q)Q=2Q(\psi(\tau)Q)-Q(Q\psi(\tau)).
			\end{aligned}
		\end{equation}
		We picking $k=1,2$ to prove this relation.
		
		Firstly note that 
		\begin{equation} 
			\begin{aligned}
				&[\psi(\tau), Q]=
				[
				\frac{1}{2}\int\int d\tau_{1}\tau_{2} \mathcal{C}_{\tau_{1}\neq \tau,\tau_{2}\neq \tau}
				\psi^{\dag}(\tau_{1})\psi^{\dag}(\tau_{2})\psi(\tau)\\
				&+
				\frac{1}{2}\int d\tau_{2} \mathcal{C}_{\tau,\tau_{2}\neq \tau}
				\psi^{\dag}(\tau_{2})
				+
				\frac{1}{2}\int\int d\tau_{1}\tau_{2} \mathcal{C}_{\tau,\tau_{2}\neq \tau}
				\psi^{\dag}(\tau)\psi^{\dag}(\tau_{2})\psi(\tau)\\
				&-
				\frac{1}{2}\int d\tau_{2} \mathcal{C}_{\tau_{1}\neq \tau,\tau}
				\psi^{\dag}(\tau_{1})
				+
				\frac{1}{2}\int\int d\tau_{1}\tau_{2} \mathcal{C}_{\tau_{1}\neq \tau,\tau}
				\psi^{\dag}(\tau_{1})\psi^{\dag}(\tau)\psi(\tau)\\
				&+
				\frac{1}{2}\int d\tau_{2} \mathcal{C}_{\tau,\tau}
				\psi^{\dag}(\tau)\psi^{\dag}(\tau)\psi(\tau)
				]
				-Q\psi(\tau)\\
				&=\frac{1}{2}\int d\tau_{2} \mathcal{C}_{\tau,\tau_{2}\neq \tau}
				\psi^{\dag}(\tau_{2})
				-
				\frac{1}{2}\int d\tau_{2} \mathcal{C}_{\tau_{1}\neq \tau,\tau}
				\psi^{\dag}(\tau_{1})\\
				&=\int d\tau_{2} \mathcal{C}_{\tau,\tau_{2}\neq \tau}
				\psi^{\dag}(\tau_{2}).
			\end{aligned}
		\end{equation}
		The same result can be obtained by constructing two anticommutators for the product of both orders
		($\psi(\tau)Q$ and $Q\psi(\tau)$),
		but we using the method that handling the order of operators in left-ordered product only,
		and subtracted by the right-ordered one.
		Thus we have
		\begin{equation} 
			\begin{aligned}
				&\psi(\tau)Q^{k}
				=Q^{k}\psi(\tau)
				+C^{2k}_{1}\mathcal{C}_{\tau;\tau_{2},\cdots,\tau_{2k}\neq\tau}
				\psi^{\dag}(\tau_{2})\cdots \psi^{\dag}(\tau_{2k})\\
				&+C^{2k}_{3}\mathcal{C}_{\tau,\tau,\tau;\tau_{4},\cdots,\tau_{2k}\neq\tau}
				\psi^{\dag}(\tau)\psi^{\dag}(\tau)
				\psi^{\dag}(\tau_{4})\cdots \psi^{\dag}_{2k}\\
				&+\cdots\\
				&+C^{2k}_{2k-1}\mathcal{C}_{\tau,\cdots,\tau;\tau_{2k}\neq\tau}
				[\psi^{\dag}(\tau)]^{2k-2}\psi^{\dag}_{2k},\\
				&Q^{k}\psi(\tau)
				=\psi(\tau)Q^{k}
				-C^{2k}_{1}\mathcal{C}_{\tau;\tau_{2},\cdots,\tau_{2k}\neq\tau}
				\psi^{\dag}(\tau_{2})\cdots \psi^{\dag}(\tau_{2k})\\
				&-C^{2k}_{3}\mathcal{C}_{\tau,\tau,\tau;\tau_{4},\cdots,\tau_{2k}\neq\tau}
				\psi^{\dag}(\tau)\psi^{\dag}(\tau)
				\psi^{\dag}(\tau_{4})\cdots \psi^{\dag}_{2k}\\
				&-\cdots\\
				&-C^{2k}_{2k-1}\mathcal{C}_{\tau,\cdots,\tau;\tau_{2k}\neq\tau}
				[\psi^{\dag}(\tau)]^{2k-2}\psi^{\dag}_{2k},\\
			\end{aligned}
		\end{equation}
		where the combination formula reads $C^{2k}_{n}=\frac{(2k)!}{n!(2k-n)!}$,

		For $k=1$,
		\begin{equation} 
			\begin{aligned}
				Q(\psi(\tau) Q)
				&=Q
				[
				\int d\tau_{2} \mathcal{C}_{\tau,\tau_{2}\neq \tau}
				\psi^{\dag}(\tau_{2})
				+Q\psi(\tau)
				]\\
				&=Q
				[\int d\tau_{2} \mathcal{C}_{\tau,\tau_{2}\neq \tau}
				\psi^{\dag}(\tau_{2})]
				-Q
				[\int d\tau_{2} \mathcal{C}_{\tau,\tau_{2}\neq \tau}
				\psi^{\dag}(\tau_{2})]
				+(Q \psi(\tau)) Q;
			\end{aligned}
		\end{equation}
		For $k=2$,
		\begin{equation} 
			\begin{aligned}
				Q^{2}(\psi(\tau) Q^{2})
				&=Q^{2}
				[
				\int\int\int  d\tau_{2} d\tau_{3} d\tau_{4} \mathcal{C}_{\tau;\tau_{2},\tau_{3},\tau_{4}\neq \tau}
				\psi^{\dag}(\tau_{2})\psi^{\dag}(\tau_{3})\psi^{\dag}(\tau_{4})
				+
				\int d\tau_{4} \mathcal{C}_{\tau,\tau,\tau;\tau_{4}\neq \tau}
				\psi^{\dag}(\tau)\psi^{\dag}(\tau)\psi^{\dag}(\tau_{4})
				]\\
				&-
				Q^{2}
				[
				\int\int\int  d\tau_{2} d\tau_{3} d\tau_{4} \mathcal{C}_{\tau;\tau_{2},\tau_{3},\tau_{4}\neq \tau}
				\psi^{\dag}(\tau_{2})\psi^{\dag}(\tau_{3})\psi^{\dag}(\tau_{4})
				+
				\int d\tau_{4} \mathcal{C}_{\tau,\tau,\tau;\tau_{4}\neq \tau}
				\psi^{\dag}(\tau)\psi^{\dag}(\tau)\psi^{\dag}(\tau_{4})
				]\\
				&+(Q^{2}\psi(\tau)) Q^{2}\\
				&=(Q^{2}\psi(\tau)) Q^{2}.
			\end{aligned}
		\end{equation}
		Similarly, for $k=3$, where $Q$ contains six $\psi^{\dag}(\tau)$ operators,
		two set of products each of whose amount is $(C^{6}_{1}+C^{6}_{3}+C^{6}_{5})$ will be cancelled out 
		and leads to $Q^{3}(\psi(\tau) Q^{3})=(Q^{3}\psi(\tau)) Q^{3}$.
		
		Specifically,
		for $Q[\psi(\tau),Q]Q=Q^{2}[\psi(\tau),Q]=[\psi(\tau),Q]Q^{2}$
		(we remind the reader that the $Q$ can be replaced by $r$ to connect to the content of above subsection),
		according to above results,
		we have 
		\begin{equation} 
			\begin{aligned}
				Q[\psi(\tau),Q]Q
				=Q(\psi(\tau)Q)Q-Q(Q\psi(\tau))Q
				=(Q\psi(\tau))Q^{2}-Q^{2}(\psi(\tau)Q),
			\end{aligned}
		\end{equation}
		where we using $Q(Q\psi(\tau))Q=Q[(Q\psi(\tau))Q]=Q[(Q\psi(\tau))Q]$
		(which can be obtained through the above procedure),
		and thus (see also Eq.(\ref{12192}))
		\begin{equation} 
			\begin{aligned}
				&(Q\psi(\tau))Q^{2}=2Q^{2}(\psi(\tau)Q)-Q^{2}(Q\psi(\tau)),\\
				&(\psi(\tau)Q)Q^{2}=2(Q\psi(\tau))Q^{2}-Q^{2}(\psi(\tau)Q)
				=3Q^{2}(\psi(\tau)Q)-Q^{2}(Q\psi(\tau)),\\
				&Q^{2}(\psi(\tau)Q)=2(Q\psi(\tau))Q^{2}-(\psi(\tau)Q)Q^{2},\\
				&Q^{2}(Q\psi(\tau))=2Q^{2}(\psi(\tau)Q)-(Q\psi(\tau))Q^{2}
				=3(Q\psi(\tau))Q^{2}-2(\psi(\tau)Q)Q^{2},
			\end{aligned}
		\end{equation}
		which leads to
		\begin{equation} 
			\begin{aligned}
				&(Q\psi(\tau))Q^{2}=\frac{1}{3}Q^{3}\psi(\tau)+\frac{2}{3}\psi(\tau)Q^{3},\\
				&Q^{2}(\psi(\tau)Q)=\frac{1}{3}\psi(\tau)Q^{3}+\frac{2}{3}Q^{3}\psi(\tau).
			\end{aligned}
		\end{equation}
		Similarly, we can further obtain
		\begin{equation} 
			\begin{aligned}
				&(Q\psi(\tau))Q^{3}=\frac{1}{4}Q^{4}\psi(\tau)+\frac{3}{4}\psi(\tau)Q^{4},\\
				&Q^{3}(\psi(\tau)Q)=\frac{1}{4}\psi(\tau)Q^{4}+\frac{3}{4}Q^{4}\psi(\tau),
			\end{aligned}
		\end{equation}
		which can be cast into more general form
		\begin{equation} 
			\begin{aligned}
				&(Q\psi(\tau))Q^{k}-Q^{k}(\psi(\tau)Q)=\frac{k-1}{k+1}[\psi(\tau),Q^{k+1}]
				=[\psi(\tau),Q](k-1)Q^{k}.
			\end{aligned}
		\end{equation}

		\subsection{effective field operator in terms of the functional and the two-fold
			pseudo-DOF}

		Base on the functional $Q$,
		we rewrite the commutator $[\psi(\tau),r^{m}]$ as $[\psi_{\alpha}(\tau),L[Q]]$ 
		by introducing the functional $L[Q]$ (effective field operator) as
		\begin{equation} 
			\begin{aligned}
				L[Q]=L[0]+\sum_{m=1}^{\infty}\frac{Q^{m}}{m!}\frac{d^{(m)}L[Q]}{d Q^{(m)}}\bigg|_{Q=0},
			\end{aligned}
		\end{equation}
		where $Q$ is a functional which also plays the role of test function here.
		This expression can be rewritten as
		\begin{equation} 
			\begin{aligned}
				\label{1111}
				L[Q]=e^{Q\mathcal{H}}L[0]e^{-Q\mathcal{H}}
				&=L[0]+\sum^{\infty}_{m=1}\frac{Q^{m}}{m!}
				\underbrace{[\mathcal{H},[\cdots,[\mathcal{H}}_\text{m}
				,L[0]]]]\\
				&=L[0]+\sum^{\infty}_{m=1}\int d\tau_{1}\cdots d\tau_{m}
				\frac{\delta ^{m}L[0]}{\delta \psi^{\dag}_{\alpha}(\tau_{1})\cdots \delta \psi^{\dag}_{\alpha}(\tau_{m})},
			\end{aligned}
		\end{equation}
		where $\mathcal{H}=\int d\tau_{1} \psi_{\alpha}(\tau_{1})$
		can be expressed in the form of Gor’kov equation\cite{Kita} by endows the unshifted functional $L[0]$
		a imaginary variable $\tau_{2}(\neq \tau_{1})$, i.e., $L[0;\tau_{2}]$
		\begin{equation} 
			\begin{aligned}
				\label{12223}
				\mathcal{H}=\int d\tau_{1} L^{\dag}[\tau_{1}] \mathcal{H}(\tau_{1})L[\tau_{1}]
				+\frac{1}{2}\int d\tau_{1}\int d\tau_{1}'
				V_{|\tau_{1}-\tau_{1}'|}
				L^{\dag}[\tau_{1}]L^{\dag}[\tau_{1}']L[\tau_{1}']L[\tau_{1}],
			\end{aligned}
		\end{equation}
		where $ \mathcal{H}(\tau_{1})$ can be expressed in the form of kinetic operator $(-i\partial_{\tau_{1}}-A_{1})^{2}$.
		
		Then we have
		\begin{equation} 
			\begin{aligned}
				[\psi_{\alpha},L[Q]]
				=&[\psi_{\alpha}, L[0]]+\sum_{m=1}^{\infty}\frac{1}{m!}\frac{d^{(m)}L[Q]}{d Q^{(m)}}\bigg|_{Q=0}
				[\psi_{\alpha}, Q^{m}]\\
				=&[\psi_{\alpha}, L[0]]+[\psi_{\alpha},Q]\frac{\partial L[Q]}{\partial Q}\\
				=&[\psi_{\alpha}, L[0]]+\sum_{m=1}^{\infty}\frac{1}{(m-1)!}\frac{d^{(m)}L[Q]}{d Q^{(m)}}\bigg|_{Q=0}
				[\psi_{\alpha}, Q]Q^{m-1},
			\end{aligned}
		\end{equation}
		where
		\begin{equation} 
			\begin{aligned}
				\label{12221}
				&\frac{\partial L[Q]}{\partial Q}
				=e^{Q\mathcal{H}}\left[\frac{d L[Q]}{d Q}\bigg|_{Q=0} \right]e^{-Q\mathcal{H}}\\
				&=e^{Q\mathcal{H}}[\mathcal{H},L[0;\tau_{2}]]e^{-Q\mathcal{H}}\\
				&=
				e^{Q\mathcal{H}}
				\left[-\mathcal{H}(\tau_{2})L[0;\tau_{2}]
				+\frac{1}{2}\int_{\tau_{1}\neq \tau_{2}} d\tau_{1} V_{|\tau_{1}-\tau_{2}|}
				L^{\dag}[\tau_{1}]L[0;\tau_{2}]L[0;\tau_{1}]\right.\\
				&\left.-\frac{1}{2}\int_{\tau_{1}'\neq \tau_{2}} d\tau_{1}' V_{|\tau_{2}-\tau_{1}'|}
				L^{\dag}[\tau_{1}']L[0;\tau_{1}']L[0;\tau_{2}]\right]
				e^{-Q\mathcal{H}}\\
				&=
				e^{Q\mathcal{H}}
				\left[
				-\mathcal{H}(\tau_{2})L[0;\tau_{2}]
				-\int d\tau_{1} V_{|\tau_{1}-\tau_{2}|}
				L^{\dag}[0;\tau_{1}]L[0;\tau_{1}]L[0;\tau_{2}]
				\right]
				e^{-Q\mathcal{H}}\\
				&=
				\left[
				-\mathcal{H}(\tau_{2})L[Q;\tau_{2}]
				-\int d\tau_{1} V_{|\tau_{1}-\tau_{2}|}
				L^{\dag}[Q;\tau_{1}]L[Q;\tau_{1}]L[Q;\tau_{2}]
				\right],
			\end{aligned}
		\end{equation}
		where $e^{Q\mathcal{H}}L[0;\tau_{2}]e^{-Q\mathcal{H}}=L[Q;\tau_{2}]$,
		with $\tau_{2}$ the parameter generated by shifted $\tau_{1}$ within the $\mathcal{H}$,
		and thus $\partial_{\tau_{2}}\mathcal{H}\neq 0$ as $\tau_{2}$ here is a "gradience"-dependent quantity now
		(like the $V_{\tau_{1}-\tau_{2}}$).
		
		For a comparasion, the above
		derivative $\frac{\partial L[Q]}{\partial Q}$ can also be expressed by
		\begin{equation} 
			\begin{aligned}
				&\frac{\partial L[Q]}{\partial Q}
				=e^{Q\mathcal{H}}\left[\frac{d L[Q]}{d Q}\bigg|_{Q=0}\right] e^{-Q\mathcal{H}}\\
				&=\frac{d L[Q]}{d Q}\bigg|_{Q=0}+\sum_{m=1}^{\infty}
				\frac{Q^{m}}{m!}
				\underbrace{[\mathcal{H},[\cdots,[\mathcal{H}}_\text{m},
				\frac{d L[Q]}{d Q}\bigg|_{Q=0}]]]\\
				&=e^{Q\mathcal{H}}[\mathcal{H},L[0;\tau_{2}]]e^{-Q\mathcal{H}}\\
				&=[\mathcal{H},L[0;\tau_{2}]]+\sum_{m=1}^{\infty}
				\frac{Q^{m}}{m!}
				\underbrace{[\mathcal{H},[\cdots,[\mathcal{H}}_\text{m},
				[\mathcal{H},L[0;\tau_{2}]]]]]\\
				&=\frac{d L[Q]}{d Q}\bigg|_{Q=0}+\sum_{m=1}^{\infty}
				\frac{Q^{m}}{m!}\frac{d ^{(m+1)}L[Q]}{d Q^{(m+1)}}\bigg|_{Q=0}.
			\end{aligned}
		\end{equation}
		Thus using Eq.(\ref{1111}),
		we also have
		\begin{equation} 
			\begin{aligned}
				&\frac{\partial e^{Q\mathcal{H}}L[0] e^{-Q\mathcal{H}}}{\partial Q}
				=&e^{Q\mathcal{H}}\frac{d}{dQ}  (L[0]+\sum_{m=1}^{\infty}
				\frac{Q^{m}}{m!}\frac{d ^{(m)}L[Q]}{d Q^{(m)}}\bigg|_{Q=0})\bigg|_{Q=0} e^{-Q\mathcal{H}}.
			\end{aligned}
		\end{equation}

		Using the relation
		\begin{equation} 
			\begin{aligned}
				[\mathcal{H},L[0;\tau_{2}]]\
				&=-\frac{\delta \mathcal{H}}{\delta L^{\dag}[0;\tau_{2}]}\\
				&=
				[-\mathcal{H}(\tau_{1})L[0;\tau_{1}]
				-\frac{1}{2}\int_{\tau_{1}\neq \tau_{2}} d\tau_{1} V_{|\tau_{1}-\tau_{2}|}
				L^{\dag}[0;\tau_{1}]L[0;\tau_{1}]L[0;\tau_{2}]\\
				&-\frac{1}{2}\int_{\tau_{1}'\neq \tau_{2}} d\tau_{1}' V_{|\tau_{2}-\tau_{1}'|}
				L^{\dag}[0;\tau_{1}']L[0;\tau_{1}']L[0;\tau_{2}]],
			\end{aligned}
		\end{equation}
		we can express the $\mathcal{H}$ as 
		\begin{equation} 
			\begin{aligned}
				\mathcal{H}
				&=\int d\tau_{2}L^{\dag}[0;\tau_{2}]\frac{\delta \mathcal{H}}{\delta L^{\dag}[0;\tau_{2}]}\\
				&=
				\int d\tau_{2}L^{\dag}[0;\tau_{2}]\mathcal{H}(\tau_{2})L[0;\tau_{2}]
				+\int_{\tau_{2}<\tau_{2}'} d\tau_{2} \int_{\tau_{2}'\neq \tau_{2}} d\tau_{2}' V_{|\tau_{2}-\tau_{2}'|}
				L^{\dag}[0;\tau_{2}']L[0;\tau_{2}']L[0;\tau_{2}],
			\end{aligned}
		\end{equation}
		which is equivalent to Eq.(\ref{12223}).

		Similar to Eq.(\ref{12221}), for $L^{\dag}[Q]$ we have
		\begin{equation} 
			\begin{aligned}
				\label{12226}
				&\frac{\partial L^{\dag}[Q]}{\partial Q}=
				e^{Q\mathcal{H}} [\mathcal{H},L^{\dag}[0;\tau_{2}]] e^{-Q\mathcal{H}}\\
				&=e^{Q\mathcal{H}} \frac{\delta \mathcal{H}}{\delta L[0;\tau_{2}]} e^{-Q\mathcal{H}}\\
				&=
				e^{Q\mathcal{H}}
				\left[
				L^{\dag}[0;\tau_{2}]\mathcal{H}(\tau_{2})
				+\frac{1}{2}\int d\tau_{1} V_{|\tau_{2}-\tau_{1}'|}
				L^{\dag}[0;\tau_{2}]L^{\dag}[0;\tau_{1}']L[0;\tau_{1}']\right.\\
				&\left.+\frac{1}{2}\int d\tau_{1} V_{|\tau_{1}-\tau_{2}|}
				L^{\dag}[0;\tau_{1}]L^{\dag}[0;\tau_{2}]L[0;\tau_{1}]
				\right]
				e^{-Q\mathcal{H}}\\
				&=\left[
				L^{\dag}[Q;\tau_{2}]\mathcal{H}(\tau_{2})
				+\int d\tau_{2}' V_{|\tau_{2}-\tau_{2}'|}
				L^{\dag}[Q;\tau_{2}]L^{\dag}[Q;\tau_{2}']L[Q;\tau_{2}']
				\right]\\
				&=
				\left[
				\mathcal{H}^{\dag}(\tau_{2})L^{\dag}[Q;\tau_{2}]
				+\int d\tau_{2}' V_{|\tau_{2}-\tau_{2}'|}
				L^{\dag}[Q;\tau_{2}]L^{\dag}[Q;\tau_{2}']L[Q;\tau_{2}']
				\right],
			\end{aligned}
		\end{equation}
		where using integration by parts we have
		\begin{equation} 
			\begin{aligned}
				\label{12225}
				\int d\tau_{2}
				\left(L^{\dag}[0;\tau_{2}]\mathcal{H}(\tau_{2})
				-\mathcal{H}^{\dag}(\tau_{2})L^{\dag}[Q;\tau_{2}]\right)
				L[0;\tau_{2}]=0.
			\end{aligned}
		\end{equation}
		
		Then the two-fold sublattice-like (pseudo) degrees-of-freedom
		can be obtained by combining Eq.(\ref{12221}) and Eq.(\ref{12226}),
		\begin{equation} 
			\begin{aligned}
				&\frac{\partial L^{a}[Q]}{\partial Q}
				&=(-1)^{a}
				\left[
				\mathcal{H}^{a}(\tau_{2})L^{a}[Q;\tau_{2}]
				+\int d\tau_{2}' V_{|\tau_{2}-\tau_{2}'|}
				L^{\dag}[Q;\tau_{2}']L[Q;\tau_{2}']L^{a}[Q;\tau_{2}]
				\right],
			\end{aligned}
		\end{equation}
		where $a=2$ corresponds to the creation operator while $a=1$ corresponds to 
		anihilation operator.
		That is to say, the creation or anihilation operators correspond to the
		fermion fields of different sites, respectively (in the sublattice).
		
		Thus, in terms of the functional $L[Q]$, which plays the role of effective field operator in the Gor’kov equation,
		the effect of topological modes can be counted by considering their effective particle-number operator.
		And the two-fold pseudo-DOF can be shown by the equation of motion of the effective field operator
		$L[Q]$.
		Further, although it originates from the statistical properties (of scalar products) in
		bosonic chain,
		it follows some pattern of symmetry like the $Z_{2}$-fermionic parity.

		\subsection{Calculations using the two-fold DOF character and the Jacobian}

		As will be shown in the next subsection,
		the functional defined here can be related to the original commutator between fermion field $\psi(\tau)$
		and the operator $r$.
		This can be related to the effective overlap (replacement) between fermion bilinear and interacting fermion
		in $\alpha$-sector as we mention below,
		where we introducing the fermionic field matrix constituted by the fermionic indices $n=1,\cdots,N$
		and the sublattice (or pseudospin) indices $s=a,b$
		\begin{equation} 
			\begin{aligned}
				\Psi=
				\begin{pmatrix}
					\psi_{1}^{a} & \psi_{1}^{b}\\
					\psi_{2}^{a} & \psi_{2}^{b}\\
					\cdots\\
					\psi_{N}^{a} & \psi_{N}^{b}\\
				\end{pmatrix}.
			\end{aligned}
		\end{equation}
		Then according to Weinstein–Aronszajn identity,
		we have a relation similar to Eq.(\ref{192})
		\begin{equation} 
			\begin{aligned}
				{\rm Det}[:\Psi\Psi^{\dag}:-{\bf I}_{N}]=(-1)^{N-2}{\rm Det}[:\Psi^{\dag}\Psi:-{\bf I}_{2}].
			\end{aligned}
		\end{equation}
		Next, according to the matrix determinant lemma,
		we know the trace of a certain matrix can be seem as the directional derivative of the determinant,
		and as long as $\Psi$ has two columns ($s=a,b$),
		we have
		\begin{equation} 
			\begin{aligned}
				{\rm Det}[:\Psi\Psi^{\dag}:+{\bf I}_{N}]=1+{\rm Det}[:\Psi^{\dag}\Psi:]+{\rm Tr}[:\Psi\Psi^{\dag}:],\\
				{\rm Det}[:\Psi\Psi^{\dag}:-{\bf I}_{N}]={\rm Det}[:\Psi^{\dag}\Psi:]+1-{\rm Tr}[:\Psi\Psi^{\dag}:].
			\end{aligned}
		\end{equation}
		When the DOF representing the flavors of fermion bilinear terms are omiited
		to conserving the total number of DOFs,
		in which case $\psi_{n}^{s}=\psi^{s}$,
		we have
		\begin{equation} 
			\begin{aligned}
				&{\rm Det}[:\Psi\Psi^{\dag}:+{\bf I}_{N}]=1+{\rm Tr}[:\Psi\Psi^{\dag}:]
				=1+{\rm Tr}[:\Psi^{\dag}\Psi:]=1+N\psi_{i}^{\dag}\psi_{i}+N\psi_{j}^{\dag}\psi_{j},\\
				&{\rm Det}[:\Psi\Psi^{\dag}:-{\bf I}_{N}]=1-{\rm Tr}[:\Psi\Psi^{\dag}:],\\
				&{\rm Det}[:\Psi\Psi^{\dag}:]={\rm Det}[:\Psi^{\dag}\Psi:]=0.
			\end{aligned}
		\end{equation}

		Now we know that the DOF of current densities
		represented by two sectors $\alpha$ and $\beta$
		results in the sublattice (or pseudospin) DOF in the local Majorana fermions of the sector $\alpha$.
		That is the reason why we must using the definition of $\chi_{i,j}$ instead of $\gamma_{i,j}$ in the $\alpha$-sector,
		when considering the global behaviors of the whole system.
		In terms of this, due to the effect of gauge field-modified single particle operator 
		(Eq.(\ref{201})), the distinguishable creation and annihilation operators can be replaced by the 
		the DOF described by sublattice indices $s=i,j$,
		i.e., $\psi_{n}^{\dag}\psi_{n}\rightarrow :\psi^{\dag}_{s}\psi_{s'}:$,

		We can using the Lagrange multipliers in replica theory to describe such additional DOF,
		in terms of a partition function 
		\begin{equation} 
			\begin{aligned}
				Z_{r}=&\int \prod_{i}^{N}\prod_{s,s'=a,b}D[\psi^{\dag}_{i,s},\psi_{i,s'}]
				{\rm Exp}
				[-\int d\tau\int d\tau' \psi^{\dag}_{i,s'}(\tau')\mathcal{H}_{s,s'}(\tau,\tau')\psi_{i,s}(\tau)]\\
				=&\left[\int \prod_{s,s'=a,b}D[\psi^{\dag}_{i,s},\psi_{i,s'}]
				{\rm Exp}
				[-\int d\tau\int d\tau' \psi^{\dag}_{i,s'}(\tau')\mathcal{H}_{s,s'}(\tau,\tau')\psi_{i,s}(\tau)]\right]^{N}\\
				=&\pi^{N}{\rm Det}[\mathcal{H}_{s,s'}(\tau,\tau')]^{-N/2},
			\end{aligned}
		\end{equation}
		where $\mathcal{H}_{s,s'}(\tau,\tau')$ here denotes a $2\times 2$ imaginary symmetric matrix,
		and in last step we using the formula of Gaussian integral.
		The above partition function can be expressed in terms of
		the bosonic fluctuation as a function of the coordinate $\phi^{F}(r)$,
		$Z_{r}=\int {\rm Exp}[-\int dr \frac{1}{2}(\partial_{\tau}\phi^{F}(r))^{2}]$,
		where the action can be normalized by adding an another term 
		$\frac{1}{2}(\phi^{F}(r))^{2}$ into the exponential part\cite{JK2}.
		The field $\phi^{F}(r)$ can be written in terms of an orthonormal basis
		$F_{m}(\tau)$ and the saddle point $r_{0}$\cite{JK,JK2}
		\begin{equation} 
			\begin{aligned}
				\phi^{F}(r)=\psi_{i,s}(r-r_{0})+\sum_{m=1}^{\infty}
				\phi_{m}^{F}F_{m}(r),
			\end{aligned}
		\end{equation}
		where the variation of $\phi^{F}(r)$ with respect to $r_{0}$ is orthogonal
		to that with respect to $\phi_{m}^{F}$,
		and in to make two sublattice flavors be mutually independent
		(to conserve the number of degrees of freedom),
		we have
		\begin{equation} 
			\begin{aligned}
				\label{1992}
				\int dr F_{m}(r)\partial_{r}\psi_{i,s}(r) =0.
			\end{aligned}
		\end{equation}
		In terms of the functional Taylor expansion\cite{FFk},
		we can know that the $r$-independent term $\phi_{m}^{F}$
		is indeed a Lagrangian $\mathcal{L}[\phi^{F}(\dot{r}),\phi^{F}(r)]$,
		and the result of Eq.(\ref{1992}) is due to
		\begin{equation} 
			\begin{aligned}
				\phi^{F}(r)=\sum_{m=1}^{\infty}\int^{r}_{0} dr_{1} \cdots \int^{r}_{0} dr_{m}
				\frac{\partial^{m}\phi^{F}(r)}{\partial u(r-r_{1})\cdots
					\partial u(r-r_{m})}u(r-r_{1})\cdots u(r-r_{m}),
			\end{aligned}
		\end{equation}
		where $\dot{r}$ is a function of $u(r-r')$,
		and that results in the transition of dependence on $r$ to $r'$
		for the $\phi_{m}^{F}$ term.

		Here we note that the vanishing result of Eq.(\ref{12225})
		can be related to the orthonormality of the saddle points around $\tau_{2}$.
		If we represent the commutator $[\mathcal{H},L^{\dag}[0;\tau_{2}]]$
		in terms of an orthonormal set of the function $f(\tau_{2})$,
		\begin{equation} 
			\begin{aligned}
				[H,L[0;\tau_{2}]]=\sum_{\tau_{1}}^{\infty}F_{\tau_{1}}f_{\tau_{1}}(\tau_{2})
				=F(\tau_{2})-F_{c}(\tau_{2}-\tau_{2}^{0}),
			\end{aligned}
		\end{equation}
		where $\tau_{2}^{0}$ denotes all the degenerate saddle points around $\tau_{2}$.
		The functional $F(\tau_{2})$ is a result of summation over all possible fluctuation directions,
		and $\tau_{1}$ denotes all the possible fluctuation directions other that the one
		to which the saddle point $\tau_{2}^{0}$ belongs.
		Then we can divide $F(\tau_{2})$ into two parts by making sure $f_{\tau_{1}}(\tau_{2})$ 
		forms an orthonormal basis
		\begin{equation} 
			\begin{aligned}
				F(\tau_{2})=F_{c}(\tau_{2}-\tau_{2}^{0})+\sum_{\tau_{1}}^{\infty}F_{\tau_{1}}f_{\tau_{1}}(\tau_{2}),
			\end{aligned}
		\end{equation}
		which means the function $F(\tau_{2})$ is a summation of components (functional)
		in different directions that modified by the mutually independent weights (i.e., the saddle points
		whose amount corresponds to the dimension of Euclidean space-time).
		Thus the variation of $F(\tau_{2})$
		with respect to $\tau_{2}^{0}$ and $F_{\tau_{1}}$ (saddle points in all possible fluctuation directions) are nonzero.
		In terms of the mechanism,
		this is similar to the transition of area-law entanglement 
		(the summation boundary of fluctuation directions here) to the volume-law entanglement
		(weight distribution for each direction here) in a multipartite system with non-Abelian symmetry\cite{Protopopov}.

		The term $F_{c}(\tau_{2}-\tau_{2}^{0})$ describes the fluctuation of target function $F(\tau_{2})$
		around the saddle-point $\tau_{2}^{0}$,
		while the summation (commutator) $\sum_{\tau_{1}}^{\infty}F_{\tau_{1}}f_{\tau_{1}}(\tau_{2})$
		corresponds to all the other possible fluctuations of target function $F(\tau_{2})$
		around the saddle-points $\tau_{1}\neq \tau_{2}^{0}$.
		Due to the mutual independence between saddle points,
		the variation of $F(\tau_{2})$ with respect to $\tau_{2}^{0}$ are orthogonal to that with respect to
		$F_{\tau_{1}}$\cite{Zinn-Justin},
		we have 
		\begin{equation} 
			\begin{aligned}
				\label{12231}
				\int d\tau_{2}
				f_{\tau_{1}}(\tau_{2})
				\partial_{\tau_{2}^{0}} F_{c}(\tau_{2}-\tau_{2}^{0})
				=f_{\tau_{1}}(\tau_{2})F_{c}(\tau_{2}-\tau_{2}^{0})\bigg|_{\tau_{2}}
				=0,
			\end{aligned}
		\end{equation}
		which shows that the product consists of the normalized basis and the fluctuating function around
		a saddle-point $\tau_{2}^{0}$ is an even function of $\tau_{2}$
		(in the interval where $f_{\tau_{1}}(\tau_{2})$ is a set of orthonormal functions).
		As $F_{\tau_{1}}$ represents the group made up by the saddle-points 
		that each one of them from a fluctuation direction distinct from all the others,
		and the summation range of fluctuation directions in the term $\sum_{\tau_{1}}^{\infty}F_{\tau_{1}}f_{\tau_{1}}(\tau_{2})$ is related to the direction to which the saddle point $\tau_{2}^{0}$ belongs ($\tau_{1}'$).
		Thus the critical term $F_{c}(\tau_{2}-\tau_{2}^{0})$
		equivalents to $\sum_{\tau_{1}'} F_{\tau_{1}'} f_{\tau_{1}'}(\tau_{2})$ above
		expression can be rewritten as
		\begin{equation} 
			\begin{aligned}
				\int d\tau_{2} f_{\tau_{1}}(\tau_{2}) f_{\tau_{1}'}(\tau_{2})=0,
			\end{aligned}
		\end{equation}
		
		Using Jacobian the integral over all possible (shifted) $F(\tau_{2})$ 
		can be written as
		\begin{equation} 
			\begin{aligned}
				\label{12241}
				dF(\tau_{2})
				&=\mathcal{J} d\tau_{2}^{0} \prod^{\infty}_{\tau_{1}} dF_{\tau_{1}},
			\end{aligned}
		\end{equation} 
		with 
		\begin{equation} 
			\begin{aligned}
				\mathcal{J}&=
				\left[ \int d\tau_{2} (\partial_{\tau_{2}^{0}}F_{c}(\tau_{2}-\tau_{2}^{0}))^{2} \right]^{1/2}
				\left[ \prod_{\tau_{1}\neq\tau_{2}^{0}} 
				\int d\tau_{2} (\partial_{F_{\tau_{1}}}F_{c}(\tau_{2}-\tau_{1}))^{2} \right]^{1/2}\\
				&=\left[ \int d\tau_{2} (\partial_{\tau_{2}^{0}}F_{c}(\tau_{2}-\tau_{2}^{0}))^{2} \right]^{1/2},
			\end{aligned}
		\end{equation} 
		where the integral over $\tau_{2}$ after each squared derivatives of $F_{c}$ with respect to 
		saddle points guarantees the single-saddle-point-dependence and make sure the 
		Jacobian is diagonal,
		and the second term in above expression reduces to one due to the orthonormality 
		of functions $f_{\tau_{1}}(\tau_{2})$,
		$\int d\tau_{2} f_{\tau_{1}}(\tau_{2})f_{\tau_{1}'}(\tau_{2})=\delta_{\tau_{1},\tau_{1}'}$. 
		Note that in Eq.(\ref{12231}),
		the range of integral over $\tau_{2}$ is restricted by the saddle-point within the function $F_{c}$.
		Eq.(\ref{12241}) turns the integral from the functional measurment
		($dF(\tau_{2})$) to the saddle points (each one of them of different fluctuation directions).
		
		If we consider $d$ values of the saddle point (denoted as $\tau_{2;\mu}^{0}$ with $\mu=1,\cdots,
		m $)
		in the $\tau_{1}'$ fluctuation direction, which corresponds to
		the $d$-dimensional (Euclidean) space-time in field theory,
		we have
		\begin{equation} 
			\begin{aligned}
				\mathcal{J}
				&=\left[ \prod_{\mu=1}^{m} \int d^{m}\tau_{2}
				(\partial_{\tau_{2;\mu}^{0}}F_{c}(\tau_{2}))^{2} \right]^{1/2}\\
				&=\left[(\prod^{m}_{\mu=1}\lambda_{\mu})^{-\frac{m-1}{m}}
				\int d^{m}\tau_{2} \left[ \frac{\partial^{(m)} F_{c}(\tau_{2})}{\partial 
					\tau_{2;\mu}^{0} {}^{(m)}}\right]^{2} \right]^{m/2}\\
				&=\left[(\prod^{m}_{\mu=1}\lambda_{\mu})^{-\frac{m-1}{m}}
				\theta_{m}
				\int d^{m}\tau_{2} \left[ \prod_{\mu}^{m}
				\frac{\partial F_{c}(\tau_{2})}{\partial 
					\tau_{2;\mu}^{0} }\right]^{2} \right]^{m/2},
			\end{aligned}
		\end{equation} 
		where $d^{m}\tau_{2}=\prod_{\mu=1}^{m} d(\tau_{2}-\tau_{2;\mu}^{0})$,
		$\theta_{m}=m^{m+\frac{m}{2}}$ for even $m$ and $\theta_{m}=m^{m+\frac{m-1}{2}}\sqrt{m}$ for odd $m$,
		and $\lambda_{\mu}$ is the corresponding weight for each eigenvalue
		(next we simply denote $\tau_{2;\mu}^{0}$ as $y_{\mu}$).
		The function 
		$F_{c}(\tau_{2})$ represented in terms of the Gaussian distribution,
		\begin{equation} 
			\begin{aligned}
				F_{c}(\tau_{2})=\prod_{\mu=1}^{m} e^{-y_{\mu}^{2}\lambda_{\mu}},
			\end{aligned}
		\end{equation} 
		where $\lambda_{\mu}$ satisfies
		\begin{equation} 
			\begin{aligned}
				\partial_{y_{\mu}} F_{c}(\tau_{2})=2 F_{c}(\tau_{2})\lambda_{\mu}y_{\mu}.
			\end{aligned}
		\end{equation} 
		Similar results can be obtained also by using the joint probability distribution instead of the individual weights,
		where we will find that
		the eigenvalue density 
		reads
		\begin{equation} 
			\begin{aligned}
				&\rho(y_{1})
				=e^{-\frac{y_{1}^{2}}{\sigma^{2}}}\int e^{-\frac{y_{1}^{2}+y_{2}^{2}+\cdots+y_{m}^{2}}{\sigma^{2}}}
				\prod_{\mu=2}^{m}|y_{1}-y_{\mu}|^{2}\prod_{i<j;i,j\neq 1}|y_{i}-y_{j}|^{2}\\
				&=(m-1)!e^{-\frac{y_{1}^{2}}{\sigma^{2}}}{\rm det}
				\begin{pmatrix}
					\rho_{0,0} & \rho_{0,1} & \cdots &\rho_{0,m-2}\\
					\rho_{1,0} & \rho_{1,1} & \cdots &\rho_{1,m-2}\\
					\cdots\\
					\rho_{m-2,0} & \rho_{m-2,1} & \cdots &\rho_{m-2,m-2}
				\end{pmatrix},
			\end{aligned}
		\end{equation} 
		where $\rho_{i,j}=\int dy y^{i+j}|y_{1}-y|^{2}e^{-\frac{y^2}{\sigma^{2}}}$.

		\subsection{functional in terms of the topological modes}
		
		The functional defined here can be related to the topological modes through the following relation
		\begin{equation} 
			\begin{aligned}
				&\psi_{\alpha}L[Q]=h\chi_{2}\chi_{3}+\chi_{1}\chi_{h+2}+\sum_{j=m-h+1}^{m-1}r^{j}[\psi(\tau),r]
				=(m+\frac{h(h+1)}{2}-1)\chi_{2}\chi_{3}-\sum_{j=1}^{h-1}(h-j)\chi_{2}\chi_{3+j},\\
				&L[Q]\psi_{\alpha}=\psi_{\alpha}(\tau)L^{*}[Q]
				=\sum_{j=1}^{h-1}(h-j)\chi_{2}\chi_{3+j}.
			\end{aligned}
		\end{equation}
		Through Eq.(\ref{13kk}),
		the above relation can be rewritten as
		\begin{equation} 
			\begin{aligned}
				&\psi_{\alpha}L[Q]=h\chi_{2}\chi_{3}+\chi_{1}\chi_{h+2}+\sum_{j=m-h+1}^{m-1}r^{j}[\psi(\tau),r]
				=(m+\frac{1}{r}+\frac{1}{r^{2}}+\cdots \frac{1}{r^{h-1}})\chi_{2}\chi_{3}\\
				&=(m+\frac{1}{r}+\frac{1}{r^{2}}+\cdots \frac{1}{r^{h-1}})r^{m-1}[\psi(\tau),r],\\
				&L[Q]\psi_{\alpha}=\psi_{\alpha}(\tau)L^{*}[Q]
				=(\frac{1}{r}+\frac{1}{r^{2}}+\cdots \frac{1}{r^{h-1}})r^{m-1}[\psi(\tau),r],
			\end{aligned}
		\end{equation}
		where we can choose
		\begin{equation} 
			\begin{aligned}
				&\psi_{\alpha}
				=(m+\frac{1}{r}+\frac{1}{r^{2}}+\cdots +\frac{1}{r^{h-1}}),\\
				&L[Q]=r^{m-1}[\psi(\tau),r].
			\end{aligned}
		\end{equation}
		Here the nonzero commutation relation between $\psi_{\alpha}$ and $L[Q]$
		can be understood by treating $r$ with negative powers as operators,
		e.g., from Eq.(\ref{13kk}) we have
		\begin{equation} 
			\begin{aligned}
				[\frac{1}{r},r^{m-1}[\psi(\tau),r]]
				=r^{m-2}[\psi(\tau),r]-r^{m-1}\psi(\tau)+r^{m}\psi(\tau)\frac{1}{r},
			\end{aligned}
		\end{equation}
		and thus 
		\begin{equation} 
			\begin{aligned}
				&[\psi_{\alpha},L[Q]]=m r^{m-1}[\psi(\tau),r]
				=\psi_{\alpha}(L[Q]-L^{*}[Q])\\
				&=\sum_{j=2}^{h}\left( r^{m-j}[\psi(\tau),r]-r^{m-1}\psi(\tau)\frac{1}{r^{j-2}}
				+r^{m}\psi(\tau)\frac{1}{r^{j-1}}\right).
			\end{aligned}
		\end{equation}

		From Eq.(\ref{13kk}),
		we know the the definition of the operator $r$ has nonzero commutator with itself in different powers, 
		That results in the commutator $[\psi(\tau),r]$ acted by $r$ with higher powers can be obtained
		by the one with lower powers through adding a $r$ operator in the left-side or right-side.
		For example, in the following expression,
		\begin{equation} 
			\begin{aligned}
				r^{m-j}[\psi(\tau),r]r=[\psi(\tau),r]r^{m-j+1},
			\end{aligned}
		\end{equation}
		the operator $r$ is defined according to the term $r^{m-j}[\psi(\tau),r]$,
		\begin{equation} 
			\begin{aligned}
				r=\frac{r^{m-j}[\psi(\tau),r]-\chi_{2}\chi_{3}+\chi_{2}\chi_{j+2}}
				{r^{m-j}[\psi(\tau),r]}.
			\end{aligned}
		\end{equation}
		More generally,
		we have
		\begin{equation} 
			\begin{aligned}
				\frac{\chi_{2}\chi_{4+k}}{\chi_{2}\chi_{3}}=1+r^{-k}-r^{-(k+1)}.
			\end{aligned}
		\end{equation}
		This is guaranteed by the fact that the term $r^{m-j}[\psi(\tau),r]$ must contains all the modes
		appear in $r^{m-j+k}[\psi(\tau),r]$ with $k$ the positive integer,
		which can be verified by Eq.(\ref{13kk}). While the opposite case is not true.
		That also results in 
		\begin{equation} 
			\begin{aligned}
				\label{1112}
				[\psi(\tau),r]r^{m-j}=[\psi(\tau),r]r^{m-j-k}r^{k},
			\end{aligned}
		\end{equation}
		as long as $(m-j-k)\ge 0$,
		which can also be expanded as
		\begin{equation} 
			\begin{aligned}
				(\psi(\tau)r) r^{m}=(\psi(\tau)r r^{m-k}) r^{k},
				(r\psi(\tau)) r^{m}=(r\psi(\tau) r^{m-k}) r^{k}.
			\end{aligned}
		\end{equation}

		But it is false to write
		\begin{equation} 
			\begin{aligned}
				\label{1113}
				[\psi(\tau),r]r^{m-j}=[\psi(\tau),r]r^{m-j+k}\frac{1}{r^{k}},
			\end{aligned}
		\end{equation}
		since unlike the $r$ operators in positive powers,
		it has $[r^{m-j+k},\frac{1}{r^{k}}]=0$ which leads to 
		\begin{equation} 
			\begin{aligned}
				[\psi(\tau),r]r^{m-j+k}\frac{1}{r^{k}}\neq 
				[\psi(\tau),r](r^{m-j+k}\frac{1}{r^{k}}).
			\end{aligned}
		\end{equation}

		\subsection{Randomness and the translational invariance}
		
		As we illustrated in the above subsections,
		the effective two-fold DOF represented by the equation of motion of the appropriately constructed functional
		$L[Q]$, which equivalents to the functional derivative of the Gor'kov euqation $\mathcal{H}$ in mean field 
		approximation (which satisfies the transilational invariance before the functional derivatives
		with respect to effective field operator).
		
		As can be seen from Sec.2.5,
		the two-fold DOF is represented in terms of the equation-of-motion of the functional $L[Q]$
		which can be expressed in terms of the effective field operator $L[0]$
		by $L[Q]=e^{Q\mathcal{H}} L[0]e^{-Q\mathcal{H}}$,
		where the functional derivative with respect to the effective field operator breaks the translational invariance
		of the Gor'kov euqation $\mathcal{H}$
		by forcing a distinction between creation and anihilation operators.
		If without the functional derivative, it is obvios from Eq.(\ref{12223})
		that the first term of Gor'kov euqation $\mathcal{H}$ (realistic\cite{Berkooz2,Hu} kinetic term)
		follows the scaling invariance (nonchiral)
		and the effective potential in second term depends only on the relative shiftment between each field operators,
		
		Thus similar to the generalization in terms of the tensor models\cite{Narayan} that can realize
		the large-$N$ physics without including the randomness,
		e.g., the gauge invariant four
		point function can generates an analogue to the maximally chaotic one\cite{Hu,Narayan},
		an explicit breaking of translational invariance (in other word, the chirality)
		can be evidented by comparing the Eq.(\ref{1112}) and Eq.(\ref{1113}),
		which can be known (from below discussions) that originates from the distinctions
		between $\alpha$ and $\beta$ sectors of current density.
		Also, such breaking of translational invariance cannot be restored by the ensemble (or disorder) average
		over the different sites along the 1d spacial direction,
		but we note that it should be accessible to restoring the translational invariance
		by constructing proper filter for each operator
		that accommodates with the gauge invariance,
		but this beyond the topic of this article and we refer readers to Refs.\cite{Berkooz M,Berkooz2} for more informations about this method.

		\section{Theoretical preparation: IR cutoff in imaginary time component of the propagator}
		
		For fermionic fields the anticommution relation is strictly obeyed in terms of the time order 
		product ($\mathcal{T}[\cdot]$), but its no always the case for a normal ordering product ($:\cdot:$)
		when consider the Feynman propagator which need to satisfies the Lorentz invariance,
		\begin{equation} 
			\begin{aligned}
				\mathcal{T}[\psi_{i}(\tau)\psi_{j}^{\dag}(\tau')]=:\psi_{i}(\tau)\psi_{j}^{\dag}(\tau'):
				+\langle 0|\mathcal{T}[\psi_{i}(\tau)\psi_{j}^{\dag}(\tau')]|0 \rangle,
			\end{aligned}
		\end{equation}
		where $\langle 0|\cdot|0 \rangle$ measures the vacuum expectation value,
		and this term becomes zero when the product inside it has already be in the normal ordering,
		in which case $\mathcal{T}[\cdot]=\mathcal{T}[:\cdot:]=:\cdot:$.
		Due to the restriction of Lorentz invariance, the Feynman propagator
		reads $\langle 0|\mathcal{T}[\psi_{i}(\tau)\psi_{j}^{\dag}(\tau')]|0 \rangle
		=\langle 0|\psi_{i}(\tau)\psi_{j}^{\dag}(\tau')|0 \rangle$ when $\tau>\tau'$,
		and $\langle 0|\mathcal{T}[\psi_{i}(\tau)\psi_{j}^{\dag}(\tau')]|0 \rangle
		=\langle 0|-\psi_{j}^{\dag}(\tau')\psi_{i}(\tau)|0 \rangle$ when $\tau'>\tau$.
		Thus we can rewrite the above relation as (set $\tau'=\tau^{-}$)
		\begin{equation} 
			\begin{aligned}
				\mathcal{T}[\psi_{i}(\tau)\psi_{j}^{\dag}(\tau^{-})]
				=\mathcal{T}[-\psi_{j}^{\dag}(\tau^{-})\psi_{i}(\tau)]
				=:\psi_{i}(\tau)\psi_{j}^{\dag}(\tau^{-}):
				+\langle 0|\psi_{i}(\tau)\psi_{j}^{\dag}(\tau^{-})|0 \rangle,\\
				\mathcal{T}[\psi_{i}(\tau^{-})\psi_{j}^{\dag}(\tau)]
				=\mathcal{T}[-\psi_{j}^{\dag}(\tau)\psi_{i}(\tau^{-})]
				=:\psi_{i}(\tau^{-})\psi_{j}^{\dag}(\tau):
				+\langle 0|-\psi_{j}^{\dag}(\tau)\psi_{i}(\tau^{-})|0 \rangle,
			\end{aligned}
		\end{equation}
		where the field product are anticommute within time ordering bracket,
		and
		\begin{equation} 
			\begin{aligned}
				:\psi_{i}(\tau)\psi_{j}^{\dag}(\tau^{-}):=
				\psi_{j}^{\dag}(\tau)\psi_{i}(\tau^{-}),\\
				:\psi_{i}(\tau^{-})\psi_{j}^{\dag}(\tau):=
				\psi_{j}^{\dag}(\tau^{-})\psi_{i}(\tau),
			\end{aligned}
		\end{equation}
		which results in nonzero vacuum expectation values.
		
		Next we introduce the most essential basement for this article,
		the two-particle self-consistency theory\cite{Allen S} which is of the nonperturbative regime.
		By assuming $\tau$ close enough to the $\tau'$, i.e., setting a IR cutoff in the time domain,
		we have the following results for the fermionic Green's function in Nambu representation (
		we set the expectation value of number operators have here $(i,j)$ can be treated as pseudospin
		indices, $(n_{i}+n_{j})=1$,
		and the subscripts can be replaced by the up- and down-spin, i.e., 
		$n_{\uparrow}+n_{\downarrow}=1$):
		For the diagonal elements of Nambu Green's function matrix, we have
		\begin{equation} 
			\begin{aligned}
				G_{ii}(\tau,\tau^{+})=n_{i}=-\langle c_{i}(\tau)c_{i}^{\dag}(\tau^{+}),\\
				G_{jj}(\tau,\tau^{+})=-n_{j}=\langle c_{j}(\tau)c_{j}^{\dag}(\tau^{+})
				=G_{ii}(\tau,\tau^{-})=-\langle c_{i}(\tau)c_{i}^{\dag}(\tau^{-})
				=-G_{ii}(\tau^{+},\tau)=\langle c_{i}(\tau^{+})c_{i}^{\dag}(\tau),
			\end{aligned}
		\end{equation}
		where we conclude that, within a product,
		the exchange of fermion indices $i\rightleftharpoons j$
		equavalents to the exchange of imaginary time $\tau \rightleftharpoons \tau'$
		(which are infinitly close to each other).
		
		Then, instead of the general anticommutation relation for $\tau'=\tau$,
		which is $\{c_{i},c_{i}^{\dag}\}=1$,
		we have (according to $n_{i}-(-n_{j})=1$; and we omit the expectation notation hereafter)
		\begin{equation} 
			\begin{aligned}
				\label{141}
				&-c_{i}(\tau)c_{i}^{\dag}(\tau^{+})+c_{i}(\tau)c_{i}^{\dag}(\tau^{-})
				=-c_{i}(\tau)c_{i}^{\dag}(\tau^{+})-c_{i}(\tau^{+})c_{i}^{\dag}(\tau)\\
				&=-c_{j}(\tau)c_{j}^{\dag}(\tau^{+})+c_{j}(\tau)c_{j}^{\dag}(\tau^{-})
				=-c_{j}(\tau)c_{j}^{\dag}(\tau^{+})-c_{j}(\tau^{+})c_{j}^{\dag}(\tau)
				=1,
			\end{aligned}
		\end{equation}
		which is equivalent to
		\begin{equation} 
			\begin{aligned}
				n_{i}
				=-c_{i}(\tau)c_{i}^{\dag}(\tau^{+})
				=c_{i}(\tau^{-})c_{i}^{\dag}(\tau),\\
				n_{j}=1-n_{i}
				=-c_{i}(\tau^{+})c_{i}^{\dag}(\tau)
				=c_{i}(\tau)c_{i}^{\dag}(\tau^{-}).
			\end{aligned}
		\end{equation}
		
		Next we define
		\begin{equation} 
			\begin{aligned}
				\label{151}
				\{c_{i}(\tau),c_{i}^{\dag}(\tau^{+})\}=\delta_{\tau,\tau^{+}},\\
				\{c_{i}(\tau^{+}),c_{i}^{\dag}(\tau)\}=\delta_{\tau^{+},\tau},\\
				\{c_{i}(\tau),c_{i}^{\dag}(\tau^{-})\}=\delta_{\tau,\tau^{-}},\\
				\{c_{i}(\tau^{-}),c_{i}^{\dag}(\tau)\}=\delta_{\tau^{-},\tau},\\
			\end{aligned}
		\end{equation}
		It can be verified that all those delta functions are neither zero or one
		(by, e.g., inserting $c_{i}(\tau)c_{i}^{\dag}(\tau^{+})=-c_{i}^{\dag}(\tau^{+})c_{i}(\tau)$
		with assumption $\tau^{+}=\tau$ or
		$c_{i}(\tau)c_{i}^{\dag}(\tau^{+})=1-c_{i}^{\dag}(\tau^{+})c_{i}(\tau)$
		with assumption $\tau^{+}\neq \tau$ into Eq.(\ref{141})).
		We also have
		\begin{equation} 
			\begin{aligned}
				\label{152}
				c_{i}^{\dag}(\tau^{+})c_{i}(\tau)=\delta_{\tau,\tau^{+}}+n_{i},\\
				c_{i}^{\dag}(\tau)c_{i}(\tau^{+})=\delta_{\tau^{+},\tau}+n_{j},\\
				c_{i}^{\dag}(\tau^{-})c_{i}(\tau)=\delta_{\tau,\tau^{-}}-n_{j},\\
				c_{i}^{\dag}(\tau)c_{i}(\tau^{-})=\delta_{\tau^{-},\tau}-n_{i},
			\end{aligned}
		\end{equation}
		where the conservation law enfore
		\begin{equation} 
			\begin{aligned}
				\label{153}
				\delta_{\tau,\tau^{+}}+\delta_{\tau^{+},\tau}=-2,\\
				\delta_{\tau,\tau^{-}}+\delta_{\tau^{-},\tau}=2.
			\end{aligned}
		\end{equation}

		Next we expressing the above relations in terms of the Majorana fermions.
		We introduce the Majorana operators
		$\gamma_{2i}=\overline{\sigma_{i}^{x}}=c^{\dag}_{i}+c_{i}$,
		$\gamma_{2i-1}=\overline{\sigma_{i}^{y}}=-i(c_{i}-c_{i}^{\dag})$.
		But in the following section,
		we will using another definition of Majorana operators ($\chi_{i}$)
		which can better describe the phase feature of the system.
		We note that,
		two sectors ($\alpha$ and $\beta$) are defined in this system.
		In $\alpha$-sector,
		there is no imaginary time difference between the two Majorana operators,
		and due to the gauge-invariance restriction,
		their topological phase-dependent behaviors
		follows 
		\begin{equation} 
			\begin{aligned}
				\label{159}
				\chi_{i}\chi_{j}=\chi_{i}\chi_{i+n}
				=\gamma_{2(i-1)}\gamma_{2(i-1+n)-1},
			\end{aligned}
		\end{equation}
		for $i\ge 2$ and $n\ge 1$ is an positive integer,
		that means the system would always of the bosonic Majorana chain (and in the nontrivial phase) instead of the fermion one
		until $i=1$ (thus in the trivial phase).
		While in the $\beta$-sector where finite time difference exist,
		we have the definition just follows the above one
		$\chi_{k}(\tau)\chi_{l}(\tau')=\gamma_{2j-1}\gamma_{2j}$ (preserves the fermionic parity symmetry).
		But indeed, the unsual definition for the Majorana product in $\alpha$-sector is only due to the 
		gauge conservation-enforeced restriction 
		of the system, i.e., their topological phase exhibit unusual features due to the intrisic DOF
		of the system, and thus the resulting topological phase also affected by the inner DOFs correlations
		(like the DOFs in $\beta$-sector).
		Thus for $\beta$-sector which can be considered
		as fermionic Majorana chain, we obtain a conclusion which will be used below,
		\begin{equation} 
			\begin{aligned}
				\label{155}
				&\chi_{k}(\tau)\chi_{l}(\tau^{-})=-i
				[c^{\dag}_{i}(\tau)c_{i}(\tau^{-})-c_{i}(\tau)c^{\dag}_{i}(\tau^{-})]
				=-i[\delta_{\tau^{-},\tau}-1]\\
				&=\chi_{l}(\tau^{-})\chi_{k}(\tau)=-i
				[c_{i}(\tau^{-})c_{i}^{\dag}(\tau)-c_{i}^{\dag}(\tau^{-})c_{i}(\tau)]
				=-i[1-\delta_{\tau,\tau^{-}}].
			\end{aligned}
		\end{equation}
		Similarly,
		for $\tau'=\tau^{+}$,
		\begin{equation} 
			\begin{aligned}
				\label{156}
				&\chi_{k}(\tau)\chi_{l}(\tau^{+})=-i
				[c^{\dag}_{i}(\tau)c_{i}(\tau^{+})-c_{i}(\tau)c^{\dag}_{i}(\tau^{+})]
				=-i[\delta_{\tau,\tau^{+}}+1]\\
				&=\chi_{l}(\tau^{+})\chi_{k}(\tau)=-i
				[c_{i}(\tau^{+})c_{i}^{\dag}(\tau)-c_{i}^{\dag}(\tau^{+})c_{i}(\tau)]
				=-i[-1-\delta_{\tau^{+},\tau}],
			\end{aligned}
		\end{equation}
		which are consistent with Eqs.(\ref{151}-\ref{153}).
		It can also be verified that
		\begin{equation} 
			\begin{aligned}
				\label{157}
				\chi_{k}(\tau)\chi_{l}(\tau')=-\chi_{l}(\tau)\chi_{k}(\tau'),
			\end{aligned}
		\end{equation}
		while the effect for exchange of imaginary time cannot be easily identified
		\begin{equation} 
			\begin{aligned}
				\chi_{k}(\tau)\chi_{l}(\tau^{-})=-i(\delta_{\tau^{-},\tau}-1),\\
				\chi_{k}(\tau^{-})\chi_{l}(\tau)=-i(\delta_{\tau,\tau^{-}}-1),
			\end{aligned}
		\end{equation}
		but this relation is not needed in the following.

		\section{System}
		\subsection{Bosonic supercharge}

		To illustrating the restriction effect from the gauge field, we firstly define the bosonic supercharge
		and temporarily incorporating the real wave functions (which describe the eigenstates created by Majorana fermions) into each Majorana fermion operator,
		\begin{equation} 
			\begin{aligned}
				Q_{b}=i\sum_{i<k}\sum_{\alpha\beta}\chi_{i}(\alpha,\tau_{1})\chi_{k}(\beta,\tau_{2}),
			\end{aligned}
		\end{equation}
		where only the Majorana fermion operators of the same sector satisfy the anticommutation relation,
		i.e., $\{\chi_{i}(\alpha),\chi_{j}(\alpha)\}=\delta_{ij}$,
		$\{\chi_{k}(\beta),\chi_{l}(\beta)\}=\delta_{kl}$.
		And the Majorana fermions of different sectors satisfy $i<k,j<l$.
		While it always commutate with each other between operators in $\alpha$-sector and $\beta$-sector.
		Note that the following discussions will still be valid if we set
		$\tau_{1}=\tau_{2}$ but $\tau\neq \tau'$.

		In the absence of gauge invariance
		(both charges are of the same sector),
		this supercharge commutes with the Hamiltonian
		\begin{equation} 
			\begin{aligned}
				H&=Q^{2}_{b}\\
				&=
				(i\sum_{i<k}\sum_{\alpha}\chi_{i}(\alpha,\tau_{1})\chi_{k}(\alpha',\tau_{2}))
				(i\sum_{j<l}\sum_{\alpha}\chi_{j}(\alpha,\tau_{1})\chi_{l}(\alpha',\tau_{2})),
			\end{aligned}
		\end{equation}
		and the system becomes of the usual four-DOF regime.
		Note that the cases $i=k$ (or $j=l$) are precluded by the nonzero correlations
		between $i$ and $j$ (or $k$ and $l$) which are of the same sector and thus required to be antisymmetry with respect to each other.
		This also requires two supercharges satisfying such correlating configuration should
		not be the replica part like $Q_{b}^{*}Q_{b}$,
		instead, their correlation should be the same type with that between two U(1) gauge fields:
		\begin{equation} 
			\begin{aligned}
				\label{AA}
				[A_{\alpha\beta},A_{\alpha'\beta'}]
				=\sum_{\gamma<\nu}[\delta_{\alpha\alpha'}(\delta_{\beta\gamma}\delta_{\beta'\nu}-\delta_{\beta\nu}\delta_{\beta'\gamma})
				-\delta_{\beta\beta'}(\delta_{\alpha\gamma}\delta_{\alpha'\nu}-\delta_{\alpha\nu}\delta_{\alpha'\gamma})]A_{\gamma\nu}.
			\end{aligned}
		\end{equation}
		Note that we set $\alpha(\alpha')<\beta(\beta')$ to make $A_{\alpha\beta}$ and $A_{\alpha'\beta'}$ of the same sign.
		The antisymmetry property of gauge field also signals the broken time-reversal symmetry.
		The summation over all gauge fields with different $(\alpha,\beta)$ indices is invariant.

		If we restrict the group of imaginary time evolution $\mathcal{M}_{\tau\rightarrow \tau'}$
		has a fixed number of elements, 
		this will impose some restrictions to the selection of the indices $i,j,k,l$.
		For example, by looking at the part with degree-of-freedom $\beta$,
		the evolution from $\tau$ to $\tau'$,
		has $M^{2}$ elements given by the random combinations of $\chi_{j}(\beta,\tau_{2})$ and $\chi_{l}(\beta,\tau'_{2})$,
		and we assume this is the maximal size of the group $\mathcal{M}_{\tau\rightarrow \tau'}$
		which consists $M^{2}$ elements $m_{j\tau\rightarrow l\tau'}$,
		and cannot be enlarged in the $\alpha$ subspace.
		Every single mapping in $\beta$ subspace $m_{j\tau\rightarrow l\tau'}$
		should mostly has $M$ "copies" in $\alpha$ subspace,
		which follow the same mapping.
		Thus to make sure the group $\mathcal{M}_{\tau\rightarrow \tau'}$ still has $M^2$ irreducible
		elements,
		there can only be $M$ elements in $\alpha$ subspace,
		which can be represented by $m_{i\tau\rightarrow i\tau'}$, or equivalently, $m_{i\tau\rightarrow k\tau}$.
		According to the above definition of current density with gauge-fixing restrictions,
		the time evolution group $\mathcal{M}_{\tau\rightarrow \tau'}$ with fixed number of elements
		is a direct result of independence between
		the operator evlution speed
		$\partial_{\tau}\chi(\tau)=\chi(\tau')$ and the variance after each time step $\chi(\tau)$.

		\subsection{Supersymmetries of gauge invariant system}
		
		For a supercharge product satisfying the gauge invariance introduced by the U(1) gauge field $A_{\alpha\beta}$
		(see Eq.(\ref{AA})), 
		there should be only three mutually independent Majorana indices,
		which is also equivalents to the case with four mutually independent Majorana indices
		but only one sectors owns the full evolution group $\tau\rightarrow \tau'$
		whose size is constrained by the gauge invariance.
		Now the supercharge product reads
		\begin{equation} 
			\begin{aligned}
				\label{158}
				Q^{2}_{b}
				=&
				(i\sum_{i<k}\chi_{i}(\alpha,\tau)\chi_{k}(\beta,\tau))
				(i\sum_{j<l}\chi_{j}(\alpha,\tau)\chi_{l}(\beta,\tau')\\
				=&-\sum_{kl}\sum_{\{\overline{k},\overline{l}\}=1}^{M}
				[\sum_{\{\overline{i}\overline{j}\}}^{\{\overline{k}\overline{l}\}-1}
				\varphi_{\overline{i}}(\alpha,\tau)\varphi_{\overline{j}}(\alpha,\tau)
				\chi_{i}\chi_{j}]
				\chi_{k}(\beta,\tau)\chi_{j}\beta,\tau')\\
				=&-\sum_{kl}\sum_{\beta}^{\beta'-1}
				[-i\sum_{\alpha}^{\beta-1}c_{\alpha}^{\dag}c_{\alpha}]\chi_{k}(\beta,\tau)\chi_{l}(\beta,\tau')\\
				=&\sum_{kl}\sum_{\beta}^{\beta'-1}
				\frac{\Phi(\beta)}{2\pi}\chi_{k}(\beta,\tau)\chi_{l}(\beta,\tau')\\
				=&i\sum_{kl}\sum_{\beta}g_{kl,\beta}\chi_{k}\chi_{l},
			\end{aligned}
		\end{equation}
		where $\{\overline{i}\overline{j}\}$ denotes the selected chiral Majorana fermion fields
		which form the multiple boson field.
		We define the summations
		\begin{equation} 
			\begin{aligned}
				&\sum_{\{\overline{i},\overline{j}\}}^{\{\overline{k},\overline{l}\}-1}
				:=
				\sum_{\overline{i}<\overline{j}}\epsilon_{\overline{i}\overline{j}\overline{k}\overline{l}}\theta(\overline{k}-\overline{l})
				+\sum_{\overline{i}>\overline{j}}\epsilon_{\overline{j}\overline{i}\overline{l}\overline{k}}\theta(\overline{l}-\overline{k}),\\
				&\sum_{\beta}\sum_{\alpha}^{\beta-1}
				:=\sum_{\{\overline{k},\overline{l}\}=1}^{M}\sum_{\{\overline{i},\overline{j}\}}^{\{\overline{k},\overline{l}\}-1}.
			\end{aligned}
		\end{equation} 
		For a chiral current,
		the Majorana fermions in $\alpha$-sector has only the reduced degrees-of-freedom 
		(thus the summation over $(i,j)$ can be incorporated into the summation over $\overline{i}\overline{j}$)
		and the boundary of summation over $(i,j)$-states is also soly determined by $\{\overline{k},\overline{l}\}$.
		The size of group $\{k,l\}$ is $\sim O(N^{2}M)$ while that of
		$\{\overline{k},\overline{l}\}$ is $O(M)$,
		which means,
		although the above $Q_{b}^{2}$ term is defined initially in a similar way with the Wishart SYK model,
		but under the restricts of gauge invariance,
		it finally turns to be the one similar to the single supercharge defined in $\mathcal{N}=1$ supersymmetry SYK model\cite{Fu W}
		but with a $N\times N\times \sqrt{N}$ antisymmetry tensor instead of the $N\times N\times N$ one.
		Also,
		as can be seen from above expression (cf. Eq.(\ref{T4}) as well as Eqs.(\ref{155}-\ref{157})
		), the Hermiticity (permutations under antisymmetry exchanges) only exist 
		between $\alpha$ and $\beta$ sectors (i.e., $\overline{i}$ and $\overline{j}$, $\overline{k}$ and $\overline{l}$).
		
		Here the multiple boson field is defined in terms of the 
		hybridizing Majorana fermions through the 1D quadratic coupling in bilinear form
		\begin{equation} 
			\begin{aligned}
				\frac{\Phi(\beta)}{2\pi}=i\sum_{\alpha}^{\beta-1}\rho_{\alpha}
				=i\sum_{\{\overline{i}\overline{j}\}}\varphi_{\overline{i}}(\alpha,\tau)\varphi_{\overline{j}}(\alpha,\tau)\chi_{\overline{i}}\chi_{\overline{j}}.
			\end{aligned}
		\end{equation} 
		Due to the Hermiticity, this boson field plays the role of antisymmetry tensor within the expression of supercharge product,
		due to the antisymmetry result under exchange
		$\varphi_{\overline{i}}(\alpha,\tau)\varphi_{\overline{j}}(\alpha,\tau)=-\varphi_{\overline{j}}(\alpha,\tau)\varphi_{\overline{i}}(\alpha,\tau)$. 
		That makes the boson field be a prefect probability amplitude factor within the product
		and it sums to be zero by its own (in absent of the selection effect of the latter terms related to $\beta$),
		$\langle \Phi(\beta)\rangle=0$.
		According to the current density defined above (Sec.2.2 and Appendix.A),
		for a certain sector ($\beta$ here),
		this boson field, which is the summation over current densities,
		can be transformed into a single current density located at the upper boundary of previous summations
		above a half filling which is a conserved quantity for a certain sector.
		Thus the above antisymmetry-induced cancellation means 
		the current density cannot exists in the absence of gauge invariance.
		If we recollect the $\chi_{j}(\alpha,\tau_{1})$ and $\chi_{k}(\beta,\tau_{2})$ to the same sector $\gamma$,
		i.e., 
		$\chi_{j}(\alpha,\tau)\chi_{k}(\beta,\tau)\delta_{jk}\rightarrow
		\chi_{j}(\gamma,\tau)\chi_{k}(\gamma,\tau')\delta_{jk}$ ($\tau\neq\tau'$).
		While the operators $\chi_{i}(\alpha,\tau)$ and $\chi_{l}(\beta,\tau)$ will also be classified into the same sector
		automatucally (denoted as $\{\eta\}$). This is because classification of sectors simply depends on the different kinds of
		interacting relations (symmetry/antisymmetry) within each sector,
		and now we have 
		$\chi_{i}(\alpha,\tau)\chi_{l}(\beta,\tau')\rightarrow
		\chi_{i}(\eta,\tau)\chi_{l}(\eta,\tau')$.
		It can be verified that this will be still the three-DOF configuration like the previous one.
		This can be verified by checking the commutation/anticommutation relations between arbitrarily two of
		those four Majoarana operators (see Appendix.D),
		where we can find three anitisymmetry parts and three symmetry parts,
		in contrast to the four-DOF one where there will be four antisymmetry parts and two symmetry parts.
		
		After the selection effect due to the relations Eqs.(\ref{155}-\ref{157}),
		there will be only the permutations meet $(i<k\le j<l)$ or $(j<l\le i<k)$ survive.
		It can be easily understood why other permutations all compensate with each other.
		Note that here we temporately omit the cases with $i=j$ and $k=l$,
		which will results in the product in form of
		\begin{equation} 
			\begin{aligned}
				\chi_{i}(\tau)\chi_{i}(\tau)\chi_{k}(\tau)\chi_{k}(\tau^{\pm})
				=\delta_{\tau,\tau^{\pm}}+1-2n_{i}.
			\end{aligned}
		\end{equation} 
		For all permutations meets $i=j$ but $k\neq l$ or $k=l$ but $i\neq j$ ,
		the cancellation happen as
		\begin{equation} 
			\begin{aligned}
				&\chi_{i=a}(\tau)\chi_{k=b}(\tau)\chi_{j=a}(\tau)\chi_{l=d}(\tau')
				+\chi_{i=a}(\tau)\chi_{k=d}(\tau)\chi_{j=a}(\tau)\chi_{l=b}(\tau')
				=0,\\
				&\chi_{i=a}(\tau)\chi_{k=b}(\tau)\chi_{j=c}(\tau)\chi_{l=b}(\tau')
				+\chi_{i=c}(\tau)\chi_{k=b}(\tau)\chi_{j=a}(\tau)\chi_{l=b}(\tau')
				=0,
			\end{aligned}
		\end{equation} 
		respectively.
		For all permutations meets $k>j$ and $l>i$ 
		the cancellation happen as
		\begin{equation} 
			\begin{aligned}
				&\chi_{i=a}(\tau)\chi_{k=b}(\tau)\chi_{j=c}(\tau)\chi_{l=d}(\tau')
				+\chi_{i=a}(\tau)\chi_{k=d}(\tau)\chi_{j=c}(\tau)\chi_{l=b}(\tau')
				=0,
			\end{aligned}
		\end{equation} 
		respectively.
		For all permutations meets $k>j$ and $l>i$ 
		the cancellation happen as
		\begin{equation} 
			\begin{aligned}
				&\chi_{i=a}(\tau)\chi_{k=b}(\tau)\chi_{j=c}(\tau)\chi_{l=d}(\tau')
				+\chi_{i=a}(\tau)\chi_{k=d}(\tau)\chi_{j=c}(\tau)\chi_{l=b}(\tau')
				=0.
			\end{aligned}
		\end{equation} 
		respectively.
		Note that due to the absence of anticommutation relation between the two sectors,
		we have $\chi_{i=a}(\tau)\chi_{k=b}(\tau')=\chi_{i=b}(\tau)\chi_{k=a}(\tau')$.
		As only the permutations of the forms $(i<k\le j<l)$ and $(j<l\le i<k)$ survive,
		and the corresponding $(k,l)$-states would also cancel each other,
		we will consider only the $(i<k\le j<l)$ case here.

		Then the summation over couplings has the following relation
		In above gauge-invariant charge product,
		the coupling is an antisymmetry tensor and the summation over couplings has the following relation
		\begin{equation} 
			\begin{aligned}
				\label{g3}
				\sum_{\{\overline{k},\overline{l}\}}g_{kl,\beta}
				=&\sum_{\{\overline{k}\overline{l}\}}\frac{\Phi(\beta)}{2\pi}\varphi_{k}(\beta,\tau)\varphi_{l}(\beta,\tau')\\
				=&i\sum_{\{\overline{k}\overline{l}\}}
				\frac{1}{2}\sum_{\{\overline{i}<\overline{j}\}}
				[\varphi_{\overline{i}}(\alpha,\tau)\varphi_{\overline{j}}(\alpha,\tau)+\varphi_{\overline{j}}(\alpha,\tau)\varphi_{\overline{i}}(\alpha,\tau)]\\
				&\times
				\frac{1}{2}	[\varphi_{\overline{k}}(\beta,\tau)\varphi_{\overline{l}}(\beta,\tau')+\varphi_{\overline{l}}(\beta,\tau')\varphi_{\overline{k}}(\beta,\tau)]
				\chi_{\overline{i}}\chi_{\overline{j}}\delta_{\overline{j},\overline{k}}\\
				=&i\sum_{\{\overline{i}<\overline{j}=\overline{k}<\overline{l}\}}
				\frac{1}{2}
				[\varphi_{\overline{i}}(\eta,\tau)\varphi_{\overline{j}}(\gamma,\tau)+\varphi_{\overline{j}}(\gamma,\tau)\varphi_{\overline{i}}(\eta,\tau)]\\
				&\times
				\frac{1}{2}	[\varphi_{\overline{k}}(\gamma,\tau')\varphi_{\overline{l}}(\eta,\tau')+\varphi_{\overline{l}}(\eta,\tau')\varphi_{\overline{k}}(\gamma,\tau')]\delta_{\overline{j},\overline{k}}\\
				=&-i\sum_{\{\overline{i}<\overline{j}=\overline{k}<\overline{l}\}}
				\varphi_{\overline{j}}(\gamma,\tau')\varphi_{\overline{k}}(\gamma,\tau')\varphi_{\overline{i}}(\eta,\tau)\varphi_{\overline{l}}(\eta,\tau')\delta_{\overline{j},\overline{k}}\chi_{i}\chi_{j}\\
				=&-i\frac{1}{2}\sum_{\overline{l}}^{\overline{l}'-1}
				[\sum_{\overline{i}}^{\overline{l}-1}\varphi_{\overline{i}}(\eta,\tau)]\varphi_{\overline{l}}(\eta,\tau')]
				\chi_{\overline{i}}\chi_{\overline{l}}\delta_{\overline{l}'-1,\overline{j}}\\
				=&-i\frac{1}{2}\sum_{\overline{i}<\overline{l}}\chi_{\overline{i}}(\eta,\tau_{2})\chi_{\overline{l}}(\eta,\tau)\\
				=&-\frac{1}{2}Q_{b},
			\end{aligned}
		\end{equation}
		which reduces to half of a single supercharge owns the full time evolution.
		As can be seen, at the special points where $\overline{j}=\overline{k}$,
		these two operators forms a noninteracting fermion,
		\begin{equation} 
			\begin{aligned}
				\lim_{\tau\rightarrow \tau'}\chi_{j}(\gamma,\tau)\chi_{k}(\gamma,\tau')\delta_{jk}=\frac{1}{2}.
			\end{aligned}
		\end{equation}
		Here the instantaneous $\tau\rightarrow \tau'$ is guaranteed by fail of low-energy approximation here
		due to the absence of strong interacting boson self-energy and disorder.
		That is also why $\tau\neq \tau'$ in the $\eta$-sector where the bosonic self-energy dominates over the Matsubara frequency.

		\subsection{Numerical results}
		
		It can be calculated that 
		the number of all survival $(i,k,j,l)$ combinations is $
		\sum_{i=2}^{N-1}(i-1)[\sum_{j=0}^{N-i-1}(N-i-j)]=
		\frac{1}{24} N (2 - N - 2 N^2 + N^3)$
		and
		the number of distinguishing survival $(k,l)$-states is $\frac{1}{2}(N-2)(N-1)\approx
		\frac{N^{2}}{2}$.
		If we further classifies those survival $(k,l)$-states by the value $|k-l|$,
		we will see that the number of each $(k,l)$-state
		(i.e., the number of corresponding bosonic Majorana $(i,j)$-states) is always
		$|k-l|(k-1)$,
		this can be seem as a weight for each distinguishing $(k,l)$-state.

		The states number for each distinguishing $(k,l)$ state are shown in Fig.1(a),
		and the many-body level spacing distribution is shown in Fig.1(c),
		the averaged spacing ratio is $\langle r\rangle \sim 0.66$,
		which is of the GSE.
		After we classifies them into groups each with distinguishing values of $|k-l|$
		(of the order of $N$),
		the corresponding number density distribution is shown in Fig.1(b).
		For each $(k,l)$-state, the distributions of the value of $|k-l|$ are shown in Fig.2,
		where we can see 
		the many-body level spacing distribution follows the GUE with $\langle r\rangle\sim 0.59$ 
		(Fig.2(b)) when we
		ignore the weight $|k-l|(k-1)$ on each state,
		and follows GOE with $\langle r\rangle\sim 0.52$ (Fig.2(d)) when the weighted distributions are considered.

		As shown in Fig.1(d),
		through the statistical properties,
		for each $(k,l)$-state of the class $|k-l|=1$,
		their corresponding $(i,j)$-states should exhibit the same topological feature.
		Using the above Majorana operator of $\alpha$-sector in Eq.(\ref{159}),
		we can explain the classification criterion of $(i,j)$-states for each $(k,l)$-state
		of the same difference $|k-l|$.
		In $\alpha$-sector, we have 
		\begin{equation} 
			\begin{aligned}
				\label{1119}
				&\chi_{1}\chi_{1+n}=\chi_{1}\chi_{n+2}+\chi_{2}\chi_{n+2},\\
				&\chi_{i}\chi_{j}=\chi_{i}\chi_{i+n}
				=\gamma_{2(i-1)}\gamma_{2(i-1+n)-1},
			\end{aligned}
		\end{equation}
		where integers satisfy $i\ge 2$, $j\ge 3$ and $n\ge 1$,
		e.g., for $n=|k-l|=1$,
		we have
		\begin{equation} 
			\begin{aligned}
				&\chi_{2}\chi_{3}=\sigma_{1}^{x}\sigma_{2}^{y}
				=\overline{\sigma_{1}^{x}}\overline{\sigma_{2}^{y}}
				=\gamma_{2}\gamma_{3},\\
				&\chi_{3}\chi_{4}=\sigma_{2}^{y}\sigma_{2}^{x}
				=\overline{\sigma_{2}^{x}}\overline{\sigma_{3}^{y}}
				=\gamma_{4}\gamma_{5},\\
				&\chi_{4}\chi_{5}=\sigma_{2}^{x}\sigma_{3}^{x}
				=\overline{\sigma_{3}^{x}}\overline{\sigma_{4}^{y}}
				=\gamma_{4}\gamma_{5},\cdots;
			\end{aligned}
		\end{equation}
		and for $n=|k-l|=2$,
		we have
		\begin{equation} 
			\begin{aligned}
				&\chi_{2}\chi_{4}=\sigma_{1}^{x}\sigma_{2}^{x}
				=\overline{\sigma_{1}^{x}}\overline{\sigma_{3}^{y}}
				=\gamma_{2}\gamma_{5},\\
				&\chi_{3}\chi_{5}=\sigma_{2}^{y}\sigma_{2}^{x}
				=\overline{\sigma_{2}^{x}}\overline{\sigma_{4}^{y}}
				=\gamma_{4}\gamma_{7},\\
				&\chi_{4}\chi_{6}=\sigma_{2}^{x}\sigma_{3}^{x}
				=\overline{\sigma_{3}^{x}}\overline{\sigma_{5}^{y}}
				=\gamma_{6}\gamma_{9},\cdots.
			\end{aligned}
		\end{equation}
		For $j>i\ge 2$,
		all $(i,j)$-states with the same value of $|i-j|$
		can be classified into the same group.
		While the states with $i=1$ corresponds to the trivial phase with conserved fermionic charge,
		and without the degenerated group state.
		For example, as shown in Fig.1(d),
		for $|k-l|=1$,
		we have
		\begin{equation} 
			\begin{aligned}
				&\chi_{1}\chi_{2}=\chi_{1}\chi_{3}+\chi_{2}\chi_{3},
			\end{aligned}
		\end{equation}
		as topologically $\chi_{2}\chi_{3}=\chi_{3}\chi_{4}$,
		we can further have
		\begin{equation} 
			\begin{aligned}
				&\chi_{1}\chi_{2}=\chi_{1}\chi_{4}+\chi_{2}\chi_{4}+\chi_{3}\chi_{4}.
			\end{aligned}
		\end{equation}
		We also notice that,
		for a certain class, the rule for $(i,j)$ parts exhibits an ergodic feature:
		no matter how many the nodes is,
		the $(i,j)$ parts always making the maximal value of $\sum |i-j|$
		under the restriction that each node can only be throughed once.
		While for $|k-l|>1$,
		there will be more $(i,j)$ parts with $i=1$.
		A explaination for the replacement between $\chi$ and $\gamma$ are presented in Sec.2.3.


		\section{Current-density flavor: the $M$-dependent statistics}
		
		\subsection{Three-DOF configuration in BDI symmetry class: $O(M)=O(N)$}
		
		Firstly we discuss the case in symmetry class BDI,
		where the many-body localization-induced symmetry protected topological states
		exist as we illustrated in Sec.2.2.
		This is a symmetry class different to the AIII or CII clsses,
		the bulk states
		Base on current density flavors defined above (Eq.(\ref{sigma})),
		as long as the Majorana chain is dominated by many-body localization,
		the chiral current densities can be classified by
		$M=N-2$ categories.
		Each one of these $M$ categories of current density with $\beta=1,2,\cdots, M$
		can be expressed in terms of 
		\begin{equation} 
			\begin{aligned}
				\label{M1}
				\frac{\Phi(\beta)}{2\pi}=\sum_{\overline{j}\ge 2}^{\beta-1}\chi_{\overline{i}=1}\chi_{\overline{j}},
			\end{aligned}
		\end{equation}
		which is the relation Eq.(\ref{1119}).
		
		For charge product $\chi_{i}\chi_{j}\chi_{k}\chi_{l}$,
		since initially $(i,j)$ and $(k,l)$ are of different sectors,
		the delta-function $\delta_{\{\overline{j}\},\{\overline{k}\}}$ in second line should means
		the upper summing boundary of $\overline{l}$ is the same with that of $\overline{i}$ 
		(or the size of group $\{\overline{l}\}$ is the same with that of $\{\overline{i}\}$).
		This is how such gauge-invariant-type correlation happen and relating two supercharges to generate a
		3-DOF configuration. As the groups $\{\overline{j}\}$ and $\{\overline{k}\}$ are of different sectors
		($\alpha$ and $\beta$, respectively),
		and have not additional extension in size (unlike the group of $\{k,l\}$),
		then there differences (relative degrees-of-freedom) can be regarded as degenerated into a single one which is of that between two the sectors
		($\{\alpha,\beta\}=\{\overline{i},\overline{j}\}\otimes \{\overline{k},\overline{l}\}=O(M)O(M)=O(M^{2})$).
		Once $\delta_{\overline{j},\overline{k}}=1$,
		the gauge invariance reduce the size of subsystem consisting of $\{\overline{i}\}$ and $\{\overline{l}\}$
		from the $O(M^{2})$ to $O(M)$.
		In this case, there will be a three-DOF configuration
		which corresponds to completely localized Majorana modes in the bulk part of the 1D chain as we 
		discussed in Sec.2.3,
		and the $(N-1)(N-2)/2$ elements in the group of $\{\overline{k},\overline{l}\}$
		are occupied by the $M\approx (N-2)$ flavors of current density,
		unevenly.
		This nonuniform occupation (with a gradient)results from the gauge invariant restriction.
		In terms of the probability distribution (see Appendix.B),
		this can be explained more clearly in the following way.
		As now each $(k,l)$-state has the same probability to be occupied by each one of the $M$ current flavor,
		the probabilities for each flavor is 
		\begin{equation} 
			\begin{aligned}
				P_{\beta=1,2,\cdots, M}=\frac{1}{M}\sum_{\overline{k}\overline{l}}\varphi_{\overline{k}}\varphi_{\overline{l}}
				=1/M,
			\end{aligned}
		\end{equation}
		which is a constant.
		While by performing the Schmidt decomposition to $(\overline{k},\overline{l})$-states,
		the corresponding probability for each "Schmidt vector" is variant
		\begin{equation} 
			\begin{aligned}
				&P_{\overline{k}=2,3,\cdots, N-1}=
				\sum_{n=1}^{N-2}\frac{1}{n},\sum_{n=2}^{N-2}\frac{1}{n},\cdots,\frac{1}{N-2},\\
				&P_{\overline{l}=3,4,\cdots, N}=
				\frac{1}{N-2},\cdots,\sum_{n=2}^{N-2}\frac{1}{n},\sum_{n=1}^{N-2}\frac{1}{n}.
			\end{aligned}
		\end{equation}
		In this case, each part of $(\overline{k},\overline{l})$ 
		has different probability ($\varphi_{\overline{k}}\varphi_{\overline{l}}$) which depends
		on the value of $|\overline{k}-\overline{l}|$.
		This probability is not commutate with the number of $(\overline{k},\overline{l})$-states,
		since $P_{\beta=1,2,\cdots, M}$ is a constant and the probabilities have to sum up to one.
		
		It can be verified that only the combinations satisfy
		$i<k\le j<l$ servive (the another possible case $j<l\le i<k$ will not be considered here).
		For the 3-DOF configuration,
		as each flavor of current density owns equal weight (represented by $(i,l)$ part here)
		with respect to each part of Majorana indec
		(represented by $(k,l)$ here),
		it requires the probability reads
		\begin{equation} 
			\begin{aligned}
				P_{k<l}=\frac{1_{M}}{M+\partial_{\Delta_{kl}}\Delta_{kl}-\Delta_{kl}},
			\end{aligned}
		\end{equation}
		where $1_{M}=1/M$ denotes the sum over all probabilities of current densities (or $(i,j)$-part)
		of the same sector,
		and that ensures the equal weight for each sector of current density.
		$\Delta_{kl}=l-k$ denotes the distance between two Majorana indices (in terms of the unit step 
		in space of Majorana DOF).

		
		From the expression of above probability we can obtain
		\begin{equation} 
			\begin{aligned}
				\partial_{\Delta_{kl}}\Delta_{kl}=
				\lim_{\Delta_{kl}\rightarrow \infty} 1
				&=\frac{1_{M}P^{-1}_{k<l}}{M+\partial_{\Delta_{kl}}\Delta_{kl}-\Delta_{kl}},
			\end{aligned}
		\end{equation}
		where $\partial_{\Delta_{kl}}\Delta_{kl}=-1$ is the only $\Delta_{kl}$-independent quantity here,
		and, as we discuss in Sec.7.3, this can be explained by introducing another framwork
		consist of new unit $1'$,
		$-1=\lim_{\Delta_{kl}\rightarrow \infty}1=\frac{1}{1'-\partial_{\Delta_{kl}}1}
		=1-\frac{\Delta_{kl}}{\lim_{\Delta_{kl}\rightarrow \infty}\Delta_{kl}}$.
		Then we can obtain 
		\begin{equation} 
			\begin{aligned}
				\partial_{\Delta_{kl}}(1_{M}P^{-1}_{k<l}
				=\partial_{\Delta_{kl}}(\Delta_{kl}-M)
				=\Delta_{kl}-M,
			\end{aligned}
		\end{equation}
		which is cnsistent with the relation
		(obtained from $\partial_{\Delta_{kl}}
		[\partial_{\Delta_{kl}}\Delta_{kl}]=\partial_{\Delta_{kl}}
		\frac{\Delta_{kl}-M}{\partial_{\Delta_{kl}}[-(\Delta_{kl}-M)]}=0$)
		\begin{equation} 
			\begin{aligned}
				-\frac{\partial_{\Delta_{kl}}(\Delta_{kl}-M)}
				{\Delta_{kl}-M}
				=\partial_{\Delta_{kl}}{\rm ln}\frac{1}{\partial_{\Delta_{kl}}[-(\Delta_{kl}-M)]}
				=-1
			\end{aligned}
		\end{equation}
		Thus the dependence of $M$ on the Majorana-index-difference $\Delta_{kl}$
		can be described by
		\begin{equation} 
			\begin{aligned}
				\partial_{\Delta_{kl}}M=-1-(\Delta_{kl}-M)
				=-1-\frac{-1-(\Delta_{kl}-M)}{-1-\frac{\Delta_{kl}-M}{-1-(\Delta_{kl}-M)}},
			\end{aligned}
		\end{equation}
		where the condition for DOFs $O(N)\sim O(M)$
		enforce $\lim_{\Delta_{kl}\rightarrow \infty}M={\rm max}\ \Delta_{kl}
		=-1-\frac{\Delta_{kl}-M}{-1-(\Delta_{kl}-M)}$ here.

		\subsection{Numerical simulations}

		The system exhibits GSE statistics when it meets the condition $O(M)\sim O(N)$.
		Since now the size of $\{k,l\}$ group is of order $O(N^2)$ and much larger than that of current densities
		$O(M)$,
		the many body localization inside the Majorana chain will naturely leads to thermalization in the 
		statistical behaviors of the chiral current densities which are
		randomly occupied by each $(k,l)$-state.
		As shown in Fig.4(a), the current density flavor distribution
		versus $(k,l)$-states plot exhibits a nearly constant slope,
		and the many-bodylevel statistics (the probability $P_{k<l}$
		exhibits GSE distribution.
		The average value of spacing ratio is $\langle r\rangle \sim 0.701$.

		When $O(M\sim O(N^2)$,
		the size of group $\{\overline{k},\overline{l}\}$ is then nearly the same with that of current density,
		and this results in a constant probability distribution in 
		both the current density group and $\{\overline{k},\overline{l}\}$ group,
		$P_{\beta=1,2,\cdots, M}=P_{\overline{k},\overline{l}}=1/M=1/N^2$.
		This will results in a four-DOF configuration with $\begin{pmatrix} N\\ 4\end{pmatrix}\sim N^4$
		mutually independent terms whose distribution follows the SYK$_{4}$ model.
		The corresponding numerical simulations are shown in Fig.5,
		where we set $O(M)\sim O(N^2)$.
		the corresponding many-body level distribution of current densities in this configuration
		still
		follows the Wigner-Dyson statistic
		and thus the DOF of chiral current density is still thermalized
		(quantum chaotically),
		and we found $\langle r\rangle \sim 0.603$.
		
		When the magnitude of $M$ exceeds $N^{2}$, which is,
		when the flavor number of current density beyonds the critical chaotic
		condition for the four-DOF configuration,
		the spectrum tends to Poisson distribution
		(completely dominated by Poisson distribution when $O(M)\sim O(N^{4})$)
		which signals the Anderson-localized insulating state for current densities
		in contrast to the disordered metal state.
		The simulation for cases with $O(M)> O(N^{2})$
		can be realized by adding a randomness factor to the previous GUE case.
		For each index $k(=2,\cdots,N-1)$,
		the randomness factors could be 
		$\updelta_{1}=\frac{(1,\cdots,N-\Delta_{kl})}{N-\Delta_{kl}}$,
		$\updelta_{2}=\frac{(1,\cdots,\Delta_{kl})}{\Delta_{kl}}$,
		$\updelta_{3}=\frac{(1,\cdots,N-k)}{N-k}$,
		$\updelta_{4}=\frac{(1,\cdots,N^2)}{N^2}$.
		Only the $\updelta_{3}$ and $\updelta_{4}$ factors correspond to the Poisson case.
		while $\updelta_{1}$ and $\updelta_{2}$ factors correspond to a
		intermediate state between GSE and Poisson statistics.
		As should in Fig.6,
		the $\updelta_{3}$ and $\updelta_{4}$ factors exhibit the same effect
		and both realizing the $O(M)=O(N^{4})$ simulation.
		This is also evidented by the averaged level spacing ratios
		obtained for these for randomness factors:
		$\langle r\rangle_{\updelta_{1}}=0.4705,\ 
		\langle r\rangle_{\updelta_{2}}=0.4535,\ 
		\langle r\rangle_{\updelta_{3}}=0.3954,\ 
		\langle r\rangle_{\updelta_{4}}=0.3853$.
		For these two case,
		we also perform the Many-body level statistics in term of $P({\rm ln}r)$
		in Fig.7,
		where the average level spacing are close to the results of Fig.6,
		$\langle r\rangle_{\updelta_{1}}=0.4715,\ 
		\langle r\rangle_{\updelta_{2}}=0.4532,\ 
		\langle r\rangle_{\updelta_{3}}=0.3975,\ 
		\langle r\rangle_{\updelta_{4}}=0.387$.

		\subsection{Automatically realized distribution beyonf Poisson distribution in
			AIII or CII symmetry classes}

		The Poisson distribution for current density spectrum here corresponds to the
		Anderson-localized 1D bulk in the AIII or CII symmetry classes.
		In the mean time, as the current density flavor number $M$ increase to of order $O(N^4)$,
		the many-body localization inside the Majorana chains is overwhelmed by the thermalization effect,
		and the characters of Majorana chain are no longer simply determined by the boundary zero modes.
		As we stated above, in the AIII or CII symmetry classes,
		the many-body localization inside the Majorana chains is overwhelmed by the thermalization effect,
		and the characters of Majorana chain are no longer simply determined by the boundary zero modes,
		but all the effective modes,
		which can be viewed as edges of several parallel chains
		(which can describe the, e.g., Anderson-localized 1D current densities where the interactions only
		exist at the 0D boundaries).
		This also leads to the absence of translational invariance and degenerate ground state
		in the 1D Majorana chain.
		Then the Eq.(\ref{M1}) is no more valid,
		and the Majorana modes in bulk are no more the frustration-free localized eigenstates
		of $Q_{b}^{2}$,
		i.e., each one of them it becomes depending on their summation upper boundary.
		For example, now the fermion bilinear term should be $\chi_{1}\chi_{2}=\chi_{2}\chi_{3}$ (in $\alpha$-sector),
		which indicates a phase transition from the topological Haldane-type nontrivial one to the trivial one.
		
		The AIII or CII symmetry classes correspond to the noninteracting 
		Anderson-localized bulk for the 1D system with a $\mathcal{Z}$ invariance
		and, respectively, processing the $\mathcal{Z}_{4}$ and $\mathcal{Z}_{2}$ classifications
		at the thermalized edges\cite{You Y Z}.
		Now we still treating each Majorana chain as the edges of parallel chains at a certain number.
		As we explained above,
		the Anderson localized current densities
		indeed equivalent to the thermalized Majorana chain which is far away from
		the critical point of quantum chaos.
		And now the translational invariance along the Majorana chain is absent.
		In this case, the category of current density can no more be distinguished simply
		by the number of edge zero modes,
		but the number of generated edge states.
		Thus the relation Eq.(\ref{sigma},\ref{sigmaalpha}) should be replaced
		by a more general one, 
		$\sigma_{i}^{y}\sigma_{i}^{x}=\sigma^{x}_{j}\sigma^{y}_{j+1}\ (i,j=1,2,3,\cdots)$.
		For example, now the fermion bilinear term should becomes $\chi_{1}\chi_{2}=\chi_{2}\chi_{3}$ (in $\alpha$-sector),
		which indicates a phase transition from the topological Haldane-type nontrivial one to the trivial one.
		In this case, the corresponding probability for each current density 
		flavor
		is determined by the value of 
		$\Delta_{kl}(|k-1|)$.


		In the former case,
		as $O(M)$ is relatively small (for a whole system),
		we have $[\partial_{\alpha}\Phi(\alpha)/2\pi,\partial_{\beta}\Phi(\beta)/2\pi]=\delta_{\alpha,\beta}=0$, and thus
		each current density in a certain sector
		is independet of their summation upper boundary (the next sector),
		which results in the relations
		$[\rho_{\alpha},\beta]=0,[\sum^{\beta-1}_{\alpha}\rho_{\alpha},\beta]\neq 0$
		(or $[\rho_{\beta},\beta']=0,[Q_{b}^{2},\beta']\neq 0$).
		While for the latter case,
		the fractionally-induced distance between two adjacent current densities
		$|\alpha-\beta|\sim 1/M$ is vanishingly small,
		we have $[\partial_{\alpha}\Phi(\alpha)/2\pi,\partial_{\beta}\Phi(\beta)/2\pi]=
		-\frac{i}{2\pi}\partial_{\alpha} \delta(\alpha-\beta)$.
		In this case,
		the states corresponding to each flavor of current density 
		are no more a frustration-free localized eigenstates of the system $Q_{b}^{2}$.
		
		As shown in fig.4, in this case the level statistic follows the 
		behaviors beyond Poisson distribution,
		and the averaged level spacing ratio is $\langle r\rangle=0.3623$,
		which is below the typical value of
		Poisson one (0.38).

		\subsection{Statistic tool}
		
		Here we illustrate the statistic tool use in the above simulations.
		For current density spectrum in terms of the $(k,l)$-states,
		by arranging the current density levels (eigenenergies
		obtained from the charge product term dominated by gauge field) in ascending order
		(repeated values due to the degeneracy are removed),
		we define the level spacing ratio as
		$r_{i}=\frac{{\rm min}[E_{i+1}-E_{i},E_{i+2}-E_{i+1}]}
		{{\rm max}[E_{i+1}-E_{i},E_{i+2}-E_{i+1}]}$,
		to make sure all obtain-ratio are below 1.
		The corresponding probabilities are calculated
		in terms of $P(r)$ or $P({\rm ln}r)$.
		The results are compared with the Wigner formulas
		\begin{equation} 
			\begin{aligned}
				&P_{w}(r)=\frac{1}{Z_{\beta}}\frac{(r+r^2)^{\beta}}{(1+r+r^2)^{1+3\beta/2}},\\
				&P_{w}({\rm ln}r)=P(r)r,
			\end{aligned}
		\end{equation}
		with $\beta=1,2,4$ ($Z_{\beta}=\frac{8}{27},\frac{4}{81}\frac{\pi}{\sqrt{3}},
		\frac{4}{729}\frac{\pi}{\sqrt{3}}$) for GOE, GUE, and GSE, respectively.
		This method works well for Poisson case and the one beyond Poisson distribution
		for both the small-$r$ ($P_{w}\sim r^{\beta}$)
		and large-$r$ ($P_{w}\sim r^{-(2+\beta)}$) regions.
		While for GOE and GUE cases,
		it is valid only for the small-$r$ ($P_{w}\sim r^{\beta}$) region.

		\section{Conclusions}
		
		In contrast with the 1D non-Hermitian quasiperiodic model\cite{P2} where most symmetry and topological 
		orders are well preserved,
		we considering mainly a Hermitian one, where the chiral current induced by the antisymmetry tensors
		plays a key role in the emergence of conformal field effect as well 
		as the topological-phase-dependent
		statistical variance of the gauge-field-dominated conserved extensive quantity,
		which describes the product of two bosonic supercharge here.
		We also reveals the relation between antisymmetry-relation-dependent
		topological phase transitions/breaking and the transitions between different statistical behaviors,
		which is, equivalently, the transition between the quantum localized 
		(non-Hermitian) and thermalized (Hermitian) phases.
		This is also meaningfull
		in exploring the measurement-induced uncertainties
		in the many-body entangled systems\cite{P1}.
		The unglobal unit quantities used in this article,
		which are defined only by the independence with the central variable,
		is similar in spirit with the definition of Fibonacci numbers,
		and appear in the form of the constant ratio between two neighboring values of a variable
		whose continuous variation is replaced by the fractionated steps,
		and the system size always depend only on the large one.
		That also corresponds to the key reason of the
		existence of hierarchy gaps of Hermitian system.

		\clearpage

		\section{Appendix.A: Current density obtained using the Jordan-Wigner transformation}
		
		We define the boson fields here in position space as
		\begin{equation} 
			\begin{aligned}
				\phi(z)=2\pi\sum_{i}^{z-1}c_{i}^{\dag}c_{i}.
			\end{aligned}
		\end{equation}
		Then using the Jordan-Wigner transformation,
		we can define the hard core bosons as 
		\begin{equation} 
			\begin{aligned}
				b_{i}^{\dag}=c_{i}^{\dag}e^{i\phi(z)},\\
				b_{i}=e^{-i\phi(z)}c_{i}.
			\end{aligned}
		\end{equation}
		It is easy to know that 
		\begin{equation} 
			\begin{aligned}
				\partial_{z}\phi(z)=2\pi c_{z}^{\dag}c_{z}=2\pi b_{z}^{\dag}b_{z}=2\pi n_{z},
			\end{aligned}
		\end{equation}
		where the boson operators here are locally anticommuting.
		Next we will denote the number of quasiparticles (participating the formation of current) by $n_{z}$,
		which is a dimensionless quantity.
		
		By choosing a certain ordering,
		we consider the boson number exhibit a power law relation with the position $z$,
		basing on the turth that the perpendicular current often shows a power-law decay with the depth of a layered material.
		For convienient,
		we choose the bottom layer as the zeroth layer,
		with the smallest boson number $b^{\dag}_{0}b_{0}=n_{0}\lesssim  1$,
		while the boson number in other layers are $n_{i}=n_{0}^{i}$ where $i>0$ is the layer index.
		In long wavelength limit with $q_{z}\rightarrow 0$ and $z\rightarrow\infty$,
		we found the following relations which are valid in long-wavelength limit
		\begin{equation} 
			\begin{aligned}
				\label{relations1}
				&e^{iq_{z}z}=\frac{\phi(z)}{2\pi},\\
				&\lim_{q_{z}\rightarrow 0}(-iq_{z})=\lim_{z\rightarrow\infty}\frac{\frac{\partial_{z}\phi(z)}{2\pi}}{\frac{\phi(z)}{2\pi}}
				=\lim_{z\rightarrow\infty}\frac{n_{0}^{z}}{\sum_{i}^{z-1}n_{0}^{i}}\approx n_{0}-1(\approx {\rm ln}n_{0})\equiv \partial_{z},\\
				&\frac{\phi(z)}{2\pi}=\lim_{z\rightarrow\infty}\frac{n_{0}^{z}}{n_{0}-1}
				=\lim_{z\rightarrow\infty}\frac{n_{0}^{z}}{{\rm ln}n_{0}},
			\end{aligned}
		\end{equation}
		which indicates a strongly nonlinear relation between the boson number and $z$ positions.
		In the following we define $\lim_{z\rightarrow\infty}b_{z}^{\dag}b_{z}=n_{z}$ as the boson number operator at position $z$,
		which, for a chirally one-dimensional system (not the translational invariant one),
		is the largest boson number.
		We will see that this is the precondition for the projection from the fermion tunneling in three-dimension space 
		to the boson interaction in one-dimensional space.
		That is to say, the chirality of bosons here are necessary, nut due to the anticommuting relation between boson operators,
		different orderings can be directly sum up, i.e., the symmetry permutations.

		Note that the key point to generate $(1+1)$-D SYK model is by changing the fermions in 3D momentum space,
		which carrier the original degrees-of-freedom (induced by the hopping matrix related to the conductivity tensor),
		to the 1D modes in position space,
		which are the quasiparticles related to the survival perpendicular current under (linear or nonlinear) reponses.
		Thus all the position notation-$z$ in this paper corresponds to the quasiparticle in position space,
		like the above-defined boson field $\phi(z)$, is indeed equivalents to the Fourier transformation to the real fermions in momentum space (with notation $q_{z}$).
		Specifically, by virtue of Jordan-Wigner transformation in 1D systems,
		this Fourier transformation is be carried out by a summation over large numer of steps (after infinitely fractionization).
		Thus the derivation operator $\partial_{z}$ with positions of quasiparticles is in fact equivalents
		to the Fourier transformation to change the representation of fermions in momentum space to the real space,
		\begin{equation} 
			\begin{aligned}
				&\partial_{z}=\int dq_{z}e^{iq_{z}z}=\frac{\phi(z)}{2\pi}\int dq_{z}=\lim_{q_{z}\rightarrow 0}(-iq_{z}),\\
				&\int dq_{z}=iq_{z}\frac{2\pi}{\phi(z)}=iq_{z}e^{-iq_{z}z},
			\end{aligned}
		\end{equation}
		where the second line is consistent with the third line of Eq.(\ref{relations1}).
		Thus in long-wavelength limit, the operators $iq_{z}$, together with $\partial_{\tau}$,
		are of the position space.
		Note that here we artificially define $\int dz=2\pi$ insteads of $\int dz=1$ since
		although $z$ is defined in position space for quasi[articles, but 
		The real fermion number operator in momentum space (1D now) is simply the inverse position of its corresponding quasiparticle
		\begin{equation} 
			\begin{aligned}
				n_{q_{z}}=n_{z}e^{-iq_{z}z}=2\pi(n_{0}-1)=2\pi \partial_{z}=\frac{1}{z},
			\end{aligned}
		\end{equation}
		which are also be obtained by performing the Fourier transformation to the real fermion number operator in position space 
		\begin{equation} 
			\begin{aligned}
				n_{q_{z}}=\int e^{-iq_{z}z}n_{z}dz=\int \partial_{z}{\rm ln}\frac{\phi(z)}{2\pi}dz=iq_{z}\int dz=-2\pi (iq_{z}).
			\end{aligned}
		\end{equation}
		Similarly, the derivation with the fermion number operator in 1D position space obtained is such way
		is equivalents to the Fourier transformation to change the quasiparticles in momentum space representation
		to the real space representation,
		which is the current density, the 1D mode we need.

		Base on the above results which are valid in the long-wavelength limit,
		the 1D quasiparticle number (e.g., the current-induced spin-wave excitations) operator in real space and momentum space read as
		\begin{equation} 
			\begin{aligned}
				&n_{z}=\frac{\partial_{z}\phi(z)}{2\pi}=\frac{n_{0}^{z}}{2\pi}=\lim_{q_{z}\rightarrow 0}(-iq_{z})\frac{\phi(z)}{2\pi},\\
				&n_{q_{z}}=2\pi(n_{0}-1)=2\pi\lim_{q_{z}\rightarrow 0}(-iq_{z}),
			\end{aligned}
		\end{equation}
		which are related by the Fourier transformation $n_{z}=\frac{1}{2\pi}\int e^{iq_{z}z}n_{q_{z}}dq_{z}$.
		
		The boson field in momentum space can be written as
		\begin{equation} 
			\begin{aligned}
				\phi(q_{z})
				=\int  e^{-iq_{z}z} \phi(z)dz
				=\int  2\pi \phi^{-1}(z)\phi(z)dz
				=2\pi.
			\end{aligned}
		\end{equation}
		By performing the Fourier transformation to the effective
		electric field term $E_{q_{z}}$ in momentum space,
		\begin{equation} 
			\begin{aligned}
				\int E_{q_{z}} e^{iq_{z}z} dq_{z}
				=\int (-iq_{z}\phi(q_{z}))e^{iq_{z}z} dq_{z}
				=\lim_{q_{z}\rightarrow 0}(-iq_{z})\phi(z)=\partial_{z}\phi(z).
			\end{aligned}
		\end{equation}
		Thus the above Fourier transform in fact turn the system in momentum space to the boson-number space.
		
		The boson number also directly related to energies of the corresponding modes,
		and these boson modes can be considered as excitations coupled by the SYK-type random all-to-all interactions.
		But such one-dimensional boson mode can be obtained only after the fourier transform
		to the states of fermions in three-dimensional momentum space.
		We can look into the inner process by taking $c_{i}^{\dag}$ as an example,
		\begin{equation} 
			\begin{aligned}
				\label{JW}
				\int c^{\dag}_{i}e^{i\phi(q_{z})} e^{iq_{z}z}dq_{z}=c_{i}^{\dag}e^{i2\pi}\frac{\phi(z)}{2\pi}
				=c_{i}^{\dag}e^{i2\pi}\frac{n_{0}^{z}}{n_{0}-1}
				=c_{i}^{\dag}e^{i\Phi(n_{z})}=B^{\dag}_{i},
			\end{aligned}
		\end{equation}
		where there is a new one-dimensional boson field defined as
		\begin{equation} 
			\begin{aligned}
				\label{Phi}
				\Phi(n_{z})=-i{\rm ln}\frac{\phi(z)}{2\pi}=-i{\rm ln}\frac{n_{0}^{z}}{n_{0}-1}.
			\end{aligned}
		\end{equation}
		This boson field reflects the variance with boson number at position $z$, instead of simply the variance with position.
		Next we will replace the boson number indices $n_{z}(n_{z'})$ simply by $\alpha(\beta)=1,\cdots, M$.
		The above boson field will results in the one-dimensional boson operators with new definition,
		like $B^{\dag}_{\alpha}$ (to distinguish with the one $b^{\dag}_{i}=c^{\dag}e^{i\phi(q_{z})}$ defined in momentum space)
		whose index $\alpha$ indicates the corresponding boson number $n_{z}$ at position $z$,
		while $B^{\dag}_{\alpha}B_{\alpha}$ corresponds to the
		current density $\rho_{\alpha} \sim J/n_{0}^{\alpha}$ which is suppressed by the fermion number $n_{0}^{\alpha}$.
		Thus we obtain the simple expression
		\begin{equation} 
			\begin{aligned}
				\label{Phi}
				\frac{\Phi(\alpha)}{2\pi}=\sum^{\alpha-1}_{\gamma}\rho_{\gamma}.
			\end{aligned}
		\end{equation}
		Then the gauge-invariant current density is obtained as the derivation of this boson field,
		\begin{equation} 
			\begin{aligned}
				\label{app}
				\frac{\partial_{\alpha} \Phi(\alpha)}{2\pi}
				=\partial_{n_{0}^{z}} \frac{-i}{2\pi}{\rm ln}\frac{\phi(z)}{2\pi}
				=\frac{-i}{2\pi}\frac{1}{n_{0}^{z}}
				=\rho_{\alpha}.
			\end{aligned}
		\end{equation}
		The last step is valid here due to the long wavelength limit in momentum space.

		\section{Appendix.B: Definitions about the gauge invariant chiral current density}
		
		Firstly we consider a continuous SU(2) symmetry in complex fermion system,
		where we have
		\begin{equation} 
			\begin{aligned}
				&c_{i\pm}=\frac{1}{\sqrt{2}}(c_{i1}\pm ic_{i2})=e^{\pm i\phi},\\
				&c^{\dag}_{i\pm}=\frac{1}{\sqrt{2}}(c^{\dag}_{i1}\mp ic^{\dag}_{i2})=e^{\mp i\phi},\\
			\end{aligned}
		\end{equation}
		which can be obtained from the usual Majorana system by adding the restrictions
		like U(1) charge conservation.
		Then we remove such restriction and back to the Majorana system,
		according to Secs.2.5-2.6, we can can replacing the DOF as
		(we using the same notation of Ref.\cite{Lian B})
		\begin{equation} 
			\begin{aligned}
				&\frac{1}{\sqrt{2}}(\chi_{i}(\tau)+i\chi_{j}(\tau))=c(\tau)=:e^{i\phi(\tau)}:,\\
				&\frac{1}{\sqrt{2}}(\chi_{i}(\tau)-i\chi_{j}(\tau))=c^{\dag}(\tau)=:e^{-i\phi(\tau)}:,
			\end{aligned}
		\end{equation}
		where the boson field $\phi(\tau)$ satisfies
		\begin{equation} 
			\begin{aligned}
				&[\phi(\tau),\phi(\tau')]=i\pi{\rm sgn}(\tau-\tau')
				={\rm ln}(\tau-\tau'-i\eta)-{\rm ln}(\tau'-\tau-i\eta),\\
				&[\partial_{\tau'}\phi(\tau'),\phi(\tau)]=
				[\partial_{\tau}\phi(\tau),\phi(\tau')]=2\pi i\delta(\tau-\tau')
				=\frac{1}{\tau-\tau'-i\eta}+\frac{1}{\tau'-\tau-i\eta}.
			\end{aligned}
		\end{equation}
		But dividing the boson field into
		$\phi(\tau)=\psi(\tau)+\psi^{\dag}(\tau)$\cite{Lian B},
		we have
		\begin{equation} 
			\begin{aligned}
				&[\psi(\tau),\psi^{\dag}(\tau')]=-{\rm ln}2\pi i(\tau'-\tau-i\eta),\\
				&
				[\partial_{\tau}\psi(\tau),\psi^{\dag}(\tau')]
				=[\partial_{\tau'}\psi(\tau'),\psi^{\dag}(\tau)]=\frac{1}{\tau'-\tau-i\eta},\\
				&
				[\partial_{\tau}\psi^{\dag}(\tau),\psi(\tau')]
				=[\partial_{\tau'}\psi^{\dag}(\tau'),\psi(\tau)]=\frac{1}{\tau-\tau'-i\eta},\\
				&
				[\partial_{\tau}\psi(\tau),\psi(\tau')]
				=-[\partial_{\tau}\psi^{\dag}(\tau),\psi^{\dag}(\tau')]
				=-[\partial_{\tau}\phi(\tau)-\partial_{\tau}\psi(\tau),\phi(\tau')-\psi(\tau')],
			\end{aligned}
		\end{equation}
		where from the last formula we can further obtain
		\begin{equation} 
			\begin{aligned}
				[\partial_{\tau}\phi(\tau),\phi(\tau')-\psi(\tau')]=
				[\partial_{\tau}\psi(\tau),\phi(\tau')-2\psi(\tau')].
			\end{aligned}
		\end{equation}
		Considering the time ordering product\cite{Lian B}
		\begin{equation} 
			\begin{aligned}
				\mathcal{T}c^{\dag}(\tau)c(\tau')
				&=:e^{-i\phi(\tau)}::e^{i\phi(\tau')}:\\
				&=\frac{1}{2\pi i (\tau'-\tau-i\eta)}
				e^{-i(\psi^{\dag}(\tau)-\psi^{\dag}(\tau')+\psi(\tau)-\psi(\tau')}\\
				&=\frac{1}{2\pi i (\tau'-\tau-i\eta)}
				\left[e^{-i(\phi(\tau)-\phi(\tau'))}
				-\left(\frac{1}{2\pi i (\tau'-\tau-i\eta)}-1 \right)
				e^{-i(\psi^{\dag}(\tau)-\psi^{\dag}(\tau')+\psi(\tau)-\psi(\tau')}\right]\\
				&=\frac{1}{2\pi i (\tau'-\tau-i\eta)}
				+\frac{\partial_{\tau}\phi(\tau)}{2\pi}-O(\tau-\tau'),
			\end{aligned}
		\end{equation}
		where in the third line we using the commutation
		\begin{equation} 
			\begin{aligned}
				[e^{-i\psi(\tau)},e^{\psi^{\dag}(\tau')}]
				=\left(\frac{1}{2\pi i (\tau'-\tau-i\eta)}-1\right)
				e^{i\psi^{\dag}(\tau')}e^{-i\psi(\tau)},
			\end{aligned}
		\end{equation}
		with the vacuum expectation term
		reads $\frac{1}{2\pi i (\tau'-\tau-i\eta)}
		e^{[\psi(\tau),\psi^{\dag}(\tau')}$,
		and for the first term within the bracket of third line,
		we expand it as 
		\begin{equation} 
			\begin{aligned}
				e^{-i(\phi(\tau)-\phi(\tau'))}
				=1-i(\phi(\tau)-\phi(\tau'))-\frac{1}{2}(\phi(\tau)-\phi(\tau'))^{2}\\
				=1-i(\partial_{\tau}\phi(\tau)(\tau-\tau')+O((\tau-\tau')^{m+1})
				+O((\tau-\tau')^{2}),
			\end{aligned}
		\end{equation}
		where we can savely approximate (with $m$ depends on $\phi(\tau)$)
		\begin{equation} 
			\begin{aligned}
				\frac{\phi(\tau)-\phi(\tau')}{\tau-\tau'}
				=\partial_{\tau}\phi(\tau)+O((\tau-\tau')^{m}),
			\end{aligned}
		\end{equation}
		since we will at the end let $(\tau-\tau')\rightarrow 0$.
		Thus we obtain the normal ordered product as
		\begin{equation} 
			\begin{aligned}
				:c^{\dag}(\tau)c(\tau):
				&=\mathcal{T}c^{\dag}(\tau)c(\tau')
				-\langle 0|\mathcal{T}c^{\dag}(\tau)c(\tau')|0\rangle\\
				&=\mathcal{T}c^{\dag}(\tau)c(\tau')
				-\frac{1}{2\pi i (\tau'-\tau-i\eta)}\\
				&=\frac{\partial_{\tau}\phi(\tau)}{2\pi}.
			\end{aligned}
		\end{equation}
		Then we have the commutation relation
		\begin{equation} 
			\begin{aligned}
				&[:\psi_{i}(\tau)\psi_{j}(\tau):,:\psi_{i}(\tau')\psi_{j}(\tau'):]
				=[:-ic^{\dag}(\tau)c(\tau):,:-ic^{\dag}(\tau')c(\tau'):]\\
				&=-[\frac{\partial_{\tau}\phi(\tau)}{2\pi},\frac{\partial_{\tau'}\phi(\tau')}{2\pi}]\\
				&=\frac{-1}{(2\pi i)^{2}}\partial_{\tau}
				\left(\frac{1}{\tau-\tau'-i\eta}+\frac{1}{\tau'-\tau-i\eta}\right),
			\end{aligned}
		\end{equation}
		which also leads to 
		\begin{equation} 
			\begin{aligned}
				[\partial_{\tau'}\phi(\tau'),\partial_{\tau}\phi(\tau)]
				=[\partial_{\tau}^{(2)}\phi(\tau),\phi(\tau')]
				=[\phi(\tau),\partial_{\tau'}^{(2)}\phi(\tau')].
			\end{aligned}
		\end{equation}
		Thus in the following,
		we can treating the term $\frac{i}{2\pi}$,
		i.e., the vacuum expectation term in $(\tau-\tau')\rightarrow 0$ limit,
		as a variable,
		and fix the value of time ordering product, then the charge depending on the normal ordered
		fermion product is related to the vacuum term,
		and vanishes in the limit of $\frac{i}{2\pi}\rightarrow\infty$,
		which will appears below (Eq.(\ref{1151})).
		
		In the above expression of antisymmetry tensor $T_{ik}^{jl}$,
		any pairs of two Majorana fermion indices (each with an overline) enforced by a nonzero delta function
		carrier the same information about the chiral current density but not means they are the same index
		(just able to combined into a multiple boson field).
		For example, for the subgroup $\overline{k}\overline{l}$ of the $\beta$-sector,
		which is the part carriering the informations about the chiral current density in group $\{k,l\}$,
		we have
		$\delta_{\overline{k}\overline{l}}=1$, and in the mean time we are able to have 
		\begin{equation} 
			\begin{aligned}
				\label{unit}
				i\chi_{\overline{k}}\chi_{\overline{l}}
				&=:i\chi_{k}\chi_{l}:\\
				&=\sum_{\beta=1}^{\beta'-1}(c^{\dag}_{\beta}c_{\beta}-\frac{\mathds{1}}{2})\\
				&=\sum_{\beta=1}^{\beta'-1}c^{\dag}_{\beta}c_{\beta}-\frac{\mathds{1}}{2}(\beta'-1)\\
				&=c^{\dag}_{\beta}c_{\beta}\bigg|_{\beta=\beta'-1}\\
				&=\partial_{\beta}\sum^{\beta-1}_{\alpha}c^{\dag}_{\alpha}c_{\alpha}\\
				&:=\rho_{\beta},
			\end{aligned}
		\end{equation}
		where we define the unit quantity (for sector $\beta$) in a projection form
		\begin{equation} 
			\begin{aligned}
				\mathds{1}:=
				\frac{2}{\beta'-1}(\sum^{\beta'-1}_{\gamma}c^{\dag}_{\gamma}c_{\gamma}-c^{\dag}_{\beta'-1}c_{\beta'-1}),
			\end{aligned}
		\end{equation}
		where $\gamma$ is an arbitrary index and during the summation we can choose initial step $\gamma=1$
		and $(\beta'-1)=M$ to make sure the size of group $\{\beta\}$ is of order $O(M)$.
		This unit quantity is a conserved (U(1)) quantity only for a group
		of variables of the sector $\beta$,
		but it is not a conserved quantity for a group containing the elements of different sectors.
		like a gauge field $A_{\alpha\beta}$.
		For example, in sector $(\beta'-2)$, 
		we have
		\begin{equation} 
			\begin{aligned}
				\sum_{\beta=1}^{\beta'-2}c^{\dag}_{\beta}c_{\beta}-\frac{\mathds{1}'}{2}(\beta'-2)
				=c^{\dag}_{\beta}c_{\beta}\bigg|_{\beta=\beta'-2},
			\end{aligned}
		\end{equation}
		with 
		\begin{equation} 
			\begin{aligned}
				\mathds{1}':=
				\frac{2}{\beta'-2}(\sum^{\beta'-2}_{\gamma}c^{\dag}_{\gamma}c_{\gamma}-c^{\dag}_{\beta'-2}c_{\beta'-2}).
			\end{aligned}
		\end{equation}

		The above definition of unit quantity is different from the usual
		definition of normal ordered U(1) charge, which is in the form
		$:i\chi_{k}\chi_{l}:=:c_{\beta}^{\dag}c_{\beta}-1/2:$,
		but we will see that, the definition of such unglobally-conserved unit quantity is enssential 
		for the gauge invariance of the system investigated in this paper.
		According to above definitions,
		we can also obtain these relations (between neighboring sectors)
		\begin{equation} 
			\begin{aligned}
				&\frac{\mathds{1}}{2}
				=\sum^{\beta'-2}_{\beta}c^{\dag}_{\beta}c_{\beta}-(\beta'-2)\frac{\mathds{1}}{2}
				=\sum^{\beta'-1}_{\beta}c^{\dag}_{\beta}c_{\beta}-c^{\dag}_{\beta'-1}c_{\beta'-1}-(\beta'-2)\frac{\mathds{1}}{2},\\
				&\partial_{\beta'}\mathds{1}=\frac{2}{\beta'-1}(c^{\dag}_{\beta}c_{\beta}-\frac{\mathds{1}}{2}),\\
				&(\beta'-2)(\frac{\mathds{1}}{2}-\frac{\mathds{1}'}{2})=c^{\dag}_{\beta}c_{\beta}\bigg|_{\beta=\beta'-2}-\frac{\mathds{1}}{2}.
			\end{aligned}
		\end{equation}

		As shown in the Sec.,
		the distinct statistic behaviors will be related to the specific limiting conditions.
		We choose the fermionic term $Q_{\tau,\tau'}:=\chi_{j}(\tau)\chi_{k}(\tau')\delta_{jk}=\frac{1}{2}\delta(\tau-\tau')$ as an example,
		and note that it can be rewritten as $
		Q_{j,k}:=\chi_{j}(\tau)\chi_{k}(\tau)
		$ in perspective of statistical behavior,
		where we have the derivative
		\begin{equation} 
			\begin{aligned}
				\label{sec7}
				\partial_{\tau}\delta(\tau-\tau')
				=\partial_{\tau}2\chi_{j}(\tau)\chi_{k}(\tau')\delta_{jk}
				=-2i\pi[Q_{j,k}(\tau),Q_{j,k}(\tau')].
			\end{aligned}
		\end{equation}
		This derivative term is equivalent to 
		\begin{equation} 
			\begin{aligned}
				\partial_{\tau}\delta(\tau-\tau')
				=1-\frac{\delta(\tau-\tau')}
				{\lim_{\tau\rightarrow\infty}
					\delta(\tau-\tau')},
			\end{aligned}
		\end{equation}
		where the limiting term $\lim_{\tau\rightarrow\infty}
		\delta(\tau-\tau')$ is independent with $\tau$ and thus satisfies
		\begin{equation} 
			\begin{aligned}
				\partial_{\tau}\lim_{\tau\rightarrow\infty}
				\delta(\tau-\tau')=\partial_{\tau}\frac{\delta(\tau-\tau')}{1+2i\pi[Q_{j,k}(\tau),Q_{j,k}(\tau')]}=0,
			\end{aligned}
		\end{equation}
		then we obtain
		\begin{equation} 
			\begin{aligned}
				\frac{\partial_{\tau}\delta(\tau-\tau')}{1-\partial_{\tau}\delta(\tau-\tau')}=-\delta(\tau-\tau')\partial_{\tau}\frac{1}{1-
					\partial_{\tau}\delta(\tau-\tau')}.
			\end{aligned}
		\end{equation}
		Considering the independence with $\tau$,
		this limiting term can also be replaced by $\lim_{\tau\rightarrow \tau'}\delta(\tau-\tau')=\mathds{1}_{\tau'}$,
		where we define $\mathds{1}_{\tau'}$ here as a $\tau'$-dependent unglobal unit quantity (independent of $\tau$).
		Thus we approximate result $\mathds{1}_{\tau'}=
		\frac{\delta(\tau-\tau')}{1-\partial_{\tau}\delta(\tau-\tau')}$,
		\begin{equation} 
			\begin{aligned}
				\lim_{\tau'\rightarrow\infty}\lim_{\tau\rightarrow \tau'}\delta(\tau-\tau')=\lim_{\tau'\rightarrow\infty}
				\mathds{1}_{\tau'}=\frac{\mathds{1}_{\tau'}}{1-\partial_{\tau'}\mathds{1}_{\tau'}}.
			\end{aligned}
		\end{equation}
		Similarly, the independence with $\tau'$ of this limiting expression results in
		\begin{equation} 
			\begin{aligned}
				\label{s00}
				\frac{\partial_{\tau'}\mathds{1}_{\tau'}}{1-\partial_{\tau'}\mathds{1}_{\tau'}}=-\mathds{1}_{\tau'}\partial_{\tau'}\frac{1}{1-
					\partial_{\tau'}\mathds{1}_{\tau'}},
			\end{aligned}
		\end{equation}
		thus the derivative of $\mathds{1}_{\tau'}$ with $\tau'$ reads
		\begin{equation} 
			\begin{aligned}
				\label{ee23}
				\partial_{\tau'}\mathds{1}_{\tau'}
				=\delta(\tau-\tau')\partial_{\tau}\frac{1}{1-
					\partial_{\tau}\delta(\tau-\tau')}(\frac{\partial \tau}{\partial \tau'}-1)
				=\frac{\partial {\rm ln}\mathds{1}_{\tau'}}{
					\partial_{\tau'}\frac{1}{1-
						\partial_{\tau'}\mathds{1}_{\tau'}}
				}+1.
			\end{aligned}
		\end{equation}
		where we can also know $\frac{\partial {\rm ln}\mathds{1}_{\tau'}}{
			\partial_{\tau'}\frac{1}{1-
				\partial_{\tau'}\mathds{1}_{\tau'}}
		}=-\frac{\mathds{1}_{\tau'}}{\lim_{\tau'\rightarrow\infty}\mathds{1}_{\tau'}}$.
		
		As a mathematical trick widly used in Sec.7, the expression Eq.\ref{s00} can be regard as a limiting result of a certain variable $k$,
		\begin{equation} 
			\begin{aligned}
				\label{s00}
				-\frac{1}{1-[\partial_{\tau'}\mathds{1}_{\tau'}]^{-1}}=
				-\lim_{k\rightarrow\infty}\sum_{m=0}^{k-1}[\partial_{\tau'}\mathds{1}_{\tau'}]^{-m}
				=&-\frac{\sum_{m=0}^{k-1}[\partial_{\tau'}\mathds{1}_{\tau'}]^{-m}}
				{1-\partial_{k}\sum_{m=0}^{k-1}[\partial_{\tau'}\mathds{1}_{\tau'}]^{-m}}\\
				=&-\frac{\sum_{m=0}^{k-1}[\partial_{\tau'}\mathds{1}_{\tau'}]^{-m}}
				{1-[\partial_{\tau'}\mathds{1}_{\tau'}]^{-k}},
			\end{aligned}
		\end{equation}
		then the detailed form of expression related to $m$ and $k$ can be obtained by relating this equation to Eqs.(\ref{s00},\ref{ee23}).

		As all the DOF's of this system originate from the different categories of interaction in different sectors,
		and the distinct Majorana fermions in each sector,
		the time difference $(\tau-\tau')$ is indeed correlated to the Majorana fermionic DOF.
		To see this,
		we using the SO(N)1 Kac-Moody algebra and write the commutate with $j=k$ as
		\begin{equation} 
			\begin{aligned}
				[\chi_{i}(\tau,\eta)\chi_{k}(\tau,\gamma),\chi_{k}(\tau',\gamma)\chi_{l}(\tau',\eta)]
				=-\delta(\tau-\tau')\chi_{i}\chi_{l}.
			\end{aligned}
		\end{equation}
		Note that here arbitrarily two operators at the same time are symmetry with each other,
		e.g., $[\chi_{k}(\tau',\gamma),\chi_{l}(\tau',\eta)]=0$.
		By treating the Majorana indices as countable quantities,
		we can taking the limit $l\rightarrow i$ and then the above commutator transforms to the previous one appearing in Eq.(\ref{sec7}).
		Similar to above procedure, in perspective of removing the dependence of Majorana-$l$,
		the limit $l\rightarrow i$ is equivalents to $l\rightarrow \infty$ during the calculation.
		Thus we have the limiting result
		\begin{equation} 
			\begin{aligned}
				\lim_{l\rightarrow \infty}(-\delta(\tau-\tau')\chi_{i}\chi_{l})
				=\frac{-\delta(\tau-\tau')\chi_{i}\chi_{l}}{1-\partial_{l}(-\delta(\tau-\tau')\chi_{i}\chi_{l})}
				=\frac{i}{2\pi}\partial_{\tau}\delta(\tau-\tau').
			\end{aligned}
		\end{equation}
		Then we obtain
		\begin{equation} 
			\begin{aligned}
				\partial_{\tau}\delta(\tau-\tau')
				=-\frac{\partial_{l}[\partial_{\tau}\delta(\tau-\tau')]}{\partial_{l}{\rm ln}\frac{i}{2\pi}}
				=-\frac{\partial l}{\partial l}(1-\frac{\partial_{\tau}\delta(\tau-\tau')}
				{\lim_{{\rm ln}\frac{i}{2\pi}\rightarrow\infty}\partial_{\tau}\delta(\tau-\tau')}),
			\end{aligned}
		\end{equation}
		where we obatin another operator transformation under limiting condition,
		\begin{equation} 
			\begin{aligned}
				\label{lim30}
				\lim_{l\rightarrow \infty}[-(\frac{\partial l}{\partial l})^{-1}]=\partial_{{\rm ln}\frac{i}{2\pi}}
				=\frac{[-(\frac{\partial l}{\partial l})^{-1}]}{1-\partial_{l}
					[-(\frac{\partial l}{\partial l})^{-1}]}.
			\end{aligned}
		\end{equation}
		Then combined with the $[\frac{i}{2\pi}\partial_{\tau}\delta(\tau-\tau'),l]=0$,
		we obtain the following expressions of $l$-dependent identity
		\begin{equation} 
			\begin{aligned}
				\label{1151}
				\mathds{1}_{l}
				&\equiv \partial_{\tau}\delta(\tau-\tau')
				(\frac{1}{\lim_{{\rm ln}\frac{i}{2\pi}\rightarrow\infty}\partial_{\tau}\delta(\tau-\tau')}
				-(\frac{\partial l}{\partial l})^{-1})\\
				&\equiv \partial_{\tau}\delta(\tau-\tau')
				(\frac{1}{\lim_{l\rightarrow\infty}\partial_{\tau}\delta(\tau-\tau')}
				-\partial_{l}{\rm ln}\frac{i}{2\pi}),
			\end{aligned}
		\end{equation}
		which can be rewritten as
		\begin{equation} 
			\begin{aligned}
				\mathds{1}_{l}
				&\equiv 
				1-\partial_{{\rm ln}\frac{i}{2\pi}}(\partial_{\tau}\delta(\tau-\tau'))
				-\frac{\partial l}{\partial l})^{-1}\partial_{\tau}\delta(\tau-\tau')\\
				&\equiv 
				1-\partial_{l}(\partial_{\tau}\delta(\tau-\tau'))
				-\partial_{l}{\rm ln}\frac{i}{2\pi}\partial_{\tau}\delta(\tau-\tau').
			\end{aligned}
		\end{equation}
		These two expressions of identity can be related by the relation
		\begin{equation} 
			\begin{aligned}
				\lim_{\partial_{{\rm ln}\frac{i}{2\pi}}
					\rightarrow \partial_{l}}
				[-(\frac{\partial l}{\partial l})^{-1}]
				=
				-\partial_{l}{\rm ln}\frac{i}{2\pi},
			\end{aligned}
		\end{equation}
		which is independent with $\partial_{{\rm ln}\frac{i}{2\pi}}$.
		By substituting Eq.(\ref{lim30}) into the above identity,
		we have $\lim_{l\rightarrow \infty}\mathds{1}_{l}=1$.

		Similar to the discussion above Eq.(\ref{244}), by
		letting $\partial_{{\rm ln}\frac{i}{2\pi}}[\partial_{{\rm ln}\frac{i}{2\pi}}
		(\partial_{\tau}\delta(\tau-\tau'))+\frac{\partial_{\tau}\delta(\tau-\tau')}
		{\lim_{{\rm ln}\frac{i}{2\pi}\rightarrow\infty} \partial_{\tau}\delta(\tau-\tau')}]
		=\partial_{{\rm ln}\frac{i}{2\pi}}1=0$,
		we can obtain more detailed form of the unglobal unit quantity 
		$-(\frac{\partial l}{\partial l})^{-1}$,
		\begin{equation} 
			\begin{aligned}
				-(\frac{\partial l}{\partial l})^{-1}
				&=\frac{\partial_{{\rm ln}\frac{i}{2\pi}}
					[\partial_{\tau}\delta(\tau-\tau')]}
				{\partial_{\tau}\delta(\tau-\tau')}\\
				&=\frac{1}{\partial_{{\rm ln}\frac{i}{2\pi}}}-\frac{1}
				{\lim_{{\rm ln}\frac{i}{2\pi}\rightarrow\infty}
					\partial_{\tau}\delta(\tau-\tau')}\\
				&=-\partial_{{\rm ln}\frac{i}{2\pi}}[{\rm ln}
				\frac{1}{\lim_{{\rm ln}\frac{i}{2\pi}\rightarrow\infty} \partial_{\tau}\delta(\tau-\tau')}].
			\end{aligned}
		\end{equation}
		Due to the same reason as we discuss in Sec.7.4,
		the negative one should be ${\rm ln}\frac{i}{2\pi}$-dependent (unlike the positive one),
		in the mean time, the chiral symmetry will not be necessarily preserved
		due to the absence of symmetry-protected-topological phases here
		(which usually appears in 1D noninteracting systems,
		where an even number of U(1) conserved fermionic charge (instead of bosonic like this article) 
		guarantees the chiral symmetry),
		and thus we have a phase-shift-dependent chiral transformation result
		$\mathcal{C}c_{i}^{\dag}c_{j}\mathcal{C}^{-1}=(-1)^{i+j}c_{i}^{\dag}c_{j}$
		where the the raising and lowering operators here are all complex and obtained by using
		the Jordan-Wigner type transformation.
		Due to this reason, the derivative 
		$\partial_{{\rm ln}\frac{i}{2\pi}}[{\rm ln}
		\lim_{{\rm ln}\frac{i}{2\pi}\rightarrow\infty} \partial_{\tau}\delta(\tau-\tau')]$
		could be possible to contains arbitrary number of $(-1)$-terms (as long as it mod $2=0$),
		thus to uniquely determines the term
		$-(\frac{\partial l}{\partial l})^{-1}$,
		the only option is by taking the limit of
		vanishing dependence of $(-1)$ with ${\rm ln}\frac{i}{2\pi}$,
		and this will spontaneously happen as
		the dependence of $\partial_{\tau}\delta(\tau-\tau')$ 
		with ${\rm ln}\frac{i}{2\pi}$ tends to vanish.
		For example,
		the last line of above expression can be reads
		\begin{equation} 
			\begin{aligned}
				-\partial_{{\rm ln}\frac{i}{2\pi}}[{\rm ln}
				\frac{1}{\lim_{{\rm ln}\frac{i}{2\pi}\rightarrow\infty} \partial_{\tau}\delta(\tau-\tau')}]
				=\partial_{{\rm ln}\frac{i}{2\pi}}[{\rm ln}
				\lim_{{\rm ln}\frac{i}{2\pi}\rightarrow\infty} \partial_{\tau}\delta(\tau-\tau')]
				\mp m \partial_{{\rm ln}\frac{i}{2\pi}}(-1){\rm ln}
				\frac{1}{\lim_{{\rm ln}\frac{i}{2\pi}\rightarrow\infty} \partial_{\tau}\delta(\tau-\tau')},
			\end{aligned}
		\end{equation}
		where $m$ is an nonzero integer and it equals to zero only in the 
		$\partial_{{\rm ln}\frac{i}{2\pi}}(-1)=0$ limit.

		\section{Appendix.C: Flavors of chiral current density in BDI symmetry class: topological phases}
		
		By now we have introduce the procedure to obtain the 1D quasiparticle representation through
		the Jordan-Wigner transformation.
		As it is well-known, the Majorana fermion is a powerful tool in bulit the models
		for many novel physical phenomenas, especially when a phase in gauge theory is been considered just like our case. Using the Majorana-fermion zero modes, the 1D system can exhibits distinct phases recognized by 
		counting the number of zero modes localized at the edge of the chain,
		as long as the number of flavors of Majorana fermion modes $O(N^2)$ is large enough
		(square of that of chiral current density $O(M)$).
		
		Here we illustrate more details about the flavors of the chiral current density,
		which is of order $O(M)$,
		and it is an independent DOF relative to each part of Majorana fermion indices.
		Firstly, as shown in Appendix.A, by
		using Jordan Wigner transformation,
		we obtain the hard core bosons as well as the complex fermions though the bosonization 
		where the gauge invariant boson field is generated by the energy gradient of chiral current.
		In this way, the current densities, as the quasiparticle excitations,
		can be expressed by the product of Majorana fermion operators or the raising/lowering operators
		(hard core bosons) which are all assigned by the local structure.
		Defining the Majorana fermion operators in the following way ($i=1,2,3,\cdots$),
		\begin{equation} 
			\begin{aligned}
				\chi_{2i-1}=\sigma^{y}_{i}=\frac{-i(c_{i}-c^{\dag}_{i})}{\sqrt{2}},\\
				\chi_{2i}=\sigma^{x}_{i}=\frac{c_{i}+c^{\dag}_{i}}{\sqrt{2}},\\
				\chi_{2i}\chi_{2i-1}=\sigma^{x}_{i}\sigma^{y}_{i}=i\sigma^{z}_{i}=(c_{i}^{\dag}c_{i}-\frac{1}{2}).
			\end{aligned}
		\end{equation}
		The $\mathcal{Z}_{2}$ fermion parity operator is then given by
		\begin{equation} 
			\begin{aligned}
				\mathcal{P}=\prod_{i}i\chi_{2i-1}\chi_{2i}.
			\end{aligned}
		\end{equation}
		As illustrated in Ref.\cite{Fidkowski L,Marra P},
		the Majoran modes in 1D chain could be in trivial phase or nontrivial phase in the absence of time-reversal symmetry (TRS),
		which are $\sigma^{y}_{i}\sigma^{x}_{i}(i=1,2,3,\cdots)$ and $\sigma^{x}_{i}\sigma^{y}_{i+1}(i=1,2,3,\cdots)$, respectively.
		Here the latter one describes the bosonic spin ground state.
		While in the presence of TRS,
		we consider the BDI symmetry class where the quadratic couplings
		(here is to reflecting the intrinsic character of current densities in fermionic bilinear form)
		between Majorana fermions reduce the $\mathcal{Z}$ invariance (like what happen
		in the Anderson-localized bulk) into 
		the $\mathcal{Z}_{8}$ invariance.
		
		The TRS will prevent the formation of fermion bilinear terms, but cannot gap out the topological-protected
		degeneracies,
		like the local single-particle state (constituted by the Majorana modes of the same fermion parity)
		or the nonlocal single-particle state (superposition of two Majorana end modes
		of opposite fermion parities)\cite{Marra P}.
		In the presence of TRS,
		there would be infinite symmetry protected topological (SPT) phases instead of just two: the trivial and nontrivial Haldane phases,
		which can be described by bosonic Majorana modes:
		$\sigma^{x}_{i}\sigma^{y}_{i+k}(i=1,2,3,\cdots)$, indicated by the topological integer index $k=2,3,4,\cdots$.
		Here the topological index $k$ also correspond to the number of edge gapless Majorana modes,
		and odd $k$ corresponds to even superposition of bosonic ground states which will gap out the dangling Majorana modes
		($\sigma^{y}_{i}\sigma^{x}_{j}$) and breaks symmetry of $\mathcal{P}$\cite{Fidkowski L}.
		This also corresponds to the $N{\rm mod}4\neq 0$ case in the SYK$_{4}$ model in the edge of Majorana chain\cite{Kim J,You Y Z}, in which case the particle-hole symmetry is anomalous.
		While an even number $k$ corresponds to the odd superpositions of bosonic ground states.
		In this case, the particle-hole symmetry is not anomalous ($N_{c}{\rm mod}8= 0$;
		Note that here $N_{c}$ is the number of chains of the same type
		but can be simply replaced by $N$ since the current density of a certain phase
		can be equivalently considered as a collection of boundary fermions of $k$ parallel chains),
		and there will be nondegenerate ground states which will be asymptotically degenerate in large-$N$ limit
		due to the TRS-protected topological order.
		The robust MBL-induced nontrivial character inside the bulk avoids the zero modes beging
		fully gapped out and
		suppress the reasonances induced by disorders
		and thermalization effects at the edge.
		The resulting stable degenerate ground state is an edge zero mode localized wthin the gap of topological phase.
		During this process, the symmetry $\mathcal{P}$ is unbroken,
		and the formation of nonzero fermion bilinear expectation value
		$\sigma^{y}_{i}\sigma^{x}_{j}$ leads to $\mathcal{Z}_{2}$ spontaneous symmetry breaking.
		We also note that, the trivial phase in conformal (low energy) limit,
		where all the anomalous components are absent, preserves both the gauge invariance and discrete
		$\mathcal{Z}_{2}$ symmetry (due to the absence of degenerate group state).

		For chiral current densities expressed in terms of summation of Majorana modes,
		it is strainghforward to conclude that
		\begin{equation} 
			\begin{aligned}
				\label{sigma}
				&\sigma^{x}_{i}\sigma^{y}_{i+m}=\sigma^{x}_{j}\sigma^{y}_{j+m}=\sigma^{y}_{k}\sigma^{x}_{k}
				=\sigma^{y}_{l}\sigma^{x}_{l}(i,j,m=1,2,3,\cdots;\ k,l=2,3,\cdots),\\
				&\sigma^{y}_{1}\sigma^{x}_{i}=\sigma^{y}_{1}\sigma^{y}_{i+1}+\sigma^{x}_{1}\sigma^{y}_{i+1}\ (i=1,2,3\cdots),\\
				&\sigma^{y}_{1}\sigma^{y}_{i}=\sigma^{y}_{1}\sigma^{x}_{i}+\sigma^{x}_{1}\sigma^{x}_{i}\ (i=2,3\cdots),\\
				&\sigma^{x}_{i}\sigma^{x}_{i+m}=\sigma^{x}_{j}\sigma^{x}_{j+m}=\sigma^{y}_{k}\sigma^{y}_{k+m}
				=\sigma^{y}_{l}\sigma^{y}_{l+m}(i,j,m=1,2,3,\cdots;\ k,l=2,3,\cdots).
			\end{aligned}
		\end{equation}
		Base on these relations, the Majorana modes of different categories are
		\begin{equation} 
			\begin{aligned}
				\label{sigmaalpha}
				\alpha=1:\ \sigma^{y}_{1}\sigma^{x}_{1}=\sigma^{y}_{1}\sigma^{y}_{2}+\sigma^{x}_{1}\sigma^{y}_{2}
				=\sigma^{y}_{1}\sigma^{x}_{2}+\sigma^{x}_{1}\sigma^{x}_{2}+\sigma^{x}_{1}\sigma^{y}_{2}\\
				=\cdots\\
				=\sum_{i=1,2,3\cdots}^{\infty}(\sigma^{x}_{1}\sigma^{x}_{i+1}+\sigma^{x}_{1}\sigma^{y}_{i+1}),\\
				\alpha=2:\ \sigma^{y}_{1}\sigma^{x}_{1}+\sigma^{y}_{1}\sigma^{y}_{2},\\
				\alpha=3:\ \sigma^{y}_{1}\sigma^{x}_{1}+\sigma^{y}_{1}\sigma^{y}_{2}+\sigma^{y}_{1}\sigma^{x}_{2},\\
				\cdots,\\
				\alpha=M:\ \sum_{i=1}^{(M+1)/2}(\sigma^{y}_{1}\sigma^{x}_{1}+\sigma^{y}_{1}\sigma^{y}_{1+i}+\sigma^{y}_{1}\sigma^{x}_{1+i}),
			\end{aligned}
		\end{equation}
		where we assume $M$ is an odd number in the last line, otherwise the summation upper boundary should be $(M+2)/2$
		if $M$ is an even number.

		\section{Appendix.D: Relevant interaction effects}
		
		To make sure the theory be conformal in the large-$M$ limit,
		we need to recognize which perturbations are marginally relevant (or irrelevant).
		It was shown that, for a non-chiral system, for example, for a system consider both the fermions with left and right flavors\cite{Berkooz M},
		the non-random interactions, which are diagonal between these two flavors,
		will be relevant only in terms of the quadratic form and for one sign of the four-fermion coupling
		(treated as a perturbation during the renormalization analysis).
		In this kind of models, the relevant component of the four-fermion interactions will leads to SYK$_{4}$ behaviors
		and exhibiting instable non-fermi-liquid features.
		This model is in fact equivalents to our model in this paper after removing the constraints given by the
		imaginary time-evolution mapping between the two initial degrees-of-freedom (which are described by two subspaces as shown in above section).

		\subsection{Without time-mapping (gauge fixing) restriction}
		
		For a clear comparison, we firstly represent some results of Ref.\cite{You Y Z},
		which is the case without the time-mapping restriction.
		In terms of the four antisymmetry random tensors $g_{ij}$,
		there are three kinds of the random coupling diagrams
		(with the indices describing the mutually independent variables not being summed)
		\begin{equation} 
			\begin{aligned}
				h_{1}=\sum_{mn}g_{ij}g_{mn}g_{nm}g_{kl}\chi_{i}\chi_{j}\chi_{k}\chi_{l},\\
				h_{2}=\sum_{mn}g_{im}g_{mn}g_{nj}g_{kl}\chi_{i}\chi_{j}\chi_{k}\chi_{l},\\
				h_{3}=\sum_{mn}g_{im}g_{mj}g_{kn}g_{nl}\chi_{i}\chi_{j}\chi_{k}\chi_{l}.
			\end{aligned}
		\end{equation}
		Only the pattern of $h_{1}$ survival in large-$N$ limit,
		and thus be marginally relevant.
		While $h_{2}$ and $h_{3}$ are subleading in large-$N$ limit.
		During this process,
		each antisymmetry random tensor can describes a product of two coupling Majorana fermions.
		To further understand this we show the following examples of the tensors in each pattern:
		For $h_{1}$:
		\begin{equation} 
			\begin{aligned}
				g_{ij}&=\sum_{\alpha}\varphi_{i}(\alpha)\varphi_{j}(\alpha),\\
				g_{mn}&=\sum_{\alpha,\beta}\varphi_{m}(\alpha)\varphi_{n}(\beta),\\
				g_{nm}&=\sum_{\alpha,\beta}\varphi_{m}(\beta)\varphi_{n}(\alpha),\\
				g_{kl}&=\sum_{\beta}\varphi_{k}(\beta)\varphi_{l}(\beta),
			\end{aligned}
		\end{equation}
		where only the tensors $g_{ij}$ and $g_{kl}$ are completely independent with other tensors,
		and the commutation relation between these two tensors is exactly like that
		between two products $\varphi_{i}(\alpha)\varphi_{j}(\alpha)$ and $\varphi_{i}(\beta)\varphi_{j}(\beta)$
		before the summation over quasiparticle-number components.
		For $h_{2}$:
		\begin{equation} 
			\begin{aligned}
				g_{im}&=\sum_{\alpha}\varphi_{i}(\alpha)\varphi_{m}(\alpha),\\
				g_{mn}&=\sum_{\alpha,\beta}\varphi_{m}(\alpha)\varphi_{n}(\beta),\\
				g_{nj}&=\sum_{\beta}\varphi_{n}(\beta)\varphi_{j}(\beta),\\
				g_{kl}&=\sum_{\alpha,\beta}\varphi_{k}(\alpha)\varphi_{l}(\beta),\ (k\neq i,m;\ l\neq n,j).
			\end{aligned}
		\end{equation}
		For $h_{3}$:
		\begin{equation} 
			\begin{aligned}
				g_{im}&=\sum_{\alpha,\beta}\varphi_{i}(\alpha)\varphi_{m}(\beta),\\
				g_{mj}&=\sum_{\alpha,\beta}\varphi_{m}(\beta)\varphi_{j}(\alpha),\\
				g_{kn}&=\sum_{\alpha,\beta}\varphi_{k}(\beta)\varphi_{n}(\alpha),\ (k\neq m,l;\ n\neq i,j)\\
				g_{nl}&=\sum_{\alpha,\beta}\varphi_{n}(\alpha)\varphi_{l}(\beta),\ (n\neq i,j;\ l\neq k,m).
			\end{aligned}
		\end{equation}

		\subsection{With time-mapping restriction}
		
		As we discuss in the above section, for one of the subspaces which has all the elements ($N^{2}$) of the time mapping group $\mathcal{M}_{\tau\rightarrow \tau'}$,
		it will restrict the two Majorana fermions in another subspace 
		own only the $N$ elements of the time mapping group,
		which means the product of these two Majorana fermions can only be of the forms $\chi_{i}(\tau)\chi_{i}(\tau')$ or $\chi_{i}(\tau)\chi_{j}(\tau)$.
		We will next examining the relevance of these two cases in a statistical ensemble are indeed the same.
		We choosing the subspace of $\alpha$ as the one own only the $N$ elements.
		Then for the case of different times, $\chi_{i}(\tau)\chi_{i}(\tau')$,
		the relevant coupling pattern is
		\begin{equation} 
			\begin{aligned}
				h'_{1}&=\sum_{mn}g_{ij}g_{mn}g_{nm}g_{jk}\chi_{i}\chi_{j}\chi_{j}\chi_{k},\\
				g_{ij}&\rightarrow g_{\tau_{1},\tau_{1}'}=\sum_{\alpha}\sum_{i}\varphi_{i}(\alpha,\tau_{1})\varphi_{i}(\alpha,\tau_{1}'),\\
				g_{mn}&=\sum_{\alpha,\beta}\varphi_{m}(\alpha)\varphi_{n}(\beta),\\
				g_{nm}&=\sum_{\alpha,\beta}\varphi_{m}(\beta)\varphi_{n}(\alpha),\\
				g_{jk}&\rightarrow g_{\tau_{2},\tau_{2}'}=\sum_{\beta}\sum_{j}\varphi_{j}(\beta,\tau_{2})\varphi_{j}(\beta,\tau_{2}').
			\end{aligned}
		\end{equation}
		For the case of the same times, $\chi_{i}(\tau)\chi_{j}(\tau)$, which allows the two Majorana fermions have independent indices,
		then the relevant coupling pattern is of the one with three degrees-of-freedom as we stated above.
		This pattern reads
		\begin{equation} 
			\begin{aligned}
				h'_{2}&=\sum_{mn}g_{ij}g_{mn}g_{nm}g_{jk}\chi_{i}\chi_{j}\chi_{j}\chi_{k},\\
				g_{ij}&=\sum_{\alpha}\varphi_{i}(\alpha,\tau)\varphi_{j}(\alpha,\tau),\\
				g_{mn}&=\sum_{\alpha,\beta}\varphi_{m}(\alpha,\tau)\varphi_{n}(\beta,\tau'),\\
				g_{nm}&=\sum_{\alpha,\beta}\varphi_{m}(\beta,\tau)\varphi_{n}(\alpha,\tau'),\\
				g_{jk}&=\sum_{\beta}\varphi_{j}(\beta,\tau')\varphi_{k}(\beta,\tau').
			\end{aligned}
		\end{equation}
		The mechanism producing the three degrees-of-freedom can also be understood in this way:
		The Majorana wave functions (random variables) within the tensors $g_{ij}$ and $g_{jk}$
		are respectively, of the same time,
		thus only $N^{3}$ elements for the evolution from $\tau$ to $\tau'$ are allowed over these two terms.

		\subsection{Gauge fixing restriction in terms of anticommutation relations of Majorana operators}
		
		The above expression of $Q_{b}^{2}$ is restricted by the finite-size of time-mapping group.
		This can also be seem from the relations between the four coupled Majorana fermions,
		$\chi_{i}(\alpha,\tau_{1}), \chi_{j}(\alpha,\tau_{1}), \chi_{i}(\beta,\tau_{2}), \chi_{j}(\beta,\tau_{2})$,
		where we can at most find four zero commutators among them
		(here we focus only on the relationships between two Majorana fermions within each bosonic charge
		and that of fermion-products between two charges),
		\begin{equation} 
			\begin{aligned}
				\{\chi_{i}(\alpha,\tau_{1}),\chi_{j}(\alpha,\tau_{1})\}=0,\\
				\{\chi_{i}(\beta,\tau_{2}),\chi_{j}(\beta,\tau_{2})\}=0,\\
				\{\chi_{i}(\alpha,\tau_{1})\chi_{j}(\alpha,\tau_{1}),\chi_{i}(\beta,\tau_{2})\chi_{j}(\beta,\tau_{2})\}=0,\\,
				\{\chi_{i}(\alpha,\tau_{1})\chi_{j}(\beta,\tau_{2}),\chi_{j}(\alpha,\tau_{1})\chi_{i}(\beta,\tau_{2})\}\neq 0,\\
				\{\chi_{i}(\alpha,\tau_{1})\chi_{i}(\beta,\tau_{2}),\chi_{j}(\alpha,\tau_{1})\chi_{j}(\beta,\tau_{2})\}=0.
			\end{aligned}
		\end{equation}
		The four coupled Majorana fermions appear in the expression of $Q_{b}^{2}$ also share this character:
		$\chi_{i}(\alpha,\tau_{1}),\ \chi_{i}(\alpha,\tau_{1}'),\ \chi_{j}(\beta,\tau_{2}),\ \chi_{k}(\beta,\tau'_{2})$,
		where we can still at most find four commutating relations among them,
		\begin{equation} 
			\begin{aligned}
				\{\chi_{i}(\alpha,\tau_{1}),\chi_{i}(\alpha,\tau'_{1})\}\neq 0,\\
				\{\chi_{j}(\beta,\tau_{2}),\chi_{k}(\beta,\tau'_{2})\}=0,\\
				\{\chi_{i}(\alpha,\tau_{1})\chi_{j}(\beta,\tau_{2}),\chi_{i}(\beta,\tau_{2})\chi_{k}(\beta,\tau'_{2})\}=0,\\,
				\{\chi_{i}(\alpha,\tau_{1})\chi_{k}(\beta,\tau'_{2}),\chi_{i}(\alpha,\tau'_{1})\chi_{j}(\beta,\tau_{2})\}= 0,\\
				\{\chi_{i}(\alpha,\tau_{1})\chi_{i}(\alpha,\tau'_{1}),\chi_{j}(\beta,\tau_{2})\chi_{k}(\beta,\tau'_{2})\}=0.
			\end{aligned}
		\end{equation}
		This is in contrast with the case without time-mapping restriction,
		where the random couplings are marginally relevant at SYK$_{4}$ fixed point,
		where the corresponding four Majorana fermions are
		$\chi_{i}(\alpha,\tau_{1}),\ \chi_{j}(\alpha,\tau_{1}'),\ \chi_{i}(\beta,\tau_{2}),\ \chi_{j}(\beta,\tau'_{2})$,
		and we can find five commutating relations among them,
		\begin{equation} 
			\begin{aligned}
				\{\chi_{i}(\alpha,\tau_{1}),\chi_{j}(\alpha,\tau_{1}')\}=0,\\
				\{\chi_{i}(\beta,\tau_{2}),\chi_{j}(\beta,\tau'_{2})\}=0,\\
				\{\chi_{i}(\alpha,\tau_{1})\chi_{j}(\alpha,\tau'_{1}),\chi_{i}(\beta,\tau_{2})\chi_{j}(\beta,\tau'_{2})\}=0,\\,
				\{\chi_{i}(\alpha,\tau_{1})\chi_{j}(\beta,\tau'_{2}),\chi_{j}(\alpha,\tau'_{1})\chi_{i}(\beta,\tau_{2})\}= 0,\\
				\{\chi_{i}(\alpha,\tau_{1})\chi_{i}(\beta,\tau_{2}),\chi_{j}(\alpha,\tau'_{1})\chi_{j}(\beta,\tau'_{2})\}=0.
			\end{aligned}
		\end{equation}

		\section{Appendix.E: Relation to the Riemann zeta function and the related mathematical deductions
			of this article}
		
		For our system, the magnitude of current density flavor $M$ and 
		that of Majorana fermions $N$ dominate the phase transitions
		between chaotic to Anderson-localized phases and
		the statistical distribution of many-body level spectrum.
		Here we illustrate that the distinct statistics can also reflected
		by the nontrivial zeros of the Riemann zeta function.
		We note that,
		for a physical system described in this article,
		we using the special definition about the unglobal unit quantity
		to reflecting the gauge-invariance,
		in the mean time, as such unglobal unit quantities are indeed complex in the carrier 
		representation in position space (after the Fourier transformation),
		we can using it in the contexts of the Riemann Zeta function instead of,
		for example, expressing the argument in complex form.
		
		In the region of GSE, the gauge symmerry is indeed preserved by the
		UV cutoff on the quasiparticle position space, which is of order
		$O(z^{-2})$.
		To define the unglobal quantities in this regime,
		we firstly consider a single current density (see Appendix.A),
		\begin{equation} 
			\begin{aligned}
				\rho=\partial_{n_{0}^{z}}{\rm ln}\frac{\phi(z)}{2\pi}
				=\frac{-i}{2\pi}\frac{1}{n_{0}^{z}}.
			\end{aligned}
		\end{equation}
		If we turn to the derivative with respect to the quasiparticle positions 
		$\gamma(<z-1)$ for the essential term $\frac{q_{z}z}{2\pi}$
		which validating to mixing the momentum space and position spaces in
		certain limiting cases,
		$\frac{q_{z}z}{2\pi}=\frac{-i}{2\pi}{\rm ln}\frac{\phi(z)}{2\pi}=
		\frac{-i}{2\pi}{\rm ln}\sum_{\gamma}n_{0}^{\gamma}=\frac{-i}{2\pi}\sum_{n_{0}^{\gamma}}\frac{1}{n_{0}^{\gamma}}$,
		and its derivative with the $n_{0}^{z}$ is related to the single current density.
		As a result of this cutoff which is valid in both the position or energy spaces,
		we have $e^{1/z}=1+\frac{1}{z}$,
		and thus we can obtain the gauge-invariant shiftment $\delta z$ 
		in the position space,
		which is the Fourier transformed version of 
		the above invariant quantity ${\rm ln}n_{0}$ which
		corresponding to the Luttinger-Ward identity $\partial_{z}(i\omega_{z})$,
		\begin{equation} 
			\begin{aligned}
				&\partial_{z}{\rm ln}\frac{\phi(z)}{2\pi}
				={\rm ln}(1+\delta z)
				=\delta z+O(z^{-2}),\\
				&\delta z=\frac{n_{0}^{z}}{\sum_{\gamma}^{z-1}n_{0}^{\gamma}}.
			\end{aligned}
		\end{equation}
		These two shiftments $\delta z$ and ${\rm ln}n_{0}$ in position and energy spaces,
		respectively,
		are connected by a ratio $\delta z/{\rm ln}n_{0}=\frac{e}{e-1}\approx 1.58$,
		which is correct as long as $O(M)=O(N)$.
		Thus we can define the unglobal unit quantities in terms of positions $z$ as
		\begin{equation} 
			\begin{aligned}
				&\mathds{1}_{z}:=e^{1/z}=1+\frac{n_{0}^{z}}{\sum_{\gamma}^{z-1}n_{0}^{\gamma}},\\
				&\mathds{1}_{z+1}:=e^{1/(z+1)}=1+\frac{n_{0}^{z+1}}{\sum_{\gamma}^{z}n_{0}^{\gamma}},\\
				&\cdots.
			\end{aligned}
		\end{equation}
		As long as the gauge symmetry is not broken,
		the unit quantities satisfy
		\begin{equation} 
			\begin{aligned}
				\label{asd}
				&\partial_{z}\mathds{1}_{z}+\frac{\mathds{1}_{z}}{z}
				=\frac{\mathds{1}_{z}}{z\mathds{1}_{z+1}},\\
			\end{aligned}
		\end{equation}
		The term $\frac{{\rm ln}z}{\sum_{\gamma}^{z-1}{\rm ln}\gamma}$ appears in last line
		is useful in distinguishing the GSE from GUE.
		As we stated above,
		in terms of quasiparticle positions,
		GSE and GUE domains correspond to $O(M)=O(N)$ 
		($n_{0}=1+1/z$ in quasiparticle position space)
		and $O(M)=O(N^2)$ ($n_{0}=1+(z-1/2)^{-1}$ in quasiparticle position space), respectively.
		Simple simulation implies 
		\begin{equation} 
			\begin{aligned}
				\lim_{z\rightarrow \infty}
				\frac{{\rm ln}z}{\sum_{\gamma}^{z-1}{\rm ln}\gamma}
				=\lim_{z\rightarrow \infty}\frac{1}{z},\\
			\end{aligned}
		\end{equation}
		in GSE case which leads to $\mathds{1}_{z}\rightarrow\mathds{1}_{z+1}$ due to 
		second line of Eq.(\ref{asd});
		and
		\begin{equation} 
			\begin{aligned}
				\lim_{z\rightarrow \infty}
				\frac{{\rm ln}z}{\sum_{\gamma}^{z-1}{\rm ln}\gamma}
				=\lim_{z\rightarrow \infty}(\frac{1}{2}+i\frac{1}{z})\frac{1}{z},\\
			\end{aligned}
		\end{equation}
		in GUE case which corresponds to the complex argument
		of Riemann Hypothesis $\zeta(\frac{1}{2}+it_{n})$ with $n\rightarrow\infty$.
		More detailed mathematical discussion for GUE in random matrix theory
		in this limit, with mean asymptotic spacing $2\pi/{\rm ln}t_{\infty}$,
		are studied in Refs.\cite{KJP}.

		In GUE limit, where the gauge-invariance is broken due to the unbounded edge 
		of Zeta function,
		the global unit quantity is $\frac{-i}{2\pi}\zeta^{-1}(1)$.
		which is $z$-independent and
		its periodicity can be expressed as $\mathds{1}_{m}=
		\mathds{1}_{m}+\frac{-i}{2\pi}{\rm ln}(\mathds{1}_{m}-\zeta^{-1}(1))$.
		With unbounded edge,
		we have $\frac{\partial_{z}\phi(z)}{\phi(z)}
		=\frac{n_{0}^{z}}{\sum_{\gamma}n_{0}^{\gamma}}=\mathds{1}_{m}$
		in this limit,
		which corresponds to the
		complex argument in Riemann hypothesis
		with the nontrivial zeros $\mathds{1}_{m}=\frac{1}{2}+it_{n}$
		with $n\rightarrow\infty$.

		\subsection{GSE: $n_{0}=1+\frac{1}{z}$}
		
		Now we already known that, for three-DOF configuration where $O(M)\sim O(N)$,
		the current density follows GUE statistic.
		As we discuss in Appendix.A,
		the flavor number of current density and Majorana fermions are
		$O(M)\sim |\alpha-\beta|^{-1}(\sim |z-z'|^{-1})$ and $O(N)\sim |n_{0}-1|^{-1}$,
		which are expressed in terms of the quasiparticles in position space
		or fermionic carriers in position space, respectively.
		Next we simply denote the maximal step number in terms of the current density flavors,
		i.e., $z_{m}=|z-z'|^{-1}\sim O(M)$ (for quasiparticles in the position space).
		Here We found that, as long as $O(N)\ge O(M)$,
		Eq.(\ref{app}) is correct and the system exactly meets the three-DOF configuration
		as we discussed above.
		
		In GSE regime, we can consider each step of quaisparticle position $\delta z=1/z$
		as a constant which is independent of $z$, and $\partial_{z}\delta z=\frac{-1}{z^2}\sim 0$
		due to the IR cutoff at order $O(z^{-1})$
		(i.e., terms of order $O(z^{-2})$ or higher are omitted) in the position space.
		Then the quasiparticle number $n_{0}^{z}$ Boson phase $\phi(z)/2\pi$,
		its derivative with $z$ can be written as
		\begin{equation} 
			\begin{aligned}
				\partial_{z}n_{0}^{z-\delta z}
				&=\frac{n_{0}^{z}-n_{0}^{z-\delta z}}{\delta z}\\
				&=n_{0}^{z-\delta z}{\rm ln}n_{0},
			\end{aligned}
		\end{equation}
		where ${\rm ln}n_{0}=1/z$ under the IR cutoff,
		and $n_{0}=1+1/z$ as required by GSE condition.
		Then the derivative with $z$ for the boson field $\phi(z)$ is
		\begin{equation} 
			\begin{aligned}
				\partial_{z}\frac{\phi(z)}{2\pi}=
				\partial_{z}\sum_{\gamma}^{z-\delta z}n_{0}^{\gamma}
				&=\partial_{z}n_{0}^{z-\delta z}+\partial_{z}n_{0}^{z-2\delta z}+\cdots
				+\partial_{z}n_{0}\\
				&=\frac{n_{0}^{z}-n_{0}^{z-\delta z}}{\delta z}
				+\frac{n_{0}^{z}-n_{0}^{z-2\delta z}}{2\delta z}+\cdots
				+0\\
				&=n_{0}^{z}[\sum_{\gamma=1}^{(z-1)/\delta z}\frac{1}{\gamma\ \delta z}-
				\sum_{\gamma=1}^{(z-1)/\delta z}\frac{n_{0}^{-\gamma\ \delta z}}{\gamma\ \delta z}]\\
				&=\frac{n_{0}^{z}}{\delta z}[\sum_{\gamma=1}^{(z-1)/\delta z}\frac{1}{\gamma}-
				\sum_{\gamma=1}^{(z-1)/\delta z}\frac{n_{0}^{-\gamma\ \delta z}}{\gamma}],
			\end{aligned}
		\end{equation}
		The above expression reduces to $n_{0}^{z}/\delta z$ in the large $O(M)=O(N)$ limit.
		We note that, for the exponential part of the quasiparticle operate at the edge,
		$(z-\delta z)$, its derivative with $z$ will be $\partial_{z}(z-\delta z)=\delta z/\delta z=1$ as long as the gauge symmetry is preserved,
		and this situation changes when the system enters the GUE regime.

		\subsection{GUE: $n_{0}=1+\frac{1}{z-1/2}$}
		
		\subsubsection{$n_{0}$}
		In GUE regime, the quasiparticle operator in initial position is defined as 
		$n_{0}=1+\frac{1}{z-1/2}$ (in position representation of quasiparticles), 
		and unlike the GSE case,
		we can theoretically set the cutoff
		to $O(z^{-n})$ with arbitrary large $n$ due to the unbounded summation edge
		in the calculations of this case.
		
		Using Fourier transformation,
		we firstly represent the $n_{0}$ in terms of the current density
		(as a function of summation boundary $k$ which is always an integer)
		\begin{equation} 
			\begin{aligned}
				n_{0}(k)=\frac{\sum_{\gamma=0}^{k-1}(\frac{1}{z})^{\gamma}}
				{\sum_{\gamma=0}^{k-1}(\frac{-1}{z})^{\gamma}},
			\end{aligned}
		\end{equation}
		which can be rewritten as
		\begin{equation} 
			\begin{aligned}
				n_{0}(k)
				=&\frac{{\rm exp}{\rm ln}
					\sum_{\gamma=0}^{k-1}(\frac{1}{z})^{\gamma}}
				{{\rm exp}{\rm ln}\sum_{\gamma=0}^{k-1}(\frac{-1}{z})^{\gamma}}\\
				=&\frac{{\rm exp}{\rm ln}
					(\prod_{\gamma=0}^{k-1}(\frac{1}{z})^{\gamma}
					\sum_{\lambda=0}^{k-1}z^{\sum_{\gamma=1}^{k-1}\gamma-\lambda})}
				{{\rm exp}{\rm ln}
					(\prod_{\gamma=0}^{k-1}(\frac{-1}{z})^{\gamma}
					\sum_{\lambda=0}^{k-1}(-z)^{\sum_{\gamma=1}^{k-1}\gamma-\lambda})}\\
				=&\frac{{\rm exp}{\rm ln}
					(\frac{z^{k}-1}{z-1}z^{1-k})}
				{{\rm exp}{\rm ln}
					((-1)^{k}z^{1-k}\frac{(-1)^{k}z^{k}-1}{z+1})}\\
				=&\frac{{\rm exp}
					[i\pi+i{\rm Arg}(z-1)-i{\rm Arg}(z^{k}-1)+{\rm ln}
					(\frac{z^{k}-1}{z-1}\frac{z}{z^{k}}]}
				{{\rm exp}[i\pi-i{\rm Arg}(e^{ik\pi}z^{k}-1)+{\rm ln}
					(\frac{(-1)^{k}z^{k}-1}{z+1}\frac{z}{z^{k}})]}\\
				=&{\rm exp}[-i{\rm Arg}((-1)^{k}z^{k}-1)+{\rm ln}\frac{(-1)^{k}z^{k}-1}{z+1}],
			\end{aligned}
		\end{equation}
		where in the last line we use the following results:
		Firstly since $z$ will be a real or complex quantity in different representations
		(in terms of quasiparticles or real fermions)
		after the necessary Fourier transformations
		(whose detailed process may be omitted in the calculations),
		we have
		\begin{equation} 
			\begin{aligned}
				i{\rm Arg}(z\pm 1)={\rm ln}\frac{z\pm 1}{|z\pm 1|}
				={\rm ln}\frac{z\pm 1}{z},\\
				i{\rm Arg}(z^{k}\pm 1)={\rm ln}\frac{z^{k}\pm 1}{|z^{k}\pm 1|}
				={\rm ln}\frac{z^{k}\pm 1}{z^{k}},
			\end{aligned}
		\end{equation}
		where 
		\begin{equation} 
			\begin{aligned}
				|z-1|=\frac{z-1}{{\rm exp}[i{\rm Arg}(z-1)]}=z.
			\end{aligned}
		\end{equation}
		Here ${\rm exp}[i{\rm Arg}(z-1)]=\frac{z-1}{z}$ is in fact a very important quantity
		and we will analysis in detailed in following,
		and the quite normal relation
		\begin{equation} 
			\begin{aligned}
				{\rm exp}[-i{\rm Arg}(z-1)](z-1)=\frac{z}{z-1}(z-1)
				=z,
			\end{aligned}
		\end{equation}
		has indeed more profound explanation:
		in terms of the current density (in quasiparticle representation)
		$(-z)$ can be expressed as a function of $(-{\rm ln}(-z))=
		-(i\pi(2m+1)+{\rm ln}z)$:
		\begin{equation} 
			\begin{aligned}
				\label{nz}
				-z:=f(-{\rm ln}(-z))
				=[\lim_{-{\rm ln}(-z)\rightarrow\infty}f(-{\rm ln}(-z))]
				(1-\partial_{-{\rm ln}(-z)}f(-{\rm ln}(-z)))=\frac{z}{z-1}(1-z).
			\end{aligned}
		\end{equation}
		The expressions in this form will appear frequently in this paper,
		e.g., in the above expression,
		$(-z)$ as a function of $(-{\rm ln}(-z))$ can be expressed as
		a summation over a certain uncontinuous variable with bounded summation
		edge $(-{\rm ln}(-z)-1)$ (Note that here $(-1)$ is also a unglobal unit
		whose actual value depends on the value of each step in the summation),
		and equavalents to the product of a limiting term 
		(independent of $(-{\rm ln}(-z)-1)$) and the derivative term
		(dependent on $(-{\rm ln}(-z)-1)$).
		This product is indeed a fourier transformation under limiting condition
		of GUE region (e.g., $m\rightarrow -\infty$ here).
		Second,
		the term ${\rm ln}[(-1)^{k}]$ can be rewritten as
		\begin{equation} 
			\begin{aligned}
				{\rm ln}[(-1)^{k}]=i\pi-i{\rm Arg}((-1)^{k}z^{k}-1)
				={\rm ln}(-1)^{2k-1}-{\rm ln}(-1)^{k-1},\\
			\end{aligned}
		\end{equation}
		with
		\begin{equation} 
			\begin{aligned}
				i{\rm Arg}((-1)^{k}z^{k}-1)
				={\rm ln}\frac{(-1)^{k}z^{k}}{|(-1)^{k}z^{k}-1|}={\rm ln}(-1)^{k-1}.
			\end{aligned}
		\end{equation}
		Then similar to above expression Eq.(\ref{nz}),
		by virtue of the limiting conditions in GUE region,
		the term ${\rm exp}[i{\rm Arg}(1-(-1)^{k}z^{k})]$ 
		can be recognized as a function of discrete variable $(-1)^{-k}$,
		\begin{equation} 
			\begin{aligned}
				&{\rm exp}[i{\rm Arg}((-1)^{k}z^{k}-1)]=(-1)^{k-1}
				=\lim_{z^{-k}\rightarrow\infty} f(z^{-k})\\
				=&\frac{|f(z^{-k})|}{1-\partial_{z^{-k}}f(z^{-k})}\\
				=&\frac{\partial_{z}|f(z^{-k})|}{\partial_{z}f(z^{-k})}\\
				=&\frac{\partial_{z}|f(z^{-k})|}{(-1)^{k}kz^{k-1}}.
			\end{aligned}
		\end{equation}
		
		Now we already know $n_{0}=\frac{z+1/2}{z-1/2}$ with complex $z$ in terms of current density
		representation
		can be rewritten as a function of discrete variable $k$ (integer),
		$n_{0}(k)={\rm exp}[i{\rm Arg}(1-(-1)^{k}z^{k})-{\rm ln}\frac{1-(-1)^{k}z^{k}}{z+1}]$,
		and becomes $z$-independent.
		This can indeed be written as
		\begin{equation} 
			\begin{aligned}
				n_{0}(k)
				=&(1-\partial_{k}n_{0}(k))\frac{z+1}{z-1}\\
				=&(1-\partial_{k}n_{0}(k))
				(1+\sum_{\gamma=1}^{\infty}\frac{1}{z^{\gamma}}(\sum^{\gamma-1}_{\lambda=0}(\frac{-1}{2})^{\lambda})
				\frac{z+\frac{1}{2}}{z-\frac{1}{2}},
			\end{aligned}
		\end{equation}
		with the summation term with unbounded edge (in $z\rightarrow\infty$ limit)
		\begin{equation} 
			\begin{aligned}
				\label{xs1}
				\sum_{\gamma=1}^{\infty}\frac{1}{z^{\gamma}}(\sum^{\gamma-1}_{\lambda=0}(\frac{-1}{2})^{\lambda})
				=1-\frac{1}{-z+\frac{1}{2}+\frac{1}{2z}},
			\end{aligned}
		\end{equation}
		and it is interesting to see that, within this expression there is a
		unbounded summation (we still define it as $f$ to signal this feature)
		\begin{equation} 
			\begin{aligned}
				f(k):=\sum^{\gamma-1}_{\lambda=0}(\frac{-1}{2})^{\lambda}
				=\lim_{k\rightarrow\infty}f(k)(1-\partial_{k}f(k))\\
				=\frac{2}{3}(1-(-1)^{k}z^{-k}),
			\end{aligned}
		\end{equation}
		whose limiting value is
		\begin{equation} 
			\begin{aligned}
				\label{0666}
				\lim_{k\rightarrow\infty}f(k)=n_{0}{\rm exp}
				[{\rm ln}(1-\partial_{[{\rm ln}(z/2)]^{-1}}(z-1/2))
				-{\rm ln}(1-\partial_{[{\rm ln}(2z)]^{-1}}(z+1/2)]=\frac{2}{3}.
			\end{aligned}
		\end{equation}
		Thus in this representation, $\partial_{z}n_{0}(k)=0$, in contrast with the result of Eq.(\ref{nas}).

		\subsubsection{$n_{0}^{z}$}
		
		Let us turn back to the boson field which determines the quasiparticle number operatore
		$n_{0}^{z}$.
		Within the expression of boson field 
		$\frac{\phi(z)}{2\pi}=\sum_{\gamma}^{k-1}n_{0}^{\gamma}$,
		considering the ${\rm ln}n_{0}$ as an invariant quantity in quasiparticle space,
		the invariance can be reflected by the constant step in the exponential part of quasiparticle number operators during the summation, which is $\delta z$:
		\begin{equation} 
			\begin{aligned}
				\frac{\phi(z)}{2\pi}=\sum_{k=1}^{z/\delta z}n_{0}^{z-k\delta z}
				=\prod_{k=1}^{\frac{z}{\delta z}-1}n_{0}^{z-k\delta z}
				\sum_{\lambda=0}^{\frac{z}{\delta z}-1}n_{0}^{-(\frac{z}{\delta z}-2)z+\delta z(\sum_{\gamma=1}^{\frac{z}{\delta z}}\gamma-1)-\lambda\delta z}
				=&\frac{n_{0}^{z}-1}{n_{0}^{\delta z}-1}.
			\end{aligned}
		\end{equation}
		Next we simplify the notion $\delta z$ as 1.
		The above result is still correct in this simplified case,
		which can be verified by
		\begin{equation} 
			\begin{aligned}
				\frac{\phi(z)}{2\pi}
				=\prod_{\gamma=0}^{z-1}n_{0}^{\gamma}
				(\sum_{\lambda=0}^{z-1}n_{0}^{\frac{z(z-1)}{2}+\lambda}
				=\frac{n_{0}^{z}-1}{n_{0}^{\delta z}-1}.
			\end{aligned}
		\end{equation}
		Expresing the $\phi(z)/2\pi$ in terms of the limiting condition $z\rightarrow\infty$,
		\begin{equation}
			\begin{aligned}
				\frac{\phi(z)}{2\pi}=\frac{n_{0}^{z}-1}{n_{0}-1}
				=\lim_{z\rightarrow\infty}\frac{\phi(z)}{2\pi}(1-\partial_{z}\frac{\phi(z)}{2\pi})
				=\frac{1}{1-n_{0}}(1-n_{0}^{z}).
			\end{aligned}
		\end{equation}
		Since the boson field in the limiting condition $z\rightarrow\infty$,
		we have 
		$\partial_{z}\lim_{z\rightarrow\infty}\frac{\phi(z)}{2\pi}
		=\partial_{z}\frac{1}{1-n_{0}}=0$.

		Unlike the GSE region where $n_{0}^{z}$ is $z$-independent,
		$\partial_{z}n_{0}^{z}\neq 0$.
		The long-wavelength limit in momentum space with $q_{z}\rightarrow 0$
		results in $-iq_{z}=\partial_{z}$.
		Thus in the limiting condition,
		we have $\lim_{q_{z}\rightarrow 0}\partial_{z}=-i0=i0^{-}={\rm ln}n_{0}$,
		\begin{equation} 
			\begin{aligned}
				\lim_{\partial_{z}\rightarrow i0^{-}((i\partial_{z})^{-1}\rightarrow\infty)}
				\partial_{z}n_{0}
				=n_{0}^{z}{\rm ln}n_{0}.
			\end{aligned}
		\end{equation}
		In the mean time, the derivative term $\partial_{z}n_{0}^{z}$ can be rewritten as
		\begin{equation} 
			\begin{aligned}
				\label{nas}
				\partial_{z}n_{0}
				=\frac{\partial (z-1)}{\partial z}n_{0}^{z}{\rm ln}n_{0}
				+n_{0}^{z-1}\frac{\partial n_{0}}{\partial z},
			\end{aligned}
		\end{equation}
		so we obtain the expression in familiar form
		\begin{equation} 
			\begin{aligned}
				n_{0}{\rm ln}n_{0}=\frac{\partial_{z}n_{0}}{1-\frac{\partial (z-1)}{\partial z}},
			\end{aligned}
		\end{equation}
		where the left-hand-side is the limiting result 
		$\lim_{(i\partial_{z})^{-1}\rightarrow\infty}\partial_{z}n_{0}$,
		and
		\begin{equation} 
			\begin{aligned}
				\frac{\partial (z-1)}{\partial z}=\partial_{(i\partial_{z})^{-1}}\partial_{z}n_{0}.
			\end{aligned}
		\end{equation}
		Similarly we have
		\begin{equation} 
			\begin{aligned}
				n_{0}{\rm ln}n_{0}=\frac{\partial_{z-1}n_{0}}{1-\frac{\partial (z-2)}{\partial (z-1)}},
			\end{aligned}
		\end{equation}
		with
		\begin{equation} 
			\begin{aligned}
				&\lim_{(i\partial_{z-1})^{-1}\rightarrow\infty}\partial_{z-1}n_{0}=n_{0}{\rm ln}n_{0},\\
				&\frac{\partial (z-2)}{\partial (z-1)}=\partial_{(i\partial_{z-1})^{-1}}\partial_{z-1}n_{0}.
			\end{aligned}
		\end{equation}
		
		A Fourier transformation is contained in these processes,
		\begin{equation} 
			\begin{aligned}
				0^{+}
				=\lim_{q_{z}\rightarrow 0}i\partial_{z}
				=&\int^{\infty}_{-\infty}
				\frac{dz}{2\pi}iz\ {\rm exp}[iz0^{-}],
			\end{aligned}
		\end{equation}
		which provides the factor $\frac{i}{2\pi}$.
		This can be seem from the expression of current density
		\begin{equation} 
			\begin{aligned}
				\rho=\frac{-i}{2\pi}\frac{1}{n_{0}^{z}}
				=\frac{i}{2\pi}\lim_{(i\partial_{z})^{-1}\rightarrow\infty}\frac{\partial z}{\partial n_{0}^{z}}
				\partial_{z}{\rm ln}\frac{\phi(z)}{2\pi},
			\end{aligned}
		\end{equation}
		where ${\rm ln}\frac{\phi(z)}{2\pi}={\rm ln}\frac{n_{0}^{z}-1}{n_{0}-1}
		=iq_{z}z\rightarrow -1$ in his limit.
		In the mean time,
		\begin{equation} 
			\begin{aligned}
				{\rm ln}n_{0}(k)
				&={\rm ln}\frac{z+1}{z-1}
				(1-n_{0}^{-1}(k)\partial_{k}n_{0}(k))\\
				&=(1-\frac{z-1}{z}\frac{1}{z^{k}-1}){\rm ln}\frac{z}{z-1}
				-(1-\frac{z+1}{z}\frac{1}{(-z)^{k}-1}){\rm ln}\frac{z}{z+1}\\
				&=(1-\frac{\partial\frac{1}{z}}{\partial {\rm Li}_{1}\frac{1}{z}}\frac{1}{z^{k}-1}){\rm Li}_{1}(\frac{1}{z})
				-(1-\frac{\partial\frac{-1}{z}}{\partial {\rm Li}_{1}\frac{-1}{z}}\frac{1}{(-z)^{k}-1}){\rm Li}_{1}(\frac{-1}{z}).
			\end{aligned}
		\end{equation}
		
		\subsection{Gauge invariance-dominated radial derivative and normalized Lebesgue measurement on complex space}
		
		Next we introduce some direct results due to the Fourier transformations under certain limiting conditions.
		Firstly,
		as an example,
		we focus on the function $f_{k}:=\sum^{k-1}_{\gamma=0}\frac{1}{z^{\gamma}}$,
		which appears as a denominator term in the above expression of $n_{0}(k)$.
		
		One of the results of gauge-invariance is that,
		when $k\rightarrow\infty$, the gauge-invariant ratio term
		$\lim_{k\rightarrow\infty}f_{k}=\frac{z}{z-1}$
		does not depend on the value of integer $k$,
		but we have the nonzero commutation $[\frac{z}{z-1},\frac{1}{z}]\neq 0$ now.
		Using the above results, we can obtain
		\begin{equation} 
			\begin{aligned}
				\label{zzq}
				\partial_{\frac{1}{z}}\frac{z}{z-1}=z\partial_{\frac{1}{z}}\frac{1}{z-1}=\frac{z^{2}}{z-1},
			\end{aligned}
		\end{equation}
		which corresponds to the limiting results $
		\lim_{z\rightarrow \infty}\frac{1}{z-1}=\frac{1}{z},
		\lim_{\frac{1}{z}\rightarrow \infty}z=z$,
		and thus $
		\partial_{z}\frac{1}{z-1}=\frac{1}{1-z},
		\partial_{\frac{1}{z}}z=0$.
		In the mean time, for unbounded case in the summation,
		the $k$-dependence of $f_{k}$ results in $[f_{k},\frac{1}{z}]
		=[\frac{z}{1-z}(1-\frac{1}{z^{k}}),\frac{1}{z}]= 0$.
		Instituding Eq.(\ref{zzq}) into $\partial_{\frac{1}{z}}f_{k}=0$,
		we obtain
		$\partial_{\frac{1}{z}}\frac{1}{z^{k}}=z(1-\frac{1}{z^{k}})$
		(which is actually the fourier-transformed form of $k\frac{1}{z^{k-1}}$.
		Choosing the invariant unit in current-density space,
		we can further write
		\begin{equation} 
			\begin{aligned}
				&\partial_{\frac{1}{z}}\frac{1}{z^{k}}=z(1-\frac{1}{z^{k}})\\
				&=-\frac{1}{z^{k}}(i{\rm Arg}(z^{k+1})-i{\rm Arg}(z))
				=\frac{1}{z^{k}}{\rm ln}\frac{1}{z},
			\end{aligned}
		\end{equation}
		which means the Fourier transformation during this process
		is equivalents to the replacement of derivative operator $\partial_{\frac{1}{z}}$ by $\partial_{k}$.
		This is a natural result of the characters of integral operator in complex space
		acting on some holomorphic functions base on the defined specific unit,
		and the normalized Lebesgue measure\cite{s1} are affected by the boundedness in such space.
		Another application is the Luttinger-Ward analysis on calculating the fermion charge operator\cite{s2}.
		
		For $f_{k}$,
		if we make the derivative with $k$ again on the above-obtained expression
		$\partial_{k}f_{k}=1-\frac{f_{k}}{\lim_{k\rightarrow \infty}f_{k}}$,
		we obtain
		\begin{equation} 
			\begin{aligned}
				\partial^{(2)}_{k}f_{k}=\partial_{k}1
				-\frac{\partial_{k}f_{k}}{\lim_{k\rightarrow \infty}f_{k}}
				-f_{k}\partial_{k}\frac{1}{\lim_{k\rightarrow \infty}f_{k}}
				=\frac{1}{z^{k}}{\rm ln}\frac{1}{z}+\frac{1}{z^{k-1}}k\partial_{k}\frac{1}{z},
			\end{aligned}
		\end{equation}
		where $1$ here is defined as a global unit according to its independence of $k$,
		then we obtain the following transformations under the limiting condition
		of vanishing $k$-dependence of the unit term $1$,
		\begin{equation} 
			\begin{aligned}
				\label{2li}
				&\lim_{\partial_{k}1\rightarrow 0}\frac{1}{z}{\rm ln}\frac{1}{z}=
				-k\partial_{k}\frac{1}{z},\\
				&\lim_{\partial_{k}1\rightarrow 0}\partial_{k}[\frac{z}{z-1}]^{-1}=(\frac{z}{z-1})^{2}
				\frac{1}{1-z^{k}},
			\end{aligned}
		\end{equation}
		where the second line leads to the result about the difference between inversed $f_{k}$ and
		its limiting result,
		\begin{equation} 
			\begin{aligned}
				\label{244}
				f_{k}^{-1}-(\lim_{k\rightarrow \infty}f_{k})^{-1}=\frac{1-\frac{1}{z}}{z^{k}-1}
				=-\lim_{k\rightarrow \infty}f_{k}\partial_{k}(\lim_{k\rightarrow \infty}f_{k})^{-1}
				=\partial_{k}{\rm ln}\lim_{k\rightarrow \infty}f_{k}
				=\partial_{k}{\rm ln}\frac{z}{z-1}.
			\end{aligned}
		\end{equation}
		Note that the above two limiting expressions Eq.(\ref{2li}) corresponds to
		the cutoff at second order of $k$-derivative, i.e., $\partial_{k}^{(2)}f_{k}
		=\frac{1}{z^{k}}{\rm ln}\frac{1}{z}
		+\frac{1}{z^{k-1}}k\partial_{k}\frac{1}{z}=0$.
		
		By writing $f_{k}$ in terms of the $k$-independent unit,
		as $f_{k}=\frac{z}{z-1}(1-\frac{1}{z^{k}})$,
		the limiting expression $\lim_{k\rightarrow \infty}f_{k}$ can be expanded again as
		\begin{equation} 
			\begin{aligned}
				\frac{z}{z-1}
				=\frac{\frac{z}{z-1}(1-\frac{1}{z^{k}})}
				{1'-\frac{z}{z-1}\partial_{k}(1-\frac{1}{z^{k}})},
			\end{aligned}
		\end{equation}
		then it is easy to see that the above limiting expressions Eq.(\ref{2li})
		as well as the cutoff at second order of $k$-derivative corresponds to
		the scaling behaviors that $\partial_{k}1$ and $\partial_{k}\frac{1}{z^{2}}$
		tend to zero in a same speed and 
		in the mean time $1'\rightarrow \frac{f_{k}}{\lim_{k\rightarrow \infty}f_{k}}$.
		This result means in the framwork consist of unit $1'$, the step of
		variable $k$ is larger than the one in previous framwork consist of unit $1$,
		alought all these units are both independent with $k$,
		and their step ratio can be simply denoted as 
		$\frac{\lim_{k\rightarrow \infty}(k-1)}{\lim_{k\rightarrow \infty}k}$,
		i.e., the ratio between summation boundary and the scaled variable itself.
		While if we keep the framwork be invariant,
		i.e, $1'\rightarrow 1$,
		then $\partial_{k}(1-\frac{1}{z^{k}})\rightarrow \frac{z-1}{z^{k+1}}$.
		
		In conclusion,
		a direct result for the scaled $f_{k}$ as defined in the begining of this subsection
		is $\partial_{k}\frac{z^{k}}{z^{k}-1}=\partial_{k}\sum_{\gamma}^{\infty-1}
		\frac{1}{z^{k\gamma}}=
		\frac{z^{k}}{z^{k}-1}\frac{z-1}{z}\frac{1}{1-z^{k}}$.
		Then there are three possible cases which corresponds to different representations.
		By expanding $\frac{z^{k}}{z^{k}-1}$ as
		\begin{equation} 
			\begin{aligned}
				\frac{z^{k}}{z^{k}-1}=1+\frac{1}{z^{k}}+(\frac{1}{z^{k}})^{2}+\cdots
				=1+\frac{1}{z^{k}}+(-1).
			\end{aligned}
		\end{equation}
		Note that here the 1 and (-1)
		in last line are not the units of the same representation.
		We will see that the three possible configurations correspond to, respectively,
		the $k$-dependence of the three terms with the above expanded $k$-scaled ratio.
		The first configuration corresponds to 
		$\partial_{k}(-1)=\frac{z-1}{z}\neq 0$ while 
		$\partial_{k}1=\partial_{k}\frac{1}{z^{k}}= 0$.This corresponds to the current density representation
		where $z\rightarrow\infty$ and the unit quantity is $z^{-k}\sim -z^{-k}
		\rightarrow 0$.
		The second configuration corresponds to 
		$\partial_{k}1=\frac{1}{z^{k}}\frac{z-1}{z}\neq 0$ while 
		$\partial_{k}(-1)=\partial_{k}\frac{1}{z^{k}}= 0$.This corresponds to the Majorana fermion representation
		where $z\rightarrow 0^{+}$ and the unit quantity is $(-1)\sim z^{k}
		\rightarrow 0$.
		The third configuration corresponds to 
		$\partial_{k}\frac{1}{z^{k}}=\frac{1}{z^{k}}\frac{1-z}{z}\neq 0$ while 
		$\partial_{k}1=\partial_{k}(-1)= 0$.This corresponds to the Majorana fermion representation
		where $z\rightarrow 0^{-}$ and the unit quantity is $(-1)\sim -z^{k}
		\rightarrow 0$.
		Specifically,
		the third configuration corresponds to the Bernoulli polynomials (as discuss below)
		where ${\rm ln}(-z)
		=-{\rm ln}(\frac{-1}{z})$ with ${\rm ln}(-1)\sim {\rm ln}1\rightarrow 0$,
		which means the imaginary part ${\rm ln}(-1)=i\pi (2m+1)$
		and available the analytical continuation for complex argument larger than 1 
		(i.e., beyond convergence $|z|=1$).
		This configuration consider the $z$-dependence in terms of the normalized Lebesgue measure.
		While both the frist and second configurations correspond to the 
		above-mentioned second order cutoff at $\partial_{k}$.

		\subsection{Lebesgue measure}
		
		Next we focus on the independence between $(\frac{1}{z})^{\gamma}$ and bounded summation result
		$\frac{z}{z-1}(1-\frac{1}{z^{k}})$.
		This independence results in 
		\begin{equation} 
			\begin{aligned}
				\label{deeq}
				\partial_{(\frac{1}{z})^{\gamma}}\frac{z}{z-1}(1-\frac{1}{z^{k}})=0.
			\end{aligned}
		\end{equation}
		Note that, in the unbounded limit ($k\rightarrow \infty$),
		the result $\frac{z}{z-1}$ in fact does not directly depends on $(1/z)^{\gamma}$
		since they are indeed not the quantities of the same representation
		(the summation over $\gamma$ can be treated as Fourier transformation here),
		thus the derivative in the left-hand-side of above equation
		requires normalized Lebesgue measure on the complex space,
		$\mathcal{S}_{\gamma}=\{\frac{1}{z}\}_{m},(m=1,2,\cdots,\gamma)$, 
		which is made up with $\gamma$ integer steps.
		The unbounded edge means the step $\delta \gamma$ is an invariant quantity,
		so the value of $\frac{z}{z-1}$ only depends on the variations of $(1/z)^{\gamma}$ 
		after each move by the step $\delta \gamma$.
		We will see that this is also base one the gauge invariance,
		which defines a unit space made up of holomorphic functions,
		e.g., $\lim_{k\rightarrow \infty}f_{k}=\frac{z}{z-1}$.
		Thus the derivative $\partial_{(\frac{1}{z})^{\gamma}}\frac{z}{z-1}$ can be written as
		\begin{equation} 
			\begin{aligned}
				\partial_{(\frac{1}{z})^{\gamma}}\frac{z}{z-1}
				=&\frac{-z}{z-1}\partial_{(\frac{1}{z})^{\gamma}}{\rm ln}\frac{z^{k}-1}{z^{k}}\\
				=&\frac{z}{z-1}\frac{z^{k}}{z^{k}-1}\partial_{(\frac{1}{z})^{\gamma}}\frac{1}{z^{k}}.
			\end{aligned}
		\end{equation}
		Here $\partial_{(\frac{1}{z})^{\gamma}}\mathds{1}=0$ under some limiting condition
		with $\mathds{1}$ denotes 
		the unglobal unit, and for integer $m<\gamma$, 
		$\partial_{(\frac{1}{z})^{m}}\mathds{1}\neq 0$. This result will be used below.
		
		In the mean time, this derivative is equivalents to
		\begin{equation} 
			\begin{aligned}
				\label{eqse}
				\partial_{(\frac{1}{z})^{\gamma}}\frac{z}{z-1}
				=&\frac{-z^{\gamma}}{{\rm ln}z}\partial_{\gamma}\frac{z}{z-1}\\
				=&\sum^{\gamma}_{m=1}(\frac{1}{z})^{m}\partial_{(\frac{1}{z})^{m}}\frac{z}{z-1},
			\end{aligned}
		\end{equation}
		where the last line is the radial derivative of holomorphic function $\frac{z}{z-1}$,
		which overcomes the problem that $\frac{z}{z-1}$ and $(\frac{1}{z})^{\gamma}$
		are in the different spaces,
		and this summation over unit steps $m$
		can be replaced by a larger one $(\gamma+1)$
		whose steps $\delta (\gamma+1)$ is no more an invariant quantity.
		In terms of steps indexed by real integer $m$,
		the $\gamma$ should no more be treated as a real integer any more
		when it anticipates the calculation with $m$.
		Considering the complex argument of $\gamma$,
		only $|\gamma|$ is of the same representation with that in the configuration of $m$.
		Then the above summation over $m$ can be expanded as
		\begin{equation} 
			\begin{aligned}
				\label{sdd}
				&\sum^{\gamma}_{m=1}(\frac{1}{z})^{m}\partial_{(\frac{1}{z})^{m}}
				[\sum_{\gamma=0}^{\infty}(\frac{1}{z})^{|\gamma|}(\frac{1}{z})^{\gamma-|\gamma|}]\\
				=&[\sum^{|\gamma|}_{m=1}\frac{1}{m}\sum_{\gamma=1}|\gamma|(\frac{1}{z})^{|\gamma|}
				+(\sum_{\gamma=0}^{\infty}\frac{1}{z^{|\gamma|}}\frac{-1}{|\gamma|}
				+\sum_{m=1}^{|\gamma|}\frac{1}{m})\frac{1}{{\rm ln}\frac{1}{z}}
				+\sum_{m=|\gamma|}^{\gamma}
				(\frac{1}{z})^{m}\partial_{(\frac{1}{z})^{m}}
				\sum_{\gamma=0}^{\infty}(\frac{1}{z})^{|\gamma|}](\frac{1}{z})^{\gamma-|\gamma|}\\
				=&[\mathds{H}_{|\gamma|}({\rm Li}_{-1}\frac{1}{z}+\frac{1}{{\rm ln}\frac{1}{z}})
				+\sum_{\gamma=0}^{\infty}\frac{1}{z^{|\gamma|}}\frac{-1}{|\gamma|}
				\frac{1}{{\rm ln}\frac{1}{z}}
				+\sum_{m=|\gamma|}^{\gamma}
				(\frac{1}{z})^{m}\partial_{(\frac{1}{z})^{m}}
				\sum_{\gamma=0}^{\infty}(\frac{1}{z})^{|\gamma|}](\frac{1}{z})^{\gamma-|\gamma|}.
			\end{aligned}
		\end{equation}
		Note that the boundary of
		summation over $m$ here, $\gamma$, 
		is not in the same representation with the integers $m=1,\cdots,|\gamma|$.

		The factor $(\frac{1}{z})^{\gamma-|\gamma|}$
		is independent with both the $(\frac{1}{z})^{m}$ ($m=1,\cdots,\gamma$)
		and $\gamma$ (or factor $(\frac{1}{z})^{\gamma}$),
		e.g.,
		for $m=\gamma$,
		we have
		\begin{equation} 
			\begin{aligned}
				\frac{\partial
					(\frac{1}{z})^{|\gamma|(\frac{\gamma}{|\gamma|}-1)}}{\partial (\frac{1}{z})^{|\gamma|}}
				=\frac{\partial
					(\frac{1}{z})^{\gamma-|\gamma|}}{\partial (\frac{1}{z})^{|\gamma|}}
				=0,
			\end{aligned}
		\end{equation}
		which is due to the fact that $\frac{\partial \gamma}{\partial |\gamma|}=1$ but 
		$\frac{\gamma}{|\gamma|}\neq 1$
		(here $\frac{\partial \gamma}{\partial |\gamma|}$ is a limiting result
		of $\frac{\gamma}{|\gamma|}$), and thus we have
		\begin{equation} 
			\begin{aligned}
				\label{sf}
				\frac{\partial
					(\frac{1}{z})^{\gamma}}{\partial (\frac{1}{z})^{-|\gamma|}}
				=-(\frac{1}{z})^{\gamma+|\gamma|}\frac{\partial \gamma}{\partial |\gamma|}
				=\frac{
					-(\frac{1}{z})^{\gamma}}{(\frac{1}{z})^{-|\gamma|}}.
			\end{aligned}
		\end{equation}
		Its independence with $(\frac{1}{z})^{-\gamma}$ can also be understood by substituting the above result
		(Eq.(\ref{sf})) into $\frac{\partial
			(\frac{1}{z})^{\gamma-|\gamma|}}{\partial (\frac{1}{z})^{\gamma}}$, which turns out to be zero.
		This $\gamma$-independent term has the following relation 
		\begin{equation} 
			\begin{aligned}
				(\frac{1}{z})^{\gamma-|\gamma|}
				=\frac{\gamma}{|\gamma|}
				\frac{1}{1-\gamma{\rm ln}\frac{1}{z}},
			\end{aligned}
		\end{equation}
		which means it is in fact a limiting result of the ratio $\frac{\gamma}{|\gamma|}$
		\begin{equation} 
			\begin{aligned}
				(\frac{1}{z})^{\gamma-|\gamma|}
				=\lim_{|\gamma|\rightarrow\infty}
				\frac{\gamma}{|\gamma|},
			\end{aligned}
		\end{equation}
		and we also have the derivative 
		$\partial_{|\gamma|}\frac{\gamma}{|\gamma|}=\gamma{\rm ln}\frac{1}{z}$.
		
		A Fourier transformation under limiting condition 
		$\gamma\rightarrow -|\gamma|$ results in 
		an essential relation which connecting different representations
		$\frac{-1}{|\gamma|}=\partial_{|\gamma|}=-\partial_{\gamma}$,
		and this is similar to the relation
		$\lim_{q_{z}\rightarrow\infty}=\partial_{z}$ longwavelength limit.
		The unit quantity here is obviously not the ones in both the representations of $\gamma$ and $|\gamma|$,
		and we have 
		$\frac{\partial \gamma}{\partial |\gamma|}=1=-\frac{\partial |\gamma|}{\partial |\gamma|}
		=-\frac{\partial \gamma}{\partial \gamma}=1$.
		
		In this new representation,
		we have $(\frac{1}{z})^{\gamma-|\gamma|}=\frac{\partial \frac{1}{z^{|\gamma|}}}{\partial \frac{1}{z^{\gamma}}}$,
		thus
		the derivative $\partial_{(\frac{1}{z})^{\gamma}}\frac{z}{z-1}$
		can be rewritten as
		\begin{equation}
			\begin{aligned}
				\label{po9}
				\partial_{(\frac{1}{z})^{\gamma}}\frac{z}{z-1}
				=\partial_{(\frac{1}{z})^{|\gamma|}}
				\sum_{\gamma=0}^{\infty}(\frac{1}{z})^{|\gamma|}.
			\end{aligned}
		\end{equation}
		Defining the function of $\gamma$ as $f_{\gamma}=\sum_{\gamma=0}^{\infty}(\frac{1}{z})^{|\gamma|}$,
		we can obtain
		\begin{equation}
			\begin{aligned}
				\partial_{|\gamma|}({\rm ln}f_{\gamma}-{\rm ln}\frac{1}{|\gamma|})=0,\\
				\partial_{\gamma}({\rm ln}f_{\gamma}-{\rm ln}\frac{1}{\gamma})=0,
			\end{aligned}
		\end{equation}
		and the function $f_{\gamma}$ can indeed be represented as
		\begin{equation}
			\begin{aligned}
				f_{\gamma}=-\gamma \partial_{\gamma}f_{\gamma}
				=-|\gamma| \partial_{|\gamma|}f_{\gamma}.
			\end{aligned}
		\end{equation}
		Similarly,
		the limiting condition $\gamma\rightarrow -|\gamma|$
		corresponds to another limiting condition $z\rightarrow 0^{+}$,
		as can be shown by the derivative of polylogarithm 
		\begin{equation}
			\begin{aligned}
				\partial_{\frac{1}{z}}{\rm Li}_{-1}(\frac{1}{z})
				=z{\rm Li}_{-2}(\frac{1}{z})
				=-z{\rm Li}_{-1}(\frac{1}{z}),
			\end{aligned}
		\end{equation}
		where the last line is due to 
		$\lim_{z\rightarrow
			0^{+}}\frac{{\rm Li}_{-2}(\frac{1}{z})}{{\rm Li}_{-2}(\frac{1}{z})}=
		\lim_{z\rightarrow
			0^{+}}\frac{z+1}{z-1}=-1$.
		
		Thus in this case,
		$(\frac{1}{z})^{\gamma-|\gamma|}=\frac{1}{{\rm ln}\frac{-\gamma}{|\gamma|}}
		=\frac{1}{i\pi(2m+1)+{\rm ln}\frac{\gamma}{|\gamma|}}\rightarrow\infty$,
		and the ratio between two representations can be rewitten as
		\begin{equation}
			\begin{aligned}
				\frac{\gamma}{|\gamma|}=
				\frac{\gamma-|\gamma|}{{\rm ln}(-\gamma)-{\rm ln}|\gamma|}(1+\gamma{\rm ln}{\rm ln}\frac{-\gamma}{|\gamma|}),
			\end{aligned}
		\end{equation}
		and $f_{\gamma}$ can be related to the polylogarithm function by
		\begin{equation}
			\begin{aligned}
				\label{po90}
				\frac{\partial f_{\gamma}}{\partial {\rm Li}_{-1}(\frac{1}{z})}
				=\frac{ f_{\gamma}}{ {\rm Li}_{-1}(\frac{1}{z})}
				\frac{1}{z\gamma}\frac{\partial \gamma}{\partial \frac{1}{z}}
			\end{aligned}
		\end{equation}
		
		The relation between $f_{\gamma}$ (in representation of $\gamma$
		whose limiting condition $|\gamma|\rightarrow -\gamma$
		corresponds to invariant quantity $\frac{1}{z}\rightarrow\infty$) and the polylogarithm functions
		(in representation of $\frac{1}{z}$ whose limiting condition is $|\gamma|\gg\gamma$
		and results in $|\gamma|$-independent factor $(\frac{1}{z})^{\gamma-|\gamma|}$),
		can be further understood by considering the following expressions in the representation of $\gamma$
		(any terms with summation boundary depends on $\gamma$ are of this representation),
		\begin{equation}
			\begin{aligned}
				\label{xxc}
				\partial_{|\gamma|}\frac{\gamma-|\gamma|}{|\gamma|}=\frac{|\gamma|-\gamma}{|\gamma|^{2}},
			\end{aligned}
		\end{equation}
		which is consistent with the limiting result $\partial_{|\gamma|}=-|\gamma|^{-1}$
		as well as the relation $[(\frac{1}{z})^{\gamma-|\gamma|},(\frac{1}{z})^{|\gamma|}]=0$
		which is valid in both two representations.
		Base on the result $\partial_{|\gamma|}|\gamma|=-1$,
		we have another expression,
		\begin{equation}
			\begin{aligned}
				\partial_{|\gamma|}{\rm Li}_{0}(e^{-|\gamma|})=-{\rm Li}_{-1}(e^{-|\gamma|})
				=-e^{-|\gamma|}\partial_{e^{-|\gamma|}}{\rm Li}_{0}(e^{-|\gamma|}),
			\end{aligned}
		\end{equation}
		thus we can obtain that the condition $|\gamma|\rightarrow -\gamma$ in representation of $\gamma$
		also equivalents to the condition $z\sim {\rm ln}(-e^{|\gamma|})
		\sim i\pi (2m+1)+|\gamma|\rightarrow 0^{+}$.

		By rewritting Eq.(\ref{xxc}) 
		as $\partial_{|\gamma|}(\frac{\gamma}{|\gamma|}\frac{\gamma-|\gamma|}{\gamma})$,
		we can obtain another $\gamma$-independent ratio $\frac{\gamma-|\gamma|}{\gamma}$,
		which reflects a direct result of taking the limit $|\gamma|\rightarrow -\gamma$,
		i.e., $\partial_{|\gamma|}=-1/|\gamma|$.
		Next we prove this in terms of polylogarithm function.
		
		\subsection{Invariant unit quantities in two representations}
		
		We can notice a fact which is a result of unglobal invariance:
		the limiting result of the factors $\frac{\gamma}{|\gamma|}$ and $\frac{\gamma-|\gamma|}{|\gamma|}$
		are in different representation,
		however, that do can be related by the above-mentioned $\gamma$-independent ratio factor
		$\frac{\gamma-|\gamma|}{\gamma}$,
		then we have
		\begin{equation}
			\begin{aligned}
				\lim_{|\gamma|\rightarrow \infty}\frac{\gamma-|\gamma|}{|\gamma|}
				=(\lim_{|\gamma|\rightarrow \infty}\frac{\gamma}{|\gamma|})\frac{\gamma-|\gamma|}{\gamma}.
			\end{aligned}
		\end{equation}
		The reason here is that although the ratio factor (just like the above $\frac{z}{z-1}$ in current density representation) here is invariant under the limiting transition of $|\gamma|$,
		it plays the role of exponential factor connecting different representations,
		and thus change the limiting result of $\frac{\gamma}{|\gamma|}$.
		
		For the first factor $\frac{\gamma}{|\gamma|}$,
		the $\gamma$-independent unit reads
		\begin{equation}
			\begin{aligned}
				&\mathds{1}=\frac{\lim_{|\gamma|\rightarrow\infty}\frac{\gamma}{|\gamma|}}
				{\frac{\gamma}{|\gamma|}}
				=\frac{1}{1-\partial_{|\gamma|}\frac{\gamma}{|\gamma|}}\\
				&=\lim_{|\gamma|\rightarrow\infty}
				\sum_{m=0}^{|\gamma|-1}[\partial_{|\gamma|}\frac{\gamma}{|\gamma|}]^{m},
			\end{aligned}
		\end{equation}
		with $\partial_{|\gamma|}\frac{\gamma}{|\gamma|}
		=\gamma{\rm ln}\frac{1}{z}$;
		while for the second factor $\frac{\gamma-|\gamma|}{|\gamma|}$,
		the $\gamma$-independent unit reads
		\begin{equation}
			\begin{aligned}
				&\mathds{1}'=\frac{\lim_{|\gamma|\rightarrow\infty}\frac{\gamma-|\gamma|}{|\gamma|}}
				{\frac{\gamma-|\gamma|}{|\gamma|}}
				=\frac{1}{1-\partial_{|\gamma|}\frac{\gamma-|\gamma|}{|\gamma|}}\\
				&=\lim_{|\gamma|\rightarrow\infty}
				\sum_{m=0}^{|\gamma|-1}[\partial_{|\gamma|}\frac{\gamma-|\gamma|}{|\gamma|}]^{m},
			\end{aligned}
		\end{equation}
		with $\partial_{|\gamma|}\frac{\gamma-|\gamma|}{|\gamma|}
		=(\gamma-|\gamma|){\rm ln}\frac{1}{z}$.
		Just like the relation between $\gamma$ and $|\gamma|$,
		we have $\mathds{1}\neq \mathds{1}'$, but $\partial_{|\gamma|}\mathds{1}
		=\partial_{|\gamma|}\mathds{1}'$.
		
		That means, for these two factors, $\frac{\gamma}{|\gamma|}$ and $\frac{\gamma-|\gamma|}{|\gamma|}$,
		the unit quantities are different due to the 
		different extent of limiting transform of $|\gamma|$,
		and both of them are larger than the unit quantity in representation of integer-$m$,
		which is $\frac{|\gamma|}{|\gamma|}=1$.
		As a result the difference between the limiting form of these two factors reads
		$\lim_{|\gamma|\rightarrow\infty}(-1)=\frac{-1}{1-\partial_{|\gamma|}(-1)}
		=\frac{-1}{1+|\gamma|{\rm ln}\frac{1}{z}}$.
		By substituting the above results,
		we can further obtain the following expressions
		\begin{equation}
			\begin{aligned}
				\label{oxc}
				\frac{\mathds{1}}{\mathds{1}'}=\frac{1-\mathds{1'}^{-1}}{1-\mathds{1}^{-1}}
				={\rm ln}(\frac{1}{z})^{-|\gamma|},\\
				\frac{\partial_{|\gamma|}\frac{\mathds{1}'}{\mathds{1}}}
				{\partial_{|\gamma|}(\frac{\mathds{1}'}{\mathds{1}}-1)}=\frac{\gamma}{|\gamma|}.
			\end{aligned}
		\end{equation}
		By defining the factor 
		\begin{equation}
			\begin{aligned}
				\mathcal{X}:=\frac{1-\partial_{|\gamma|}\frac{\gamma}{|\gamma|}}{\partial_{|\gamma|}(-1)},
			\end{aligned}
		\end{equation}
		the second line of Eq.(\ref{oxc}) can be rewritten as
		\begin{equation}
			\begin{aligned}
				\frac{\partial -{\rm Li}_{0}(\mathcal{X})}
				{\partial {\rm Li}_{0}(\frac{1}{\mathcal{X}})}=\frac{\partial |\gamma|}{\partial |\gamma|}\frac{\gamma}{|\gamma|}.
			\end{aligned}
		\end{equation}
		In terms of Bernoulli polynomials,
		the factor in denominator can be expressed as
		\begin{equation}
			\begin{aligned}
				\label{op08}
				{\rm Li}_{0}(\frac{1}{\mathcal{X}})=e^{{\rm ln}(-1)+{\rm Li}_{1}(\mathcal{X})}=
				-{\rm Li}_{0}(\mathcal{X})-B_{0}[\frac{1}{2}(1-\frac{{\rm ln}\frac{-1}{\mathcal{X}}}{i\pi})],
			\end{aligned}
		\end{equation}
		where the Bernoulli polynomials $B_{0}$
		as a function of $\frac{1}{2}(1-\frac{{\rm ln}\frac{-1}{\mathcal{X}}}{i\pi})$ requires 
		the parameter $\mathcal{X}$ to satisfies $-{\rm ln}(-\mathcal{X})={\rm ln}\frac{-1}{\mathcal{X}}$.
		We will see that this restriction indeed corresponds to,
		in representation of this Bernoulli polynomial,
		${\rm ln}\frac{-|\gamma|}{|\gamma|}={\rm ln}(-1)=-{\rm ln}(-1)$,
		or equivalently,
		$2{\rm ln}(-1)\rightarrow 0=\partial_{|\gamma|}\mathds{1}=\partial_{|\gamma|}\mathds{1}'$,
		where zero here denotes a quantity shared by the two factors in $\gamma$-representation.
		The expression within the bracket of Bernoulli polynomial also implies that,
		this restriction also results in $\mathds{1}=-\frac{|\gamma|}{\gamma}=-1$,
		which meets the above-mentioned condition $\gamma=-|\gamma|$ as a spectial case
		of $\gamma=\mathcal{1}$,
		and also, now we have ${\rm ln}\frac{1}{z}\rightarrow 0$ 
		i.e., treating $\frac{1}{z}
		={\rm Exp}[\partial_{|\gamma|}^{(2)}\frac{\gamma}{|\gamma|}]$
		as the unit quantity which cannot be further departed (or derivated), and in the mean time
		$|\gamma|\rightarrow \frac{1}{2}$,
		
		We will see that, in this case, the above-obtained result $\frac{\partial |\gamma|}{\partial |\gamma|}=-1=\gamma/|\gamma|$ can be verified once more by the result
		$\frac{\partial_{|\gamma|}\frac{\mathds{1}'}{\mathds{1}}}
		{\partial_{|\gamma|}(\frac{\mathds{1}'}{\mathds{1}}-1)}=1$
		which can be obtained by using the following limiting property of polylogarithm function,
		\begin{equation}
			\begin{aligned}
				\lim_{|Y|\rightarrow 0}\frac{{\rm Li}_{s}(Y)}{Y}=1,
			\end{aligned}
		\end{equation}
		where we define $|Y|:=\frac{{\rm ln}\frac{-1}{\mathcal{X}}}{i\pi}=\frac{Y}{e^{i{\rm Arg}(Y)}}$,
		and since the left-hand-side of this equation can be replaced by
		$\partial_{|Y|}{\rm Li}_{s+1}(Y)
		=1-\frac{{\rm Li}_{s+1}(Y)}{\lim_{|Y|\rightarrow\infty}{\rm Li}_{s+1}(Y)}$,
		we can have polylogarithm function ${\rm Li}_{s'}(Y)=\frac{1}{i\pi}=[{\rm ln}(-1)]^{-1}$
		by choosing a proper order $s'$,
		then $|Y|$ can be departed as
		\begin{equation}
			\begin{aligned}
				\lim_{|Y|\rightarrow 0}
				|Y|=\frac{1}{i\pi}(-{\rm ln}(-1)-{\rm ln}\mathcal{X})
				=\frac{{\rm Li}_{s'}Y}{\lim_{Y\rightarrow\infty}{\rm Li}_{s'}Y}
				=\frac{Y}{e^{i{\rm Arg}(Y)}}
				=1-\partial_{Y}{\rm Li}_{s'}Y=0.
			\end{aligned}
		\end{equation}
		Thus in this limit, the Bernoulli polynomial in Eq.(\ref{op08})
		reduces to 
		$B_{0}(\frac{1}{2})$
		where $\frac{1}{2}$ is the above-mentioned limting result,
		as a ratio between ${\rm ln}(-1)$ ($-1=-\frac{|\gamma|}{|\gamma|}$
		is the negative unit in representation of $m$) and ${\rm ln}1=0^{+}$
		(i.e., zero in representation of $\gamma$, whose unit quantity are defined according to
		the independence with the variable $\gamma$),
		\begin{equation}
			\begin{aligned}
				\frac{1}{2}
				=\frac{{\rm ln}(-1)}{\frac{|\gamma|}{\lim_{|\gamma|\rightarrow \infty}|\gamma|}{\rm ln}(-1)}=\frac{{\rm ln}(-1)}
				{{\rm ln}(-1)^{2}}.
			\end{aligned}
		\end{equation}
		Here we present some relations derived from the results $\frac{\partial \gamma}{\partial |\gamma|}=1=\frac{-\gamma}{|\gamma|}$,
		and $\frac{\partial |\gamma|}{\partial |\gamma|}=-1$.
		Firstly note that the representation of Bernoulli polynomial requires $\gamma=\frac{|\gamma|}{-1}$ instead of $-|\gamma|$,
		then the identity $\frac{\partial \gamma}{\partial |\gamma|}=1$ can be written in another form,
		\begin{equation}
			\begin{aligned}
				\frac{\partial \frac{|\gamma|}{-1}}{\partial |\gamma|}=\frac{1+|\gamma|\partial_{|\gamma|\frac{1}{-1}}}{(-1)^{2}}=1,
			\end{aligned}
		\end{equation}
		where the $\gamma$-dependent identities follows
		$\partial_{|\gamma|}(\pm 1)=\frac{\mp 1}{|\gamma|},\ \lim_{|\gamma|\rightarrow\infty}(\mp 1)
		=\frac{|\gamma|}{|\gamma|\pm 1}$,
		and the term $\frac{1}{-1}=(-1)^{-1}$ satisfies
		\begin{equation}
			\begin{aligned}
				|\gamma|\partial_{|\gamma|}(-1)^{-1}
				=\frac{{\rm ln}(-1)}{-1}+(-1)^{-\mathds{1}}=0,
			\end{aligned}
		\end{equation}
		where $(-1)^{-\mathds{1}}=-\frac{{\rm ln}(-1)}{-1}$, and thus $(-1)^{-\mathds{1}+1}=-{\rm ln}(-1)={\rm ln}(-1)$.
		Then it is easy to check that $\lim_{|\gamma|\rightarrow \infty}|\gamma|=\frac{1}{2}|\gamma|$,
		with $|\gamma|\partial_{|\gamma|}\frac{1}{2}=\frac{1}{2}$.
		As we can see, in this representation, all quantities are defined according to their dependence between each other.
		We again by treated $\frac{1}{2}$ here as a polynomial function at a certain order $a$, then we have
		$\frac{1}{2}={\rm Li}_{a}(|\gamma|)=|\gamma|\partial_{|\gamma|}\frac{1}{2}={\rm Li}_{a-1}(|\gamma|)$,
		and obviously this can only happen in the limit $|\gamma|\rightarrow 0$ (and infinitely large order $a$).
		It can be verified that, in this limit, the $\gamma$-dependence of the factor $\frac{|\gamma|}
		{\lim_{|\gamma|\rightarrow\infty}|\gamma|}=2$ disappears
		as a result of $\partial_{|\gamma|}\frac{|\gamma|}
		{\lim_{|\gamma|\rightarrow\infty}|\gamma|}=|\gamma|\partial_{|\gamma|}\frac{2}{|\gamma|}-\frac{2}{|\gamma|}
		=\partial_{|\gamma|}(\frac{1}{2})^{-1}=0$.
		
		Thus the restriction Bernoulli polynomial in terms of $-{\rm ln}(-\mathcal{X})\equiv {\rm ln}(\frac{-1}{\mathcal{X}})$
		indeed reflect another limiting transform $\lim_{|\gamma|\rightarrow 0}\frac{-\gamma}{|\gamma|}={\rm ln}|\gamma|$
		which does not contrast with $\lim_{|\gamma|\rightarrow -\gamma}\frac{-\gamma}{|\gamma|}=1$,
		but just corresponds to another unit quantity,
		and we can use a simple expression to generalize the key feature in this representation,
		which is $-1=\partial_{|\gamma|}|\gamma|=|\gamma|\partial_{|\gamma|}{\rm ln}|\gamma|$.
		For example,
		this limiting transform can be verified by substituting 
		\begin{equation}
			\begin{aligned}
				\partial_{|\gamma|}\frac{1}{|\gamma|}=|\gamma|^{-1-\mathds{1}}({\rm ln}|\gamma|-1)=0,
			\end{aligned}
		\end{equation}
		into the derivative $\partial_{|\gamma|}\frac{\gamma}{|\gamma|}$ which still turns out to be $\frac{-1}{|\gamma|}\frac{\gamma}{|\gamma|}$.
		Note that here we still use the notion $\mathds{1}$ to denotes the unit quantity ($\gamma$-independent) in this configuration,
		which satisfies $\gamma^{-1}=|\gamma|^{-\mathds{1}}$ and thus 
		\begin{equation}
			\begin{aligned}
				&\gamma=\frac{|\gamma|{\rm ln}|\gamma|}{-1},\\
				&\gamma+1=\frac{1}{-1}=\frac{|\gamma|{\rm ln}|\gamma|}{-1},\\
				&{\rm ln}|\gamma|=-|\gamma|^{\mathds{1}-1}.
			\end{aligned}
		\end{equation}
		In this configuration, the relation between $\gamma$ and $|\gamma|$ can be described
		by the Lambert $\mathcal{W}$-function.
		We treating $|\gamma|$ as a complex quantity $|\gamma|=\mathcal{W}(|\gamma|)e^{\mathcal{W}(|\gamma|)}$ 
		where $|\gamma|=\gamma e^{-i{\rm Arg}\gamma}\rightarrow \gamma e^{\gamma}$,
		and the $\gamma$ is defined as a Lambert $\mathcal{W}$-function of $|\gamma|$,
		which satisfies
		\begin{equation}
			\begin{aligned}
				\mathcal{W}(|\gamma|):=\gamma
				={\rm ln}(-1)-{\rm ln}{\rm ln}|\gamma|={\rm ln}|\gamma|-{\rm ln}\gamma
				=-{\rm ln}|\gamma|-1,
			\end{aligned}
		\end{equation}
		and ${\rm ln}\gamma=\frac{1-\gamma}{\gamma}$.
		Relating the formulas of Lambert function, we have more another
		expression of ${\rm ln}(-1)$,
		\begin{equation}
			\begin{aligned}
				{\rm ln}(-1)
				={\rm ln}|\gamma|-{\rm ln}{\rm ln}|\gamma|+\sum^{\infty}_{m=1}
				(\sum_{k=1}^{m}
				\frac{(-{\rm ln}{\rm ln}|\gamma|)^{k}}{k!}S[m,m-k+1])(\frac{-1}{{\rm ln}|\gamma|})^{m},
			\end{aligned}
		\end{equation}
		where $S[m,m-k+1]$ is a stirling number.
		We will see that,
		the transition $\frac{-\gamma}{|\gamma|}\rightarrow {\rm ln}|\gamma|$
		is equivalents to $-{\rm ln}\frac{\gamma}{|\gamma|}\rightarrow \gamma$,
		which is consistent with the identity of Lambert function
		$e^{\mathcal{W}(|\gamma|)}=\frac{|\gamma|}{\mathcal{W}(|\gamma|)}$,
		and the term ${\rm ln}{\rm ln}|\gamma|={\rm ln}(-1)-\gamma$ can not be soly identitied.
		By virtue of Lambert function,
		we have $\frac{-\mathcal{W}(-{\rm ln}|\gamma|)}{{\rm ln}|\gamma|}
		=\frac{{\rm ln}{\rm ln}|\gamma|+1}{{\rm ln}|\gamma|}=
		-e^{-\mathcal{W}(-{\rm ln}|\gamma|)}$ equals to the infinite power tower of $|\gamma|$.
		Then the derivative $\frac{\partial \gamma}{\partial |\gamma|}=\frac{-\gamma}{|\gamma|}$ is equivalents to
		\begin{equation}
			\begin{aligned}
				\frac{\partial \gamma}{\partial |\gamma|}={\rm ln}|\gamma|=\gamma+{\rm ln}\gamma
				=\frac{1}{e^{\gamma}(1+\gamma)}.
			\end{aligned}
		\end{equation}
		According to the integer formula of Lambert function
		\begin{equation}
			\begin{aligned}
				\label{cooo}
				\int \gamma d|\gamma|=|\gamma|(\gamma-1+\frac{1}{\gamma})+{\rm cont}.,
			\end{aligned}
		\end{equation}
		which can be verified by the derivative
		\begin{equation}
			\begin{aligned}
				\frac{\partial |\gamma|(\gamma-1+\frac{1}{\gamma})}{\partial |\gamma|}
				=|\gamma|\partial_{|\gamma|}{\rm ln}(-1)-{\rm ln}|\gamma|=\gamma.
			\end{aligned}
		\end{equation}
		This integer indeed corresponds to the limiting result of $|\gamma|\rightarrow 0$,
		where we have 
		\begin{equation}
			\begin{aligned}
				\lim_{|\gamma|\rightarrow 0}\frac{{\rm ln}|\gamma|}{\gamma}\rightarrow 1,
			\end{aligned}
		\end{equation}
		which reveals the Lambert functions here are not the principal branch.
		This limting result indeed equivalents to the integer formula
		\begin{equation}
			\begin{aligned}
				\lim_{|\gamma|\rightarrow 0}[\int\frac{-\gamma}{|\gamma|}]
				=\lim_{|\gamma|\rightarrow 0}[-\gamma-\frac{\gamma^{2}}{2}-\cdots]
				=\int {\rm ln}|\gamma|d|\gamma|=1,
			\end{aligned}
		\end{equation}
		as $\gamma\rightarrow -\infty $ as $|\gamma|\rightarrow 0$.
		So we can see that this process is to remove the dependence
		on $|\gamma|$ for the unit $1=\lim_{|\gamma|\rightarrow -\gamma}
		\frac{-\gamma}{|\gamma|}$.
		This can be clarified by writting the integer about
		${\rm ln}|\gamma|$ as a polylogarithm function of $\gamma$,
		${\rm Li}_{s+1}(\gamma)$, then $\partial_{\gamma}{\rm Li}_{s+1}(\gamma)
		=\frac{{\rm ln}|\gamma|}{\gamma}\partial_{\gamma}{\rm Li}_{s+1}(\gamma)
		=\partial_{|\gamma|}{\rm Li}_{s+1}(\gamma)$.
		Relating this to Eq.\ref{cooo},
		we obtain $\lim_{|\gamma|\rightarrow 0}(\gamma+\frac{1}{\gamma})\rightarrow -\gamma$,
		in which case $\partial_{|\gamma|}\frac{1}{|\gamma|}\rightarrow e^{-\gamma}\partial_{\gamma}(-\gamma)=0$.
		So we have $|\gamma|\partial_{|\gamma|}(\gamma-(1-\frac{1}{\gamma}))=1-\frac{1}{\gamma}$,
		then using the limit 
		\begin{equation}
			\begin{aligned}
				\lim_{\gamma\rightarrow -\infty}(1-\frac{1}{\gamma})^{-\gamma-1}=e,
			\end{aligned}
		\end{equation}
		we have
		\begin{equation}
			\begin{aligned}
				&\frac{\partial (-\gamma-1)}{\partial (-\gamma)}{\rm ln}(1-\frac{1}{\gamma})
				=\frac{\partial \gamma}{\partial |\gamma|}\frac{-\gamma-1}{e^{1-\gamma}}
				=-e^{-1},\\
				&\frac{\partial (-\gamma)}{\partial (1-\frac{1}{\gamma})}
				=(-\gamma-1)(1-\frac{1}{\gamma})^{-\gamma-2}
				=\lim_{
					\frac{\partial (-\gamma-1)}{\partial (1-\frac{1}{\gamma})}\rightarrow 0
				}
				\frac{\partial (1-\frac{1}{\gamma})^{-\gamma-1}}{\partial (1-\frac{1}{\gamma})},
			\end{aligned}
		\end{equation}
		where 
		\begin{equation}
			\begin{aligned}
				\frac{\partial (-\gamma-1)}{\partial (1-\frac{1}{\gamma})}
				=\frac{\partial -\gamma}{\partial (1-\frac{1}{\gamma})}+\frac{\partial (-1)}{\partial (1-\frac{1}{\gamma})}
				=-\gamma-(1-\frac{1}{\gamma})^{-\gamma-1}.
			\end{aligned}
		\end{equation}
		Thus we
		can also obtain
		\begin{equation}
			\begin{aligned}
				\frac{\partial (-1)}{\partial |\gamma|}
				=[(-\gamma-(1-\frac{1}{\gamma})^{-\gamma-1})-(-\gamma-1)(1-\frac{1}{\gamma})^{-\gamma-2})]
				[\frac{e^{-\gamma}}{\gamma+1}-\frac{1}{|\gamma|}(1-\frac{1}{\gamma})].
			\end{aligned}
		\end{equation}

		In the mean time, under representation of the Bernoulli polynomial,
		this limiting result has another identity,
		\begin{equation}
			\begin{aligned}
				\frac{1}{2}=\lim_{|Y|\rightarrow 0}\frac{1}{Y\partial_{Y}\frac{Y}{|Y|}},
			\end{aligned}
		\end{equation}
		and this is connected to the another
		familiar relation $\frac{\partial \gamma}{\partial |\gamma|}=\frac{-\gamma}{|\gamma|}=1$
		(as the effect of gauge-invariance can be seem everywhere)
		by a similar relation
		\begin{equation}
			\begin{aligned}
				\frac{\partial Y}{\partial |Y|}
				=\frac{1}{\frac{|Y|}{Y}+Y\partial_{Y}\frac{|Y|}{Y}}=1.
			\end{aligned}
		\end{equation}
		Instituting $|Y|=\frac{Y}{e^{-i{\rm Arg}Y}}$,
		we can further obtain the relation (still using the form we often used in above)
		\begin{equation}
			\begin{aligned}
				\partial_{Y}{\rm ln}\frac{Y}{|Y|}
				=\frac{1-e^{-i{\rm Arg}Y}}{Y e^{-i{\rm Arg}Y}}
				=1-\frac{{\rm ln}\frac{Y}{|Y|}}{\lim_{Y\rightarrow\infty}{\rm ln}\frac{Y}{|Y|}}
				=\frac{Y}{|Y|}[1-\frac{\frac{Y}{|Y|}}{\lim_{Y\rightarrow\infty}\frac{Y}{|Y|}}],
			\end{aligned}
		\end{equation}
		where the limiting term as well as the derivative term in the last line can be obtained as
		\begin{equation}
			\begin{aligned}
				\label{90u}
				\lim_{Y\rightarrow\infty}\frac{Y}{|Y|}
				=\frac{\frac{Y}{|Y|}}{1-\partial_{Y}\frac{Y}{|Y|}}
				=\frac{\frac{Y}{|Y|}}{1-(\frac{1}{Y}-\frac{|Y|}{Y^{2}})}.
			\end{aligned}
		\end{equation}
		Instituting the result ${\rm ln}(-1)=\frac{1}{2}0^{+}$ into the above limit expression
		($|Y|\rightarrow 0$)
		${\rm ln}\frac{-1}{\mathcal{X}}=-{\rm ln}(-\mathcal{X})=0$ (and thus ${\rm ln}(-\mathcal{X})=0^{-}$),
		we have ${\rm ln}(-1)+{\rm ln}\mathcal{X}=\frac{1}{2}0^{+}+{\rm ln}\mathcal{X}=0^{-}$,
		thus we can know ${\rm ln}\mathcal{X}=\frac{3}{2}0^{-}$.
		This can be explained by the identity of 
		$\mathcal{X}=\frac{1}{\mathds{1}\partial_{|\gamma|}(-1)}=\frac{1}{\mathds{1}(-|\gamma|){\rm ln}
			\frac{1}{z}}$ which is of the representation of $\gamma$,
		where $1(=\frac{|\gamma|}{|\gamma|})$ should not be treated as an invariant unit.
		As a result,
		the actual value of ${\rm ln}\mathcal{X}$ can not be uniquely identified, for example,
		${\rm ln}\mathcal{X\cdot 1}=\mathcal{X}=2{\rm ln}(-1)+\frac{3}{2}0^{-}=\frac{1}{2}0^{-}$.
		We can comparing the Eq.(\ref{90u}) and the result $\frac{{\rm ln}(-\mathcal{X})}{{\rm ln}\mathcal{X}}=\frac{2}{3}$
		with the Eq.(\ref{xs1}) and Eq.(\ref{0666}), respectively, to see this connection.

		Thus the expression with the bracket of Bernoulli polynomial 
		can be explained in a more clear way in terms of the function $f_{Y}:=\frac{1}{Y\partial_{Y}\frac{Y}{|Y|}}$,
		\begin{equation}
			\begin{aligned}
				B_{0}[\frac{1}{2}(1-\frac{{\rm ln}\frac{-1}{\mathcal{X}}}{i\pi})]=
				B_{0}[\frac{1}{2}(1-\frac{-{\rm ln}(-\mathcal{X}}{i\pi})]
				=B_{0}[f_{Y}]
				=B_{0}[\lim_{|Y|\rightarrow 0}f_{Y}
				(1-\partial_{\frac{{\rm ln}(-1)}{{\rm ln}(-1)-{\rm ln}\mathcal{X}} } f_{Y}],
			\end{aligned}
		\end{equation}
		where $\lim_{|Y|\rightarrow 0}$ is equivalents to 
		$\lim_{{\rm ln}(-1)\lim_{Y\rightarrow\infty}{\rm Li}_{s'}Y\rightarrow
			=\frac{{\rm ln}(-1)}{-{\rm ln}(-1)-{\rm ln}\mathcal{X}} \infty}$. 
		Then as a special case of Eq.(\ref{op08}) at $\mathcal{X}=-\frac{|\gamma|}{|\gamma|}=-1$,
		we have 
		\begin{equation}
			\begin{aligned}
				-{\rm Li}_{0}(-1)-B_{0}(\frac{1}{2})
				=-{\rm Li}_{0}(-1)-1
				=-e^{-{\rm Li}_{1}(-1)}=\frac{-1}{2},
			\end{aligned}
		\end{equation}
		which also reveals teh correctness of Eq.(\ref{op08}).

		In this case, ${\rm ln}\frac{1}{z}=-{\rm ln}{\rm ln}\frac{-\gamma}{|\gamma|}$
		equivalents to the double partial derivatives $\partial^{(2)}_{|\gamma|}\frac{\gamma}{|\gamma|}$,
		and the limiting result ${\rm ln}(\frac{1}{z})^{\gamma-|\gamma|}=0$
		can be explained by the operator transform ${\rm ln}(\frac{1}{z})
		=\partial^{(2)}_{|\gamma|}\frac{\gamma}{|\gamma|}=
		\partial_{|\gamma|}^{(2)}(i\pi(2m+1))=\lim{2|\gamma|\rightarrow 0^{+}}$.
		Thus the unit quantity in $m$-representation can be expressed as
		$\frac{\gamma}{|\gamma|}+\gamma \partial_{|\gamma|}^{(2)}$.
		Also,
		the two representations in the limiting case can be related
		by the above-mentioned polylogarithm functions by constructing the Kummer's function
		in proper form,
		\begin{equation}
			\begin{aligned}
				{\rm Li}_{-1}(\frac{1}{z})=
				{\rm Li}_{-1}(e^{\mathds{P}})\mathds{P}^{-2}\sum_{m=0}^{\mathds{P}-1}
				{\rm Li}_{-1}(e^{1+\frac{2\pi im}{\mathds{P}}})
			\end{aligned}
		\end{equation}
		where we define the limiting form of ${\rm ln}\frac{1}{z}$
		as
		\begin{equation}
			\begin{aligned}
				\mathds{P}=\lim_{|\gamma|\rightarrow -\gamma}{\rm ln}\frac{1}{z}=
				\partial^{(2)}_{|\gamma|}\frac{\gamma}{|\gamma|}=-{\rm ln}{\rm ln}\frac{-\gamma}{|\gamma|}.
			\end{aligned}
		\end{equation}

	\end{large}
	\renewcommand\refname{References}
	
	\clearpage
	
	\begin{figure}
		\centering
		\includegraphics[width=0.9\linewidth]{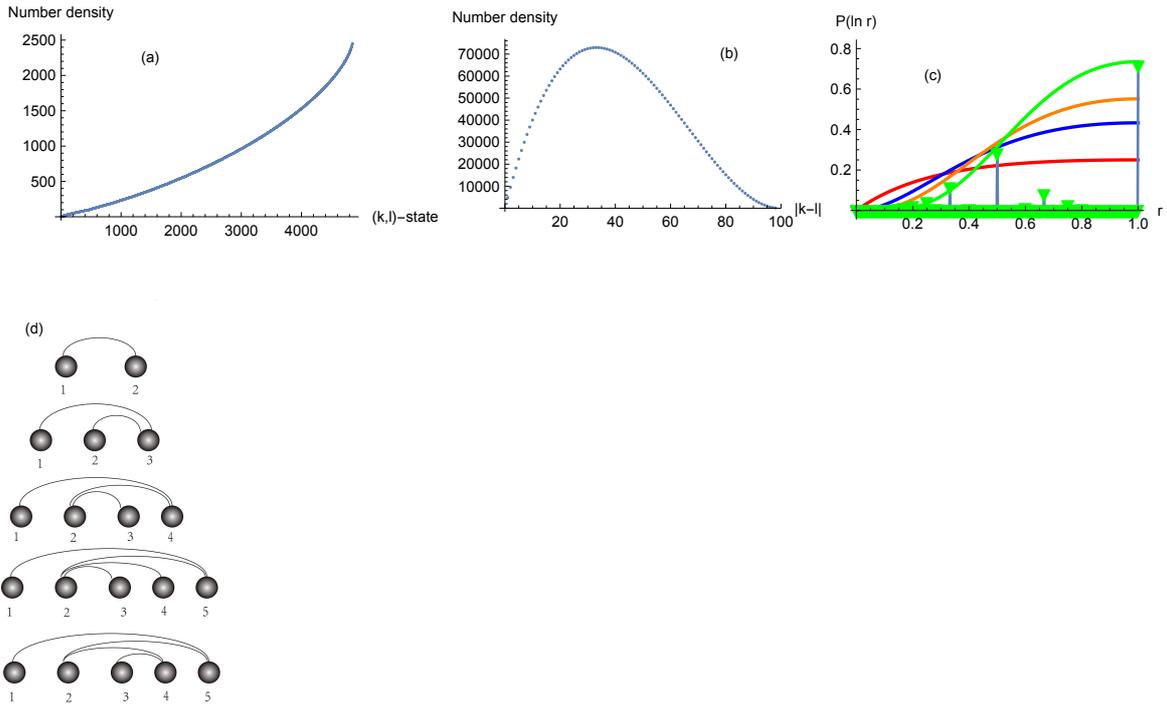}
		\caption{
			(a) States (quasiparticle) number (due to the multiplications of $(l-k)|k-1|$ $(i,j)$ states) for each $(k,l)$ state 
			(with mutually distinguishing $(k,l)$ indices).
			(b) The number density for each $|k-l|$ value,
			(i.e., classifying the $(k,l)$ states in panel (a) according to teh value of $|k-l|$).
			(c) The many-body level spacing distribution corresponds to panel (a).
			The red, blue, orange, and green lines correspond
			to the standard Poisson, GOE, GUE and GSE curves.
			(d) schematic diagram for the topological modes corresponding to the $|k-l|=1$ case as we discuss in Sec.4.3.
			The diagrams in this panel all equivalent to the mode $\chi_{1}\chi_{2}$
			($=\chi_{1}\chi_{3}+\chi_{2}\chi_{3}=\chi_{1}\chi_{4}+\chi_{2}\chi_{4}+\chi_{2}\chi_{3}
			=\chi_{1}\chi_{5}+\chi_{2}\chi_{5}+\chi_{2}\chi_{4}+\chi_{2}\chi_{3}=\cdots $),
			and note that we have $\chi_{2}\chi_{3}=\chi_{3}\chi_{4}=\cdots$ according to Sec.4.3.}
	\end{figure}

	\begin{figure}
		\centering
		\includegraphics[width=0.8\linewidth]{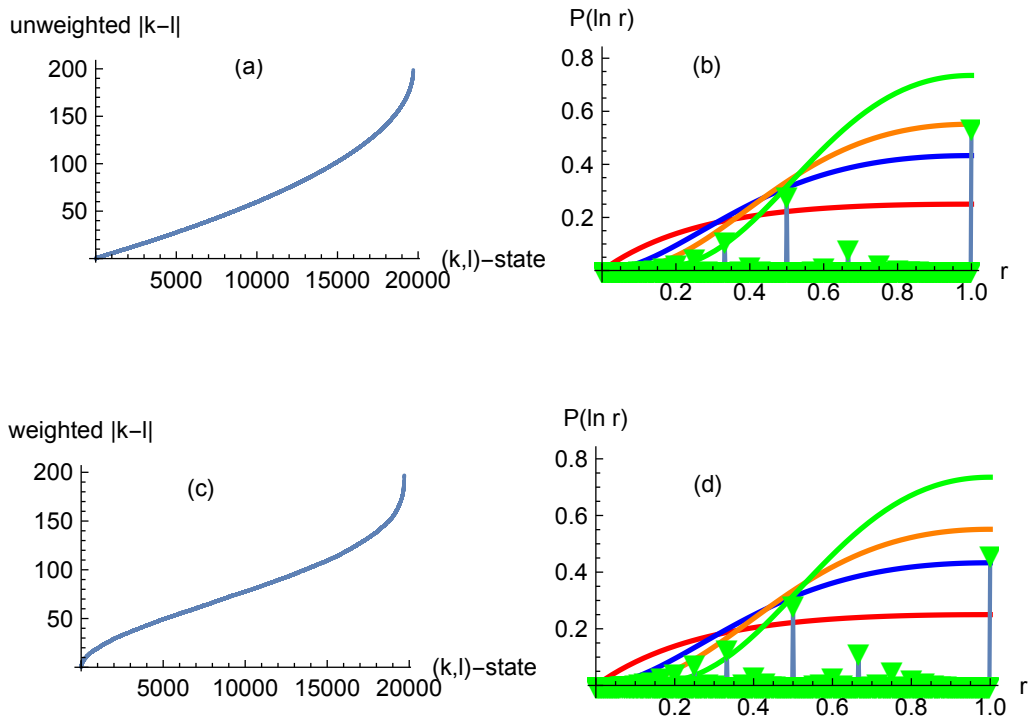}
		\caption{The value of $|k-l|$ for unweighted (a) and weighted (c)
			$(k,l)$-state. The weight distribution for each $(k,l)$ state is shown in Fig.1(a).
			(b) and (d) are the many-body level spacing distribution correspond to panel (a) and (c),
			respectively.
		}
	\end{figure}

	\begin{figure}
		\centering
		\includegraphics[width=0.6\linewidth]{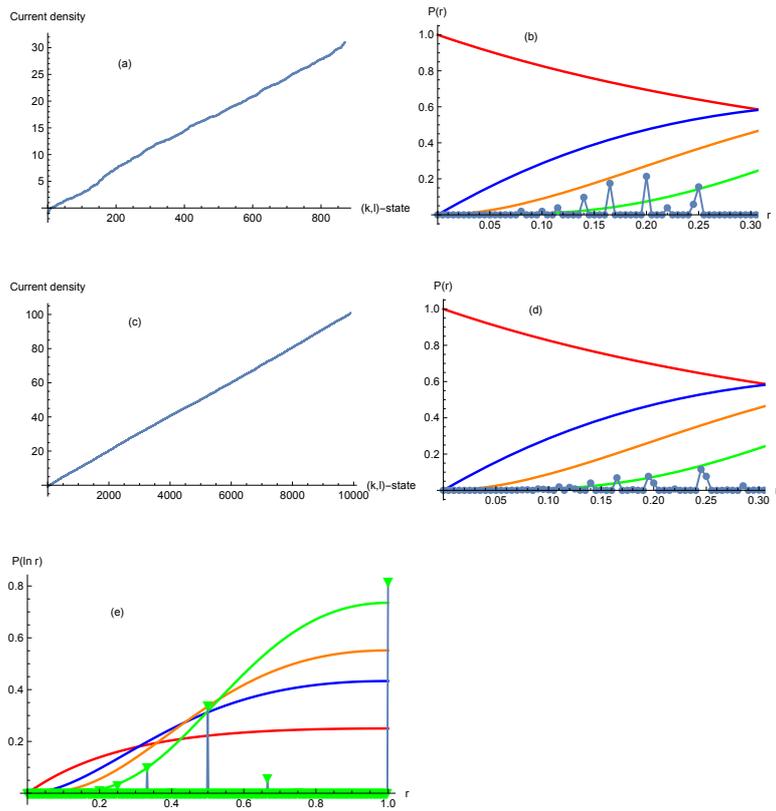}
		\caption{Current density spectrum (a,c) and GSE distribution of the level
			spacing ratios (b,d) for three-DOF configuration. We set $N=30$ and $N=100$ in (a,b)
			and (c,d), respectively. The red, blue, orange, and green lines correspond
			to the standard Poisson, GOE, GUE and GSE curves.
			The many-body level statistics in term of $P({\rm ln}r)$ which corresponds to the panel (d)
			is shown in (e).}
	\end{figure}

	\begin{figure}
		\centering
		\includegraphics[width=0.6\linewidth]{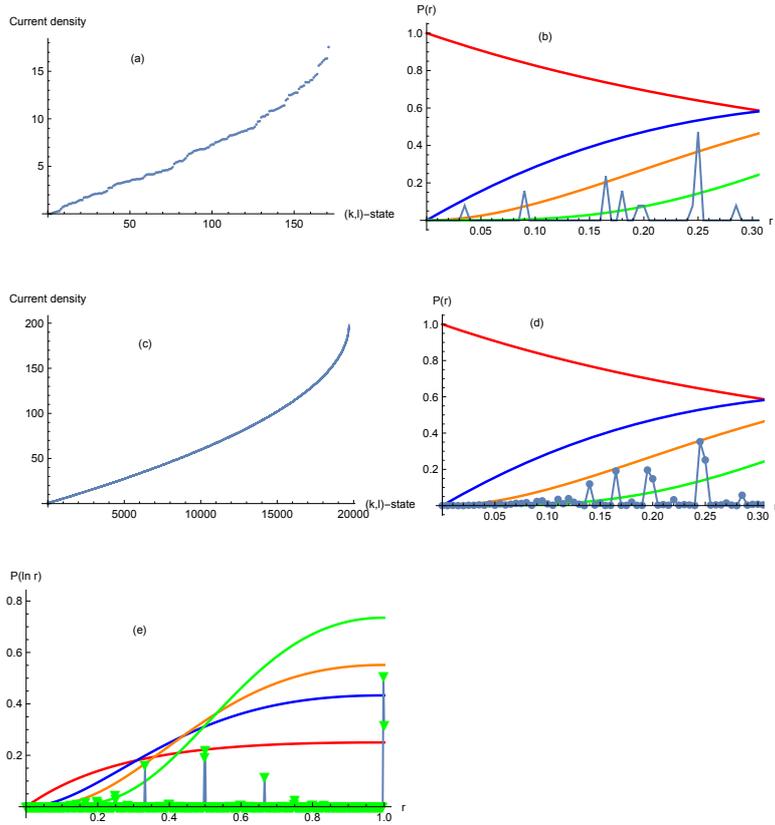}
		\caption{Current density spctrum (a,c) and GUE distribution of the level
			spacing ratios (b,d) for four-DOF configuration. We set $N=20$ and $N=200$ in (a,b)
			and (c,d), respectively.The many-body level statistics in term of $P({\rm ln}r)$ which corresponds to the panel (d)
			is shown in (e).}
	\end{figure}
	
	\clearpage
	
	\begin{figure}
		\centering
		\includegraphics[width=0.6\linewidth]{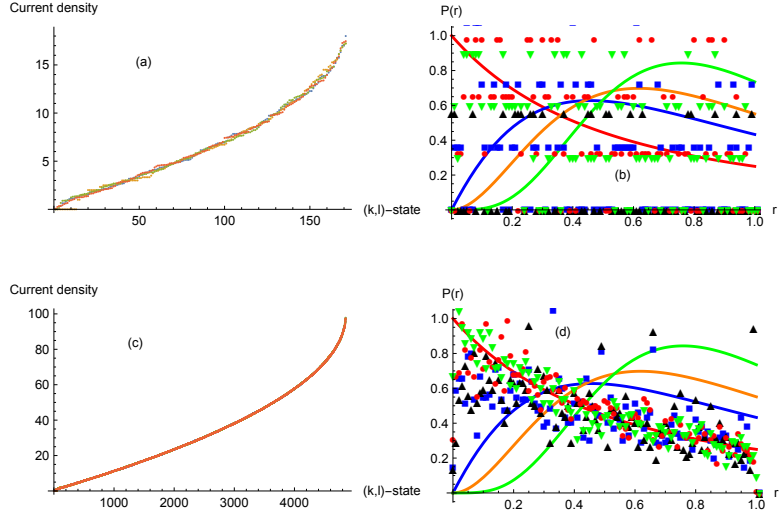}
		\caption{Current density spctrum (a,c) and Poisson distribution of the level
			spacing ratios (b,d) for localized configuration. We set $N=20$ and $N=200$ in (a,b)
			and (c,d), respectively.
			The black up-triagnle, red circle, blue square, and green down-triangle
			correspond to
			the configurations simulated using the randomness factors
			$\updelta_{1},\updelta_{2},\updelta_{2}$ and $\updelta_{4}$, respectively,
			where we can obvious see the distriution of blue squares, and green down-triangles 
			follows critically the Poisson distribution,
			and their corresponding averaged level spacing ratio are 
			$\langle r\rangle_{\updelta_{3}}=0.3954$ and 
			$	\langle r\rangle_{\updelta_{4}}=0.3853$, respectively.}
	\end{figure}

	\begin{figure}
		\centering
		\includegraphics[width=0.6\linewidth]{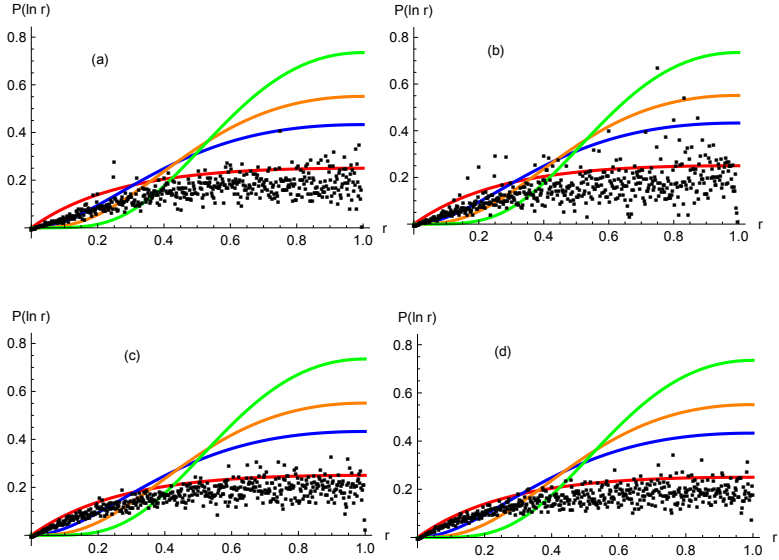}
		\caption{
			The many-body level statistics in term of $P({\rm ln}r)$ which corresponds to the last panel of Fig.6,
			where we obtain the similar result with Fig.6,
			with
			$\langle r\rangle_{\updelta_{1}}=0.4715,\ 
			\langle r\rangle_{\updelta_{2}}=0.4532,\ 
			\langle r\rangle_{\updelta_{3}}=0.3975,\ 
			\langle r\rangle_{\updelta_{4}}=0.387$.}
	\end{figure}

	\begin{figure}
		\centering
		\includegraphics[width=0.6\linewidth]{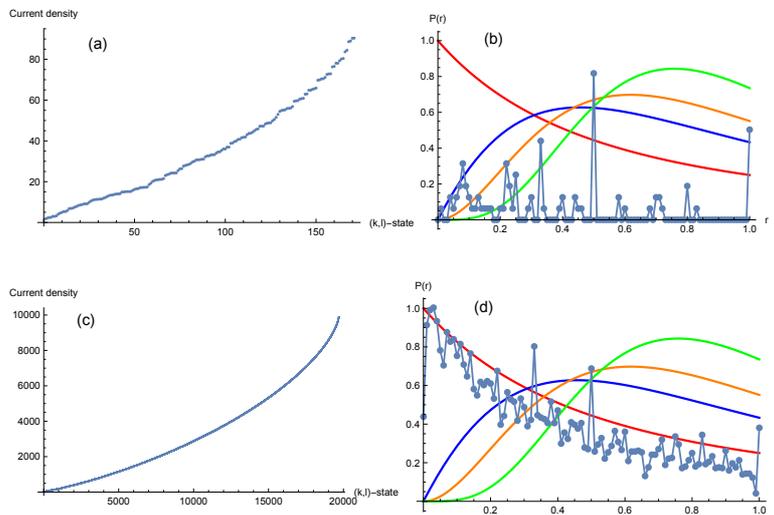}
		\caption{The automatically-realized distribution beyond Poisson ensemble
			with $O(M)\sim O(N)$. We set $N=20$ and $N=200$ in (a,b)
			and (c,d), respectively.
			The corresponding averaged level spacing ratio is 
			$\langle r\rangle=0.3623$, which is lower than the Poisson case and indicating a more localized (due to the
			strong disorder and interactions) system.}
	\end{figure}
	
\end{document}